\documentclass[12pt,preprint]{aastex}
\usepackage{epsfig,multirow,booktabs}

\newcommand{\nuc}[2]{${}^{#2} \rm #1$}

\newcommand{\text}[1]{\rm #1}

\slugcomment{}

\shorttitle{Multimessenger Astronomy of Core-Collapse Supernovae}
\shortauthors{Kotake et al.}

\begin{document}

\title{Multimessengers from core-collapse supernovae : multidimensionality as a key to bridge theory and observation}

\author{Kei Kotake\altaffilmark{1,2}, Tomoya Takiwaki\altaffilmark{2}, Yudai Suwa\altaffilmark{3}, Wakana Iwakami Nakano\altaffilmark{4}, \\ Shio Kawagoe\altaffilmark{5}, Youhei Masada\altaffilmark{6}, and Shin-ichiro Fujimoto\altaffilmark{7}}
\affil{$^1$Division of Theoretical Astronomy, National Astronomical Observatory of Japan, 2-21-1, Osawa, Mitaka, Tokyo, 181-8588, Japan}
\email{kkotake@th.nao.ac.jp}
\affil{$^2$Center for Computational Astrophysics, National
  Astronomical Observatory of Japan, Mitaka, Tokyo 181-8588, Japan}
\affil{$^3$Yukawa Institute for Theoretical Physics, Kyoto
  University, Oiwake-cho, Kitashirakawa, Sakyo-ku, Kyoto, 606-8502,
  Japan}
\affil{$^4$Department of Aerospace Engineering, Tohoku University,
6-6-01 Aramaki-Aza-Aoba, Aoba-ku, Sendai, 980-8579, Japan}
\affil{$^5$Knowledge Dissemination Unit, Oshima Lab, Institute of Industrial Science.
 The Univesity of Tokyo, 4-6-1 Komaba, Meguro-ku, Tokyo 153-8505, Japan}
\affil{$^6$Department of Computational Science, Graduate School of System Informatics, Kobe University; Nada, Kobe 657-8501}
\affil{$^7$Kumamoto National College of Technology, 
 2659-2 Suya, Goshi, Kumamoto 861-1102, Japan
}
\begin{abstract}
Core-collapse supernovae are dramatic explosions marking the
catastrophic end of massive stars. 
 The only means to get direct information about the supernova engine
is from observations of neutrinos emitted by the forming neutron star,
 and through gravitational waves which are produced when the hydrodynamic
 flow or the neutrino flux is not perfectly spherically symmetric. The
 multidimensionality of the supernova engine, which breaks the sphericity of the 
 central core such as convection, rotation, magnetic fields, and 
 hydrodynamic instabilities of the supernova shock, is attracting great attention
 as the most important ingredient to understand the long-veiled explosion mechanism.
 Based on our recent work, 
 we summarize properties of gravitational waves, neutrinos, 
and explosive nucleosynthesis obtained in a series of 
our multidimensional hydrodynamic simulations and discuss how 
the mystery of the central engines can be unraveled by deciphering 
these multimessengers produced under the thick veils of massive stars. 
\end{abstract}

\clearpage

\section{Introduction}
The majority of stars more massive than $\sim$8 $M_{\odot}$ end their lives as 
core-collapse supernovae. They have long attracted the attention of
astrophysicists because they have many facets playing important roles
in astrophysics. They are the mother of neutron stars as well as black
holes; they play an important role for acceleration of cosmic rays; they influence
galactic dynamics triggering further star formation; they are
 gigantic emitters of neutrinos and gravitational waves.
 They are also a major site for nucleosynthesis, so, naturally, any attempt 
to address human origins may need to begin
with an understanding of core-collapse supernovae (CCSNe). 

Current estimates of CCSN rates in our Galaxy predict one 
  event every $\sim 40 \pm 10$ year 
 \citep{ando_new}. When a CCSN event occurs 
in our galactic center, copious numbers of neutrinos are produced, 
some of which may be detected on the earth. Such ``supernova neutrinos'' will carry 
valuable information from deep inside the core. 
In fact, the detection of neutrinos from SN1987A 
(albeit in the Large Magellanic Cloud)
 opened up the {\it Neutrino Astronomy}, which is an alternative to 
conventional astronomy by electromagnetic waves \citep{hirata1987,bionta1987}.
 Even though there were just two dozen neutrino events from SN1987A, these 
events have 
been studied extensively (yielding $\sim$ 500 papers) and have allowed us to 
have a confidence that our basic picture of the supernova 
 physics is correct (e.g., \citet{Sato-and-Suzuki}, see
 \citet{raffelt12,raffelt_review} for a recent review).
 Recently significant progress has been made in the large water 
\u{C}herenkov detectors such as Super-Kamiokande \citep{totsuka} and 
IceCube \citep{icecube}, and also in the liquid scintillator detector as KamLAND \citep{kamland1}.
If a supernova occurs in our Galactic center ($\sim 10$ kpc), 
 about $10^5$ $\bar{\nu}_e$ events are estimated to be detectable by IceCube 
 (e.g., \citet{icecube11} and references therein).
 Those successful neutrino detections 
are important not only to study the supernova physics but also to unveil 
 the nature of neutrinos itself such as the neutrino oscillation parameters  
and the mass hierarchy (e.g., \citet{raffelt12} for a recent review).

 CCSNe are now about to start even another astronomy,
  {\it Gravitational-Wave Astronomy}.
 Currently long-baseline laser interferometers such as
LIGO (USA, e.g., \citet{firstligonew}),
VIRGO (Italy)\footnote{http://www.ego-gw.it/},
GEO600 (Germany)\footnote{http://geo600.aei.mpg.de/},
and TAMA300 (Japan, e.g., \citet{tamanew}) are currently operational and preparing 
 for the first observation (see, e.g., \citet{hough} for a recent review), 
 by which the prediction by Einstein's theory of General Relativity 
(GR) can be confirmed.
 These instruments are being updated to their {\it Advanced} status, and may start 
 taking data, possibly detecting GWs for the first time, as soon as 2015 (see 
\cite{vandel} for a recent review).
 In fact, {\it Advanced }LIGO/VIRGO, which is an upgrade of the initial LIGO and VIGRO,
 are expected to be completed by 2015 and will increase the observable detection volume
 by a factor of $\sim 1000$ \citep{harry}. The Large-scale Cryogenic Gravitational-wave
 Telescope (LCGT, \citet{lcgt}) in Japan was funded in late 2010, 
which is being built under the Kamioka mine and is expected to take its first 
data in 2016. At such a high level of precision, those GW detectors are
 sensitive to many different sources, including chirp, ring-down, and merger phases
 of black-hole and neutron star binaries (e.g., \cite{schutz,faber,duez}), neutron star normal mode 
oscillations (e.g., \cite{nils}), rotating neutron star mountains 
(e.g., \cite{horowitz}), and CCSN explosions 
(e.g., \cite{kota06,ott_rev,fryer11} for recent reviews), on the final of which 
we focus in this article.

According to the Einstein's theory of GR (e.g., \cite{shap83}),
 no GWs can be emitted if gravitational collapse of the 
supernova core proceeds perfectly spherically symmetric. 
 To produce GWs, the gravitational collapse should 
proceed aspherically and dynamically. Observational evidence gathered over
the last few decades has pointed towards CCSNe indeed being
generally aspherical (see \citet{wang08} for a recent review).
The most unequivocal example is SN1987A. 
 To explain the light-curve, a large amount of mixing 
 of Ni outward to the H-He interface and of H inward into the He-core was required 
(e.g., \citet{woosley88,bli00,utrobin04}, and \citet{woosh} for a review).
 Such mixing processes 
coming from the Rayleigh-Taylor instability at the interface
 with different compositions after shock passage have been examined 
extensively so far by 2D (e.g., \citet{arnett89,ewald91,kifo03}) and 
  3D simulations \citep{hammer10}.
  The asymmetry of iron and nickel lines in SN1987A was proposed to be explained,
 if the explosion occurs in a jet-like \citep{naga97,nagataki00} or in a 
unipolar manner  
  \citep{utrobin11}. More directly, the HST images of SN1987A are showing 
that the expanding envelope is elliptical with the long axis aligned with the rotation 
axis inferred from the ring (\citet{wang02}, see however \citet{87a} for a recent 
counter argument). 
The aspect ratio and position angle of the symmetry axis are 
consistent with those predicted earlier from the observations of speckle and 
linear polarization. 
 What is more, the linear polarization became greater as time 
passed (e.g., \citet{wang01,leonard,leon06}), a fact which has been used to argue that 
the central engine of the explosion is responsible for the
non-sphericity (e.g., \citet{kifo03,kifo06,scheck06,burrows2,burr07}).
 By performing a series of time-dependent, non-LTE(local thermodynamic equilibrium),
 radiation-transport simulations (e.g.,
 \citet{dessart10,dessart08,dessart05}\footnote{see \citet{kasen06,kasen09}
 for an alternative approach for the light-curve modeling.}), 
\citet{dessart11} recently pointed out
 that asymmetry in the ejecta can explain the increase in the continuum polarization
  observed at the nebular phases \citep{leonard06b,choro10}.
 Dense knots, indications of ejecta clumpiness, and filaments seen 
in supernova remnants by HST in the visual \citep{fesen06,fesen11} 
and by ROSAT, Chandra, and XMM-Newton \citep{aschenbach95,hughes00,
miceli08} in X-rays also provide evidence that small- and large-scale inhomogeneities
(and perhaps even fragmentation) are a common feature in supernova explosions.

 Advancing ability of the HST has enabled the direct observation of the 
progenitors of nearby CCSNe from pre-explosion images
 (see \citet{smartt09} for a review). Although observational measurements 
 of the progenitor masses are currently not many and still highly uncertain (see
 \citet{murphy11} for collective references therein),
 evidence has accumulated that SN type II-plateau (II-P) comes predominantly from 
 stars in the range about of 8 - 16 $M_{\odot}$ \citep{smartt09}.
  A generic explosion energy of the SN II-P in the mass range is 
 roughly on the order of $10^{51}$ erg \citep{utrobin11}, however 
a large diversity of the explosion signatures (i.e. explosion energy and 
 the synthesized Ni mass) have been so far observed from quite similar 
 progenitors \citep{smartt09}. For example, the inferred $^{56}{\rm Ni}$ mass and
 the kinetic energies differ by a factor of five between SNe 2005cs and 2003gd, 
 both of them are among the {\it golden events} in which 
 enough information was obtained 
 to give an accurate estimate on a color or spectral type of the 
 progenitor and the initial mass.
  More massive stars are expected to
 lose much of their mass and explode as hydrogen-stripped SNe (Ib/c and IIb).
 Among them, the type Ic-BL SNe, which are associated with long gamma-ray bursts
 \citep{woos_blom}, all show much broader lines than SNe Ic. 
Due to the large kinetic energies of $2 - 5 \times 10^{52}$ erg,
 they have been referred to as "hypernovae"  (e.g., \citet{modjaz11}
 for a recent review). These events are likely to 
come from interacting binaries in which the
 primary exploding star has a mass lower than what is usually associated with
 evolution to the massive WR phase (e.g., \citet{podsia04,podsia11,fryer07,smith11},
 see however \citet{yoon05,yoon10}). 
In addition to the two branches mentioned above,
  \citet{nomoto06} predicted yet another branch, 
in which the SNe II-P with higher progenitor  
 mass result in fainter explosions.
  By connecting to candidate SN explosion mechanisms or to 
progenitor structures, it is indeed best if one could obtain 
 a unified picture to understand the mentioned wide diversity 
which is not only dependent on the progenitor masses but also on
 the evolution scenarios (a single vs. binary evolution).
 But, this is an area for future study firstly because it is too computationally 
 expensive to perform a long-term simulation that follows 
 multidimensional (multi-D) dynamics consistently from the onset of 
 core-collapse, through explosion, up to the nebular phase\footnote{for example, see \citet{dessart11c} for the most 
up-to-date 1D modeling of spectra and light curves for binary models.},
 secondly because we are still inaccessible to multi-D stellar evolution models, 
 in which multi-D modeling has been a major undertaking (see
 \citet{meakin2011,meakin07} for recent developments).

From a theoretical point of view, clarifying what makes the dynamics of the supernova 
 engine deviate from spherical symmetry is essential in understanding the GW emission
mechanism. Here it is worth mentioning that GWs are primary observables, which 
imprint a live information of the central engine, because they carry the information 
directly to us without being affected in propagating from the stellar center to the 
earth. On the other hand, SN neutrinos are exposed to a number of 
 (external) environmental effects, including self-interaction that 
 induces collective neutrino flavor oscillations predominantly 
 in the vicinity of the neutrino sphere
\cite[see][for reviews of the rapidly growing research
field and collective references therein]{raff07,duan08,duan10,dasg10} and 
the Mikheyev-Smirnov-Wolfenstein (MSW) effect \citep{mikheev1986} both in the stellar 
 envelope and in the earth (see \citet{raffelt12} for a review).
  Although the time profiles of neutrino signals can be potentially used like a tomography 
to monitor the envelope profile (e.g.,
 \citet{kneller} for a recent review), SN neutrinos generally could provide a rather 
indirect information about the central engine compared to GWs. 

The breaking of the sphericity in the supernova engine 
has been considered as the most important 
ingredient to understand the explosion mechanism, for 
which supernova theorists have been continuously keeping their efforts 
for the past $\sim$ 40 years. Currently multi-D simulations based on refined 
numerical models have shown several promising scenarios.
 Among the candidates are the neutrino heating 
mechanism aided by convection and hydrodynamic instabilities of the supernova shock
 (e.g., \citet{jank07} for a review), the acoustic mechanism
 \citep{burr_rev}, or the magnetohydrodynamic (MHD) mechanism
 (e.g., \citet{kota06} see references therein).
 To pin down the true answer among the candidate mechanisms, GW signatures 
 albeit being a primary observable, will not be solely enough and a 
 careful analysis including neutrinos and photons should be indispensable.
 The current neutrino detectors is ready to broadcast the alert to astronomers
 to let them know the arrival of neutrinos (e.g., 
 Supernova Early Warning Systems (SNEWS) \citep{anoto}).
In addition to optical observations using largest 8-10 m telescopes such 
as VLT and Subaru telescope (e.g., \citet{patat,tanaka}), 
the planned thirty-meter-telescope (TMT)\footnote{http://www.tmt.org/} 
with a refined spectropolarimetric technique 
is expected to detect more than 20 events of the SN polarization per year\footnote{M.Tanaka
 in private communication.}. 
This should provide valuable information to understand the SN asymmetry
 with increasing statistics.
 Not only for understanding the origin of non-spherical 
ejecta morphology\footnote{especially for 
 nearby event} but more importantly for understanding the origin of heavy elements, 
 it is of crucial importance to accurately determine nucleosynthesis in the SN ejecta,
 which naturally requires a multi-D numerical modeling
(see, \citet{woosh,thielemann} for recent reviews)\footnote{It is worth mentioning that 
radioactive decay can affect the shape of the light curve for some peculiar 
SN 1987A-like events \citep{utrobin11}, in which 
explosive nucleosynthesis plays an important role for the light-curve modeling.}.

Putting things together, the multidimensionality determines 
the explosion mechanism, in turn 
we may extract the information that traces the multidimensionality by 
the SN multimessengers, which would be only possible by a careful analysis on 
GWs, neutrinos, and photons. In this review article, we hope to 
bring together various of our published and unpublished
 findings from our recent multi-D supernova simulations and the obtained 
predictions of the SN multimessengers so far (for other high-energy 
astrophysical sources such as magnetars, gamma-ray bursts, and coalescing binaries, 
see \citet{ando12,eric1,szabi,prad,aso} 
for recent reviews).
  Before we go into details from the next sections, 
we first have to draw a caution that the 
 current generation of simulation results that we report in this article should depend
 on the next generation calculations by which more sophistication can be made 
not only in determining the efficiency of neutrino-matter coupling (the so-called 
 neutrino transport calculation), but also in the treatment of general relativity.
 Therefore we provide here only a snapshot of the moving 
 (long-run) documentary film whose headline we (boldly) chose to entitle as
``multimessengers from CCSNe to bridge theory and observation''.

Among the mentioned candidate mechanisms, 
we focus on the neutrino-heating mechanism in section \ref{sec2} and the 
MHD mechanism in section \ref{sec3}, respectively.
 In each section, we first briefly summarize the properties of 
the explosion dynamics and then move on to discuss possible 
properties of the SN multimessengers paying particular attention to 
their detectability. It may be best if we can cover these 
SN messengers once for all in this article,  but unfortunately not.
 What we have studied so far is limited to GWs and explosive 
nucleosynthesis in the neutrino-heating mechanism, and to GWs 
and neutrino signals in the MHD mechanism. 
To compensate the uncovered fields, the related references will 
 be given. Although a number of excellent reviews already exist 
on various topics in this article, this one might go beyond such 
reviews by its new perspectives on the multimessenger astronomy.


\section{Neutrino-heating mechanism}\label{sec2}

 CCSN simulations have been counted as
 one of the most challenging subjects in computational astrophysics. 
 The four fundamental forces of nature are all at play; 
 the collapsing iron core bounces due to strong interactions;
 weak interactions determine the energy and lepton number loss in the core 
via the transport of neutrinos; electromagnetic interactions determine the 
properties of the stellar gas; GR plays an important role 
 due to the compactness of the proto-neutron star and also due to high velocities 
of the collapsing material outside.
 Naturally, such physical richness ranging from 
 a microphysical scale (i.e. femto-meter scale)
 of strong/weak interactions to a macrophysical 
scale of stellar explosions has long attracted the interest of researchers,
  necessitating a world-wide, multi-disciplinary collaboration to clarify
 the theory of massive stellar core-collapse and the formation 
 mechanisms of compact objects.

Ever since the first numerical simulation of such events \citep{colgate}, 
the neutrino-heating mechanism \citep{wils85,bethe85,bethe},
 in which a stalled bounce shock could be revived via neutrino absorption 
 on a timescale of several hundred milliseconds after bounce,
 has been the working hypothesis of supernova theorists for these $\sim$ 45 years.
However, the simplest, spherically-symmetric (1D) form of this mechanism
 fails to blow up canonical massive stars 
\citet{rampp00,lieb01,thom03,sumi05}.
Pushed by mounting observations of the blast morphology
\cite[e.g.,][]{wang08} mentioned above, it is now almost certain that the
breaking of the spherical symmetry is the key to solve the supernova problem.
 So far a number of multi-D hydrodynamic simulations have been reported, which  
 {\it demonstrated} that hydrodynamic motions associated with convective overturn (e.g., 
\citet{herant,burr95,jankamueller96,frye02,fryer04a}) as well as the 
Standing-Accretion-Shock-Instability (SASI, e.g., 
\citet{blon03,sche04,scheck06,Ohni05,ohnishi07,thierry,murphy08,rodrigo09,rodrigo09_2,
iwakami1,iwakami2,fern} see
 references therein) can help the onset of the neutrino-driven explosion. 

To test the neutrino heating mechanism in the multi-D context, 
it is of crucial importance to solve accurately the neutrino-matter coupling in 
  spatially non-uniform hydrodynamic environments. 
For the purpose, one ultimately needs to solve the 
six-dimensional (6D) neutrino radiation transport problem
(three in space, three in the momentum space of neutrinos), which is a 
 main reason why the supernova simulations stand out from other astrophysical 
simulations due to their complexity. In the final sentence of the last paragraph, 
we wrote "demonstrated" because the neutrino heating was given by hand as 
an input parameter
 in most of the simulations cited above (see, however \citet{frye02,fryer04a}).
  The neutrino heating proceeds 
dominantly via the charged current interactions ($\nu_{\text{e}} + \text{n} \rightleftarrows \text{e}^{-} + \text{p},~ \bar{\nu}_{\text{e}} + \text{p} \rightleftarrows \text{e}^{+} + \text{n}$) in the gain region. The neutrino heating rate in the gain region 
 can be roughly expressed as 
$Q^{+} \propto L_{\nu}\langle\mu_{\nu}\rangle^{-1}/r^2$ where $L_{\nu}$ is the neutrino 
luminosity emitted from the surface of neutrino sphere and it determines the 
 amplitude of the neutrino heating as well as cooling, and $r$ and $\langle \mu_{\nu} \rangle$ 
is the distance from the stellar center and the flux factor\footnote{This quantity 
represents the degree of anisotropy in neutrino emission; $\langle\mu_{\nu}\rangle
\sim 0.25$ near at the neutrino sphere, $\langle\mu_{\nu}\rangle=1$ in the free-streaming limit ($r \rightarrow 
\infty$).}, respectively 
(e.g., \citet{jank01}). 
For example, $L_{\nu}$ is treated as an input parameter
in the so-called `light-bulb'' approach (e.g., \citet{jankamueller96}).
 This is one of the most prevailing approximations in recent
 3D simulations \citep{iwakami1,iwakami2,annop,nordhaus} because
 it is handy to study multi-D effects on the neutrino heating mechanism 
 (albeit on the qualitative grounds). To go beyond the light-bulb scheme,
  $L_{\nu}$ should be determined in a self-consistent manner. 
 For the purpose, one needs to tackle with 
neutrino transport problem, only by which energy as well as angle 
dependence of the neutrino distribution function can be determined without any 
 assumptions. 
Since the focus of this review is on the SN multi-messengers, a detailed discussion of
 various approximations and numerical techniques taken in the recent 
radiation-hydrodynamic SN simulations cannot be provided. 
 Table 1 is not intended as a comprehensive compilation, but we just want 
 to summarize milestones that have recently reported the neutrino-driven 
{\it exploding} models so far.

In Table 1, the first column ("Progenitor") shows the progenitor model employed in each 
simulation. The abbreviation of "NH", "WHW", and "WW" means \citet{nomo88}, 
\citet{woos02}, and \citet{woos95}. 
The second column shows SN groups with the published or 
 submitted year of the corresponding work. "MPA" stands for the CCSN 
group in the Max Planck Institute for Astrophysics led by H.T. Janka 
and E. M\"uller.
 "Princeton+" stands for the group chiefly consisting of
 the staffs in the Princeton University (A. Burrows, J.Murphy), Caltech (C.D.Ott), 
Hebrew University 
 (E. Livne), Universit\'e de Provence 
(L.Dessart), and their collaborators. The SN group in the Basel university
 is led by M. Liebend\"orfer and F.K. Thielemann. "OakRidge+" stands for the 
SN group mainly 
consisting of the Florida Atlantic University (S. Bruenn) and the Oak Ridge National 
Laboratory (A. Mezzacappa, O.E.B. Messer) and their collaborators. 
Tokyo+ is the SN group chiefly consisting 
 of the staffs in the National Astronomical Observatory of Japan (myself, 
T. Takiwaki), Kyoto university (Y. Suwa), Numazu College (K. Sumiyoshi),
 Waseda University (S. Yamada),
 and their collaborators. The third column represents the mechanism of explosions 
 which are basically categorized into two (to date),
 namely by the neutrino-heating mechanism
 (indicated by "$\nu$-driven") or by the acoustic mechanism ("Acoustic").
 "Dim." in the fourth column is the fluid space dimensions which is 
 one-, two-, or three-dimension (1,2,3D). The abbreviation ``N'' stands for `Newtonian,' while ``PN''---for `Post-Newtonian'---stands for some attempt at inclusion of general relativistic effects, and ``GR'' denotes full relativity.
 $t_{\rm exp}$ in the fifth column indicates an
 approximate typical timescale when the explosion initiates and $E_{\rm exp}$ represents
 the explosion energy normalized by Bethe(=$10^{51}$ erg) given at the 
 postbounce time of $t_{\rm pb}$, both of which 
 are attempted to be sought in literatures\footnote{but if we cannot find them, 
we remain them as blank "-".}. In the final column of "$\nu$ transport", 
  "Dim" represents $\nu$ momentum dimensions and the treatment of the 
 velocity dependent term in the transport equations is symbolized by ${\cal O}(v/c)$.
  The definition of the "RBR", "IDSA", and "MGFLD" will be given soon in the following.

\vspace{- 1cm}
\begin{table}
\begin{center}
\begin{tabular}{lcccccc}
\hline
Progenitor & Group & Mechanism & Dim. & $t_{\rm exp}$ & $E_{\rm exp}$(B) & $\nu$ transport \\ 
           & (Year)  &  & (Hydro) & (ms) & @$t_{\rm pb}$ (ms)& (Dim,  ${\cal O}(v/c)$) \\ 

\hline
\multirow{3}*{8.8 {$M_{\odot}$ }} & MPA & $\nu$-driven & 1D(2D) & $\sim$200 
    & 0.1  &  Boltzmann \\ 
        & (2006,2011)      & & (PN)   & & ($\sim$800) & 2, ${\cal O}(v/c)$     \\

\cmidrule(l){2-7}
 (NH88)  & Princeton+ & $\nu$-driven & 2D & $\lesssim$125 & 0.1 &  MGFLD \\
                       & (2006) &  & (N)   & & - &1, (N)     \\    \cmidrule(l){1-7}
10 $M_{\odot}$  & Basel & $\nu$+(QCD & 1D &  255 & 0.44 &  Boltzmann \\
(WHW02)           & (2009)      & transition) & (GR)   & & (350) &2, (GR)     \\    \cmidrule(l){1-7}
11 $M_{\odot}$  & {Princeton+} & Acoustic & 2D & $\gtrsim$550 & $\sim$0.1* &  MGFLD \\
(WW95)               & (2006)  &  & (N)   & & (1000) &1, (N)     \\    \cmidrule(l){1-7}
\multirow{3}*{11.2 {$M_{\odot}$ }} & MPA & $\nu$-driven & 2D & $\sim$100 &
 $\sim 0.005${,\it 0.025}   
      &  "RBR" Boltz- \\
   & (2006,2012)  & & (PN,{\it C-GR})   &$\sim${\it 200} & $\sim$200,{\it 900} & mann, 2, ${\cal O}(v/c)$     \\    \cmidrule(l){2-7} (WHW02)   
 & Princeton+ & Acoustic & 2D & $\gtrsim$1100 & $\sim$0.1* &  MGFLD \\
      &   (2007) &  & (N)   & & (1000) &1, (N)     \\    \cmidrule(l){2-7}
      & Tokyo+ & $\nu$-driven  &  3D & $\sim$100 & 0.01  &  IDSA \\
    &  (2011) &  & (N)   & & (300) &1, (N)     \\    \cmidrule(l){1-7}
12 $M_{\odot}$    & Oak Ridge+ & $\nu$-driven & 2D & $\sim$300 & 0.3 &  "RBR" MGFLD \\
  (WHW02)  &  (2009) &  & (PN)   & & (1000) &1, ${\cal O}(v/c)$  \\    \cmidrule(l){1-7}
13 $M_{\odot}$    & Princeton+ & Acoustic & 2D & $\gtrsim$1100 & $\sim$0.3* &  MGFLD \\
  (WHW02)  &  (2007) &  & (N)   & & (1400) &1, (N)     \\    \cmidrule(l){2-7}
(NH88)    & Tokyo+ & $\nu$-driven  & 2D & $\sim$200 & 0.1  &  IDSA \\
    &  (2010) &  & (N)   & & (500) &1, (N)     \\    \cmidrule(l){1-7}
15 $M_{\odot}$    &  MPA & $\nu$-driven & 2D & $\sim$600 & 0.025,{\it 0.125} 
    &  Boltzmann \\
(WW95)  & (2009,2012) & & (PN,{\it C-GR}) & $\sim${\it 400} & ($\sim$700,{\it 800}) & 2,${\cal O}(v/c)$     \\    \cmidrule(l){2-7}
(WHW02)  & Princeton+ & Acoustic & 2D & - & - &  MGFLD \\
    &(2007)  &  & (N)   & & (-) &1, (N)     \\    \cmidrule(l){2-7}
                       & OakRidge+ &  $\nu$-driven & 2D & $\sim$300 & $\sim$ 0.3 &  "RBR" MGFLD \\
                    &  (2009) &  & (PN)   & & (600) &1,${\cal O}(v/c)$  \\    \cmidrule(l){1-7}
20 $M_{\odot}$    & Princeton+ & Acoustic & 2D & $\gtrsim$1200 & $\sim$0.7* &  MGFLD \\
  (WHW02)  &  (2007) &  & (N)   & & (1400) &1, (N)     \\    \cmidrule(l){1-7}
25 $M_{\odot}$    & Princeton+ & Acoustic & 2D & $\gtrsim$1200 & - &  MGFLD \\
  (WHW02)  &  (2007) &  & (N)   & & (-) &1, (N)     \\    \cmidrule(l){2-7} 
    & Oak Ridge+ & $\nu$-driven & 2D & $\sim$300 & $\sim$ 0.7 &  "RBR" MGFLD \\
                    &  (2009) &  & (PN)   & & (1200) &1, ${\cal O}(v/c)$  \\    \cmidrule(l){1-7}
\end{tabular}
\end{center}
\vspace{- 0.8cm}
\caption{Selected lists of recent milestones
 reported by SN groups around the world ("Group"), which obtained explosions 
by the neutrino-heating mechanism
 (indicated by "$\nu$-driven") or the acoustic mechanism ("Acoustic")
 (See text for more details).
}
\end{table}

Due to the page limit of this article, we have to start the story only 
 after 2006 (see, e.g., \citet{jank07} for a complete review, 
and also \citet{cardall} for a similar table before 2006). 
The first news of the 
exploding model was reported by the MPA group.
By performing radiation-neutrino-hydrodynamic simulations 
which includes one of the best available neutrino transfer 
approximations, they reported 1D and 2D explosions for the 8.8 $M_{\odot}$ star
 \citep{kitaura,janka08} whose progenitor has a very tenuous outer envelope with steep 
density  gradient (a characteristic property of AGB stars). 
Also in 1D, the Basel+ group reported explosions for 10 and 15 $M_{\odot}$
 progenitors of WHW02 triggered 
 by the hypothesised first-order QCD phase transition in the protoneutron 
star (PNS) \citep{sage09}. 
To date, these two are the only modern numerical results where the neutrino-driven 
mechanism succeeded in 1D. 
 In the 2D MPA simulations, they obtained explosions 
 for a non-rotating 11.2 $M_\odot$ progenitor of 
WHW02 \citep{buras06}, and then for a $15 M_{\odot}$ progenitor \citep{marek} 
of WW95 with a relatively rapid rotation imposed\footnote{by comparing 
 the precollapse angular velocity to the one predicted in a recent stellar 
 evolution calculation \cite{maeder00,hege05}.}. They newly 
 brought in the so-called ``ray-by-ray'' approach (indicated by "RBR" in the table),
 in which the neutrino transport is solved 
along a given radial direction assuming that the hydrodynamic medium for the direction 
is spherically symmetric. This method, which reduces the 2D problem partly to 1D, 
fits well with their original 1D Boltzmann solver \citep{rampp00}\footnote{Note that 
 the ray-by-ray approach has an advantage compared to other approximation schemes, such
 that it can fully take into account the available neutrino reactions
(e.g., \cite{buras1} for references therein) 
 and also give us the most accurate solution for a given angular direction.}.
 For 2D hydrodynamic simulations with the ray-by-ray transport, 
one needs to solve the 4D radiation transport problem 
(two in space and two in the neutrino momentum space).
Regarding the explosion energies obtained in the MPA simulations, 
 their values at their final simulation time are typically underpowered by one or 
two orders of magnitudes to explain the canonical supernova kinetic energy 
($\sim 10^{51}$ erg). But the explosion energies presented in their figures 
 are still growing with time, and they could be as high as 1 B if they were able to 
 follow a much longer evolution as discussed in \cite{buras06}. Very recently,
 \citet{bernhard12} reported that more energetic explosions are 
obtained for the $11.2$ and $15 M_{\odot}$ stars in their GR 2D simulations 
compared to the corresponding post-Newtonian models (e.g., \citet{marek}).
 In the table,  the data in italic character represent their most up-to-date 
 results based on the 2D GR simulations using conformally-flatness approximation 
(e.g., \citet{dimm02a,Isa2009}, indicated by "{\it C-GR}" in the table). 

 It is rather only 
recently that fully 2D multi-angle Boltzmann transport simulations become practicable 
 by the Princeton+ group \citep{ott_multi,brandt}. In this case, one needs to handle 
 the 5D problem for 2D simulations (two in space, and three in the neutrino 
 momentum space). However this scheme is very 
 computationally expensive currently to perform long-term supernova 
simulations. In fact, the most recent 2D work by \cite{brandt} succeeded in 
 following the dynamics until $\sim 400$ ms after bounce for a non-rotating 
 and a rapidly rotating 20 $M_{\odot}$ model of WHW02, but explosions seemingly 
 have not been obtained in such an earlier phase either by the neutrino-heating or 
 the acoustic mechanism.
 
In the table, "MGFLD" stands for the Multi-Group Flux-Limited Diffusion scheme
  which eliminates
 the angular dependence of the neutrino distribution function 
(see, e.g., \cite{bruenn85} for more details).
For 2D simulations, one needs to solve the 3D problem, namely 
two in space, and one in the neutrino momentum space.  By implementing 
 the MGFLD algorithm to the CHIMERA code in a ray-by-ray fashion (e.g., \cite{bruenn}), 
 Bruenn et al. (2009) obtained neutrino-driven explosions for non-rotating progenitors 
in a relatively wide range in 12, 15, 20, 25 $M_{\odot}$ of WHW02 (see table). 
These models tend to start exploding at around 300 m after bounce, and 
 the explosion energy for the longest running model of the 
 25 M$_{\odot}$ progenitor is reaching to 1 B at 1.2 s after bounce \citep{bruenn}.  

 On the other hand, the 2D MGFLD simulations implemented 
 in the VULCAN code \citep{burr06} obtained explosions for a variety of progenitors
 of 11, 11.2, 13, 15, 20, and 25 $M_{\odot}$ not by the neutrino-heating mechanism 
but by the acoustic mechanism\footnote{For the 8.8 $M_{\odot}$ progenitor,
  they obtained neutrino-driven explosions (see table 1).}. 
 The acoustic mechanism relies on the revival of the stalled bounce shock by the 
 energy deposition via the acoustic waves that the oscillating protoneutron stars (PNSs) 
would emit in a much delayed phase ($\sim 1$ second) compared to
 the conventional neutrino-heating mechanism ($\sim 300 - 600 $ milliseconds). 
 If the core pulsation energy given in Burrows et al. (2007) 
\citep{burrows2} could be used to measure 
 the explosion energy in the acoustic mechanism, they reach to 1 B after 1000 ms after
 bounce.

 By performing 2D simulations in which the spectral neutrino transport
   was solved by the isotropic diffusion source approximation (IDSA)
   scheme \citep{idsa}, the Tokyo+ group reported explosions for a
   non-rotating and rapidly rotating 13 $M_{\odot}$ progenitor of NH88. 
 They pointed out that a stronger explosion is 
obtained for the rotating model comparing to the corresponding
non-rotating model. The IDSA scheme splits the neutrino distribution
 into two components (namely the streaming and trapped neutrinos),
 both of which are solved using separate numerical techniques
 (see \citet{idsa} for more details). 
The approximation level of the IDSA
 scheme is basically the same as the one of the MGFLD.
 The main advantage of the IDSA scheme is that the fluxes in the transparent
 region can be determined by the non-local distribution of sources
 rather than the gradient of the local intensity like in MGFLD.
 A drawback in the current
 version of the IDSA scheme is that heavy lepton neutrinos
 ($\nu_x$, i.e., $\nu_{\mu}$, $\nu_{\tau}$ and their 
anti-particles) as well as the energy-coupling weak interactions 
 have yet to be implemented.
 Extending the 2D modules in \citep{suwa} to 3D,
 they recently reported explosions in the 3D models for an 11.2 $M_{\odot}$ 
progenitor of WHW02 \citep{taki11}.
 By comparing the convective motions as well as neutrino luminosities and energies between their 2D and 3D models, they pointed out whether 3D effects would help
 explosions or not is sensitive to the employed numerical resolutions.
  They argued that next-generation supercomputers are at least needed
 to draw a robust conclusion of the 3D effects.

 Having summarized a status of the current supernova simulations,
 one might easily see a number of issues that remain to be clarified.
 First of all, the employed progenitors usually rather scatter
 (e.g., Table 1).  Different SN groups seem to have a tendency to 
employ different progenitors, providing different results.
 By climbing over a wall which may have rather separated exchanges 
 among the groups,
  a detailed comparison for a given progenitor
   needs to be done seriously in the multi-D results 
(as have been conducted in the Boltzmann 1D simulations between
 the MPA, Basel+, and Oak Ridge+ groups \citep{lieb05}).

In the next section, we briefly summarize the findings obtained 
 in our 2D \citep{suwa} and 3D \citep{taki11} simulations, 
 paying particular attention to how multidimensionality such as SASI, 
convection, and rotation could affect the neutrino-driven explosions.
 
\begin{figure}[hbt]
\epsscale{0.5}
\plotone{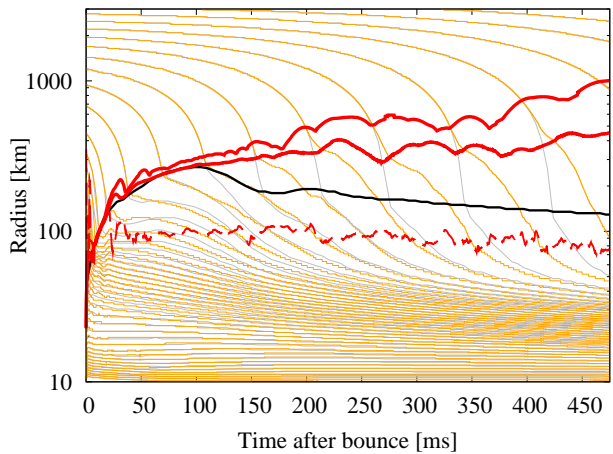}
\caption{Time evolution of  1D (thin gray lines) or 2D (thin orange lines) hydrodynamic
 simulation \citep{suwa} of a 13 $M_{\odot}$ progenitor \citep{nomo88}.
  Thick lines in red (for 2D) and black (for 1D) show
  the position of shock waves, noting for 2D that the maximum (top)
  and average (bottom) shock position are shown.  The red dashed line
  represents the position of the gain radius, which is similar to the
  1D case (not shown).}
\label{fig:1}
\end{figure}

\begin{figure}[hbtp]
\epsscale{0.8}
\begin{center}
\plottwo{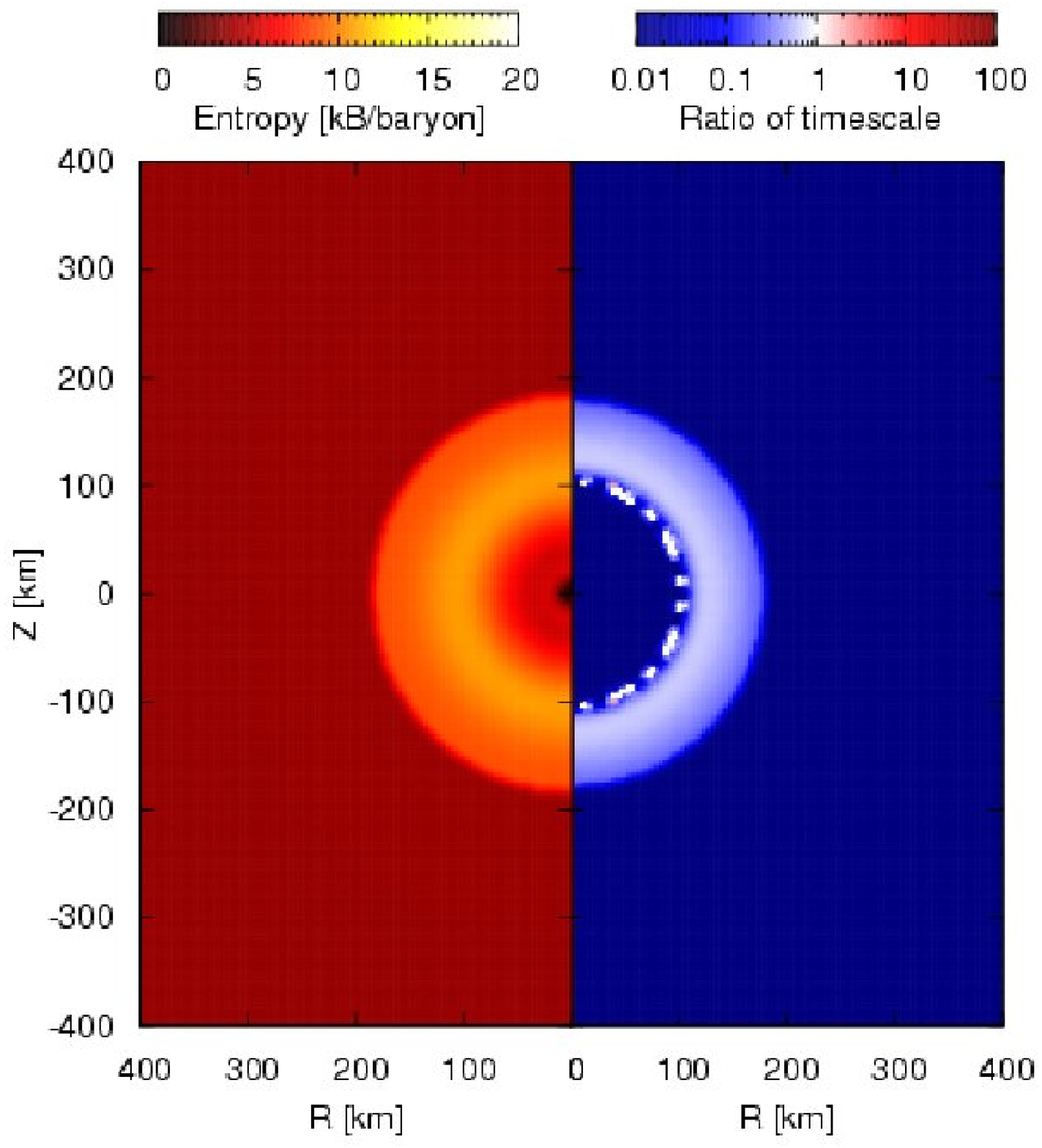}{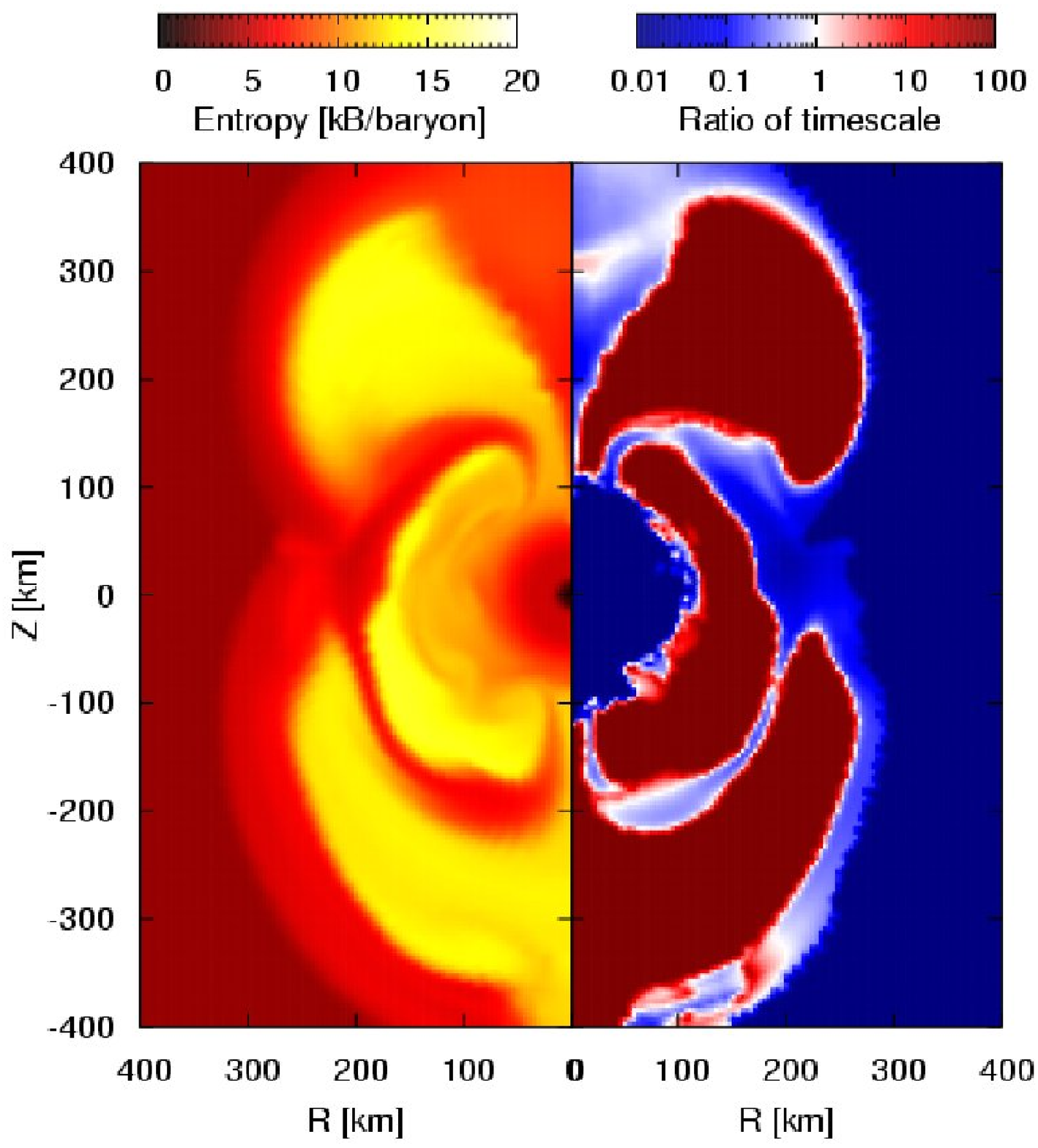}
\caption{Snapshot of the distribution of entropy (left half) and the
  ratio of the advection to the heating timescale (right half) for
  models of the 1D (left) and 2D (right) models at 200 ms after
  bounce. These figures are taken from \citet{suwa}.}\label{fig:2}
\end{center}
\end{figure}

\begin{figure}[hbtp]
\epsscale{0.5}
\plotone{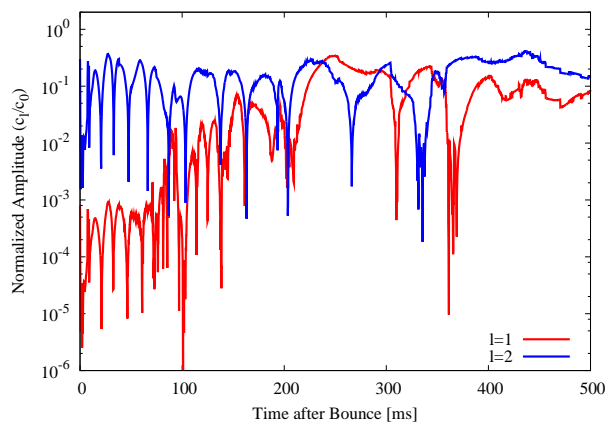}
\caption{SASI activity for the rotating model versus postbounce
  time. Shown are the coefficients of the dipole ($\ell = 1$) and quadrupole ($\ell
  = 2$) modes of the spherical harmonics of the aspherical shock
  position, normalized to the amplitude of the $\ell = 0 $ mode (taken
 from \citet{suwa}).}
\label{fig:3_added}
\end{figure}

\begin{figure}[hpbt]
  \begin{center}
    \begin{tabular}{ccc}
      \resizebox{65mm}{!}{\includegraphics{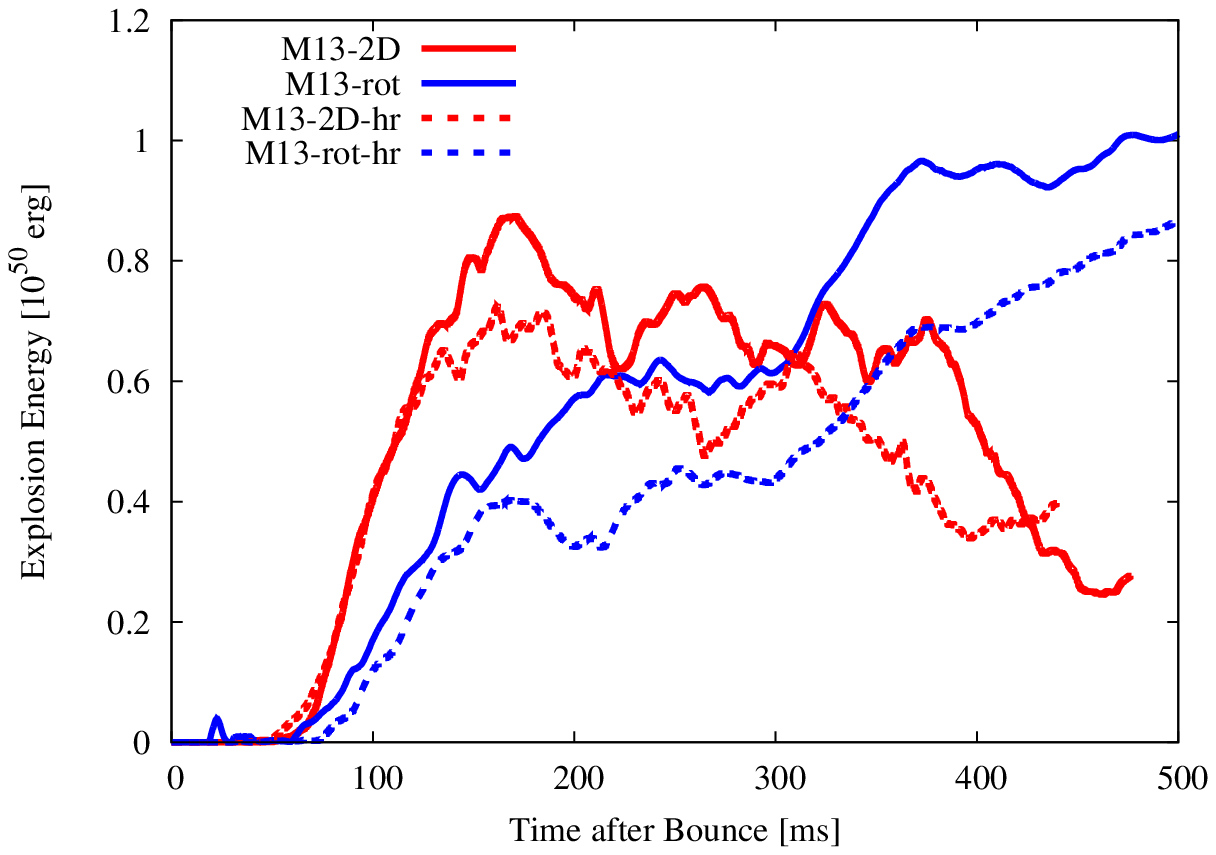}} &
      \resizebox{50mm}{!}{\includegraphics{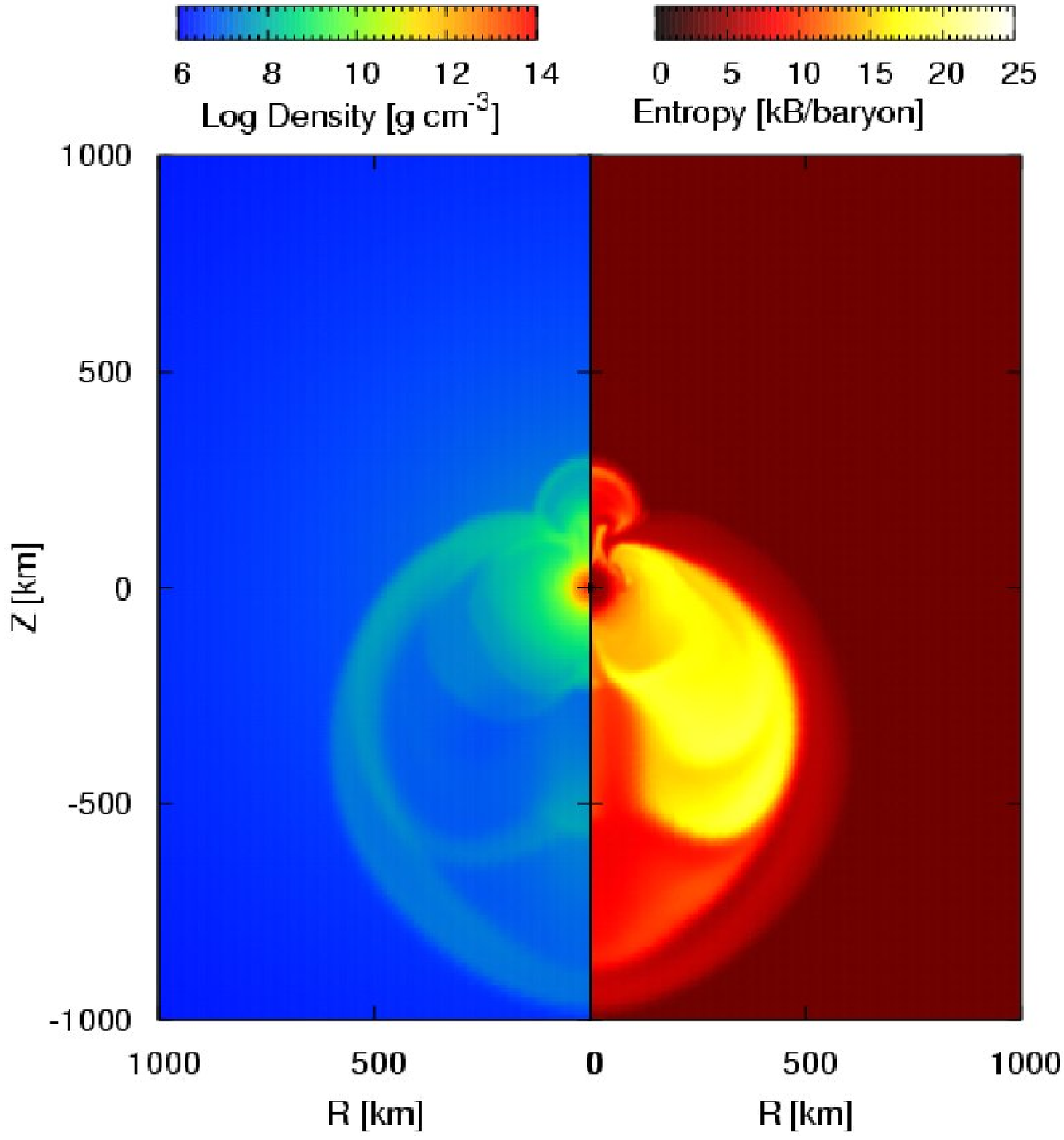}} &
      \resizebox{50mm}{!}{\includegraphics{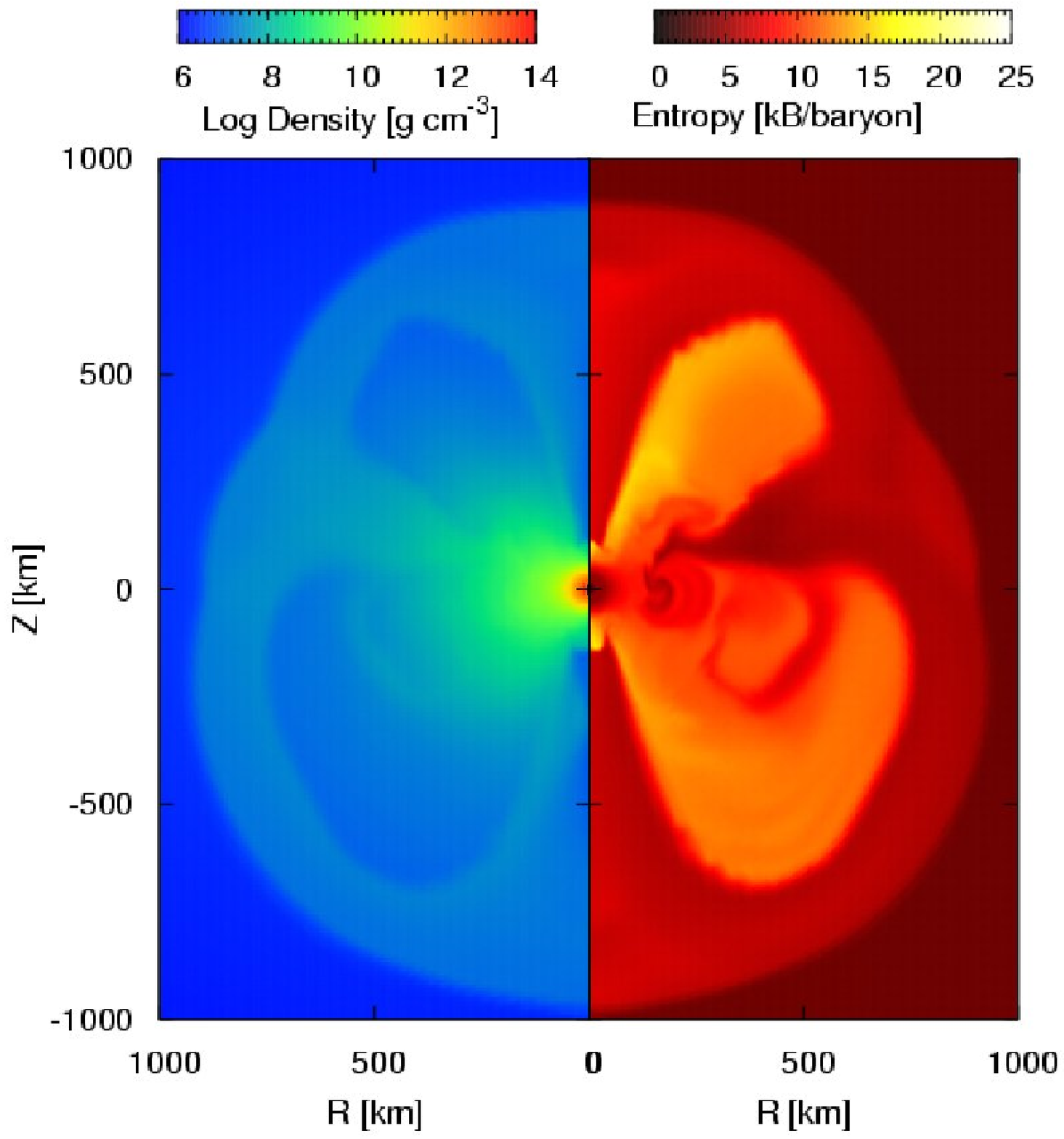}}\\
    \end{tabular}
   \caption{Time evolution of the explosion energy versus postbounce
  time for 2D models with and without rotation (left panel). 
The explosion energy is defined as the total
    energy (internal plus kinetic plus gravitational), integrated over
    all matter where the sum of the corresponding specific energies is
    positive. Models with ``-hr'' indicates the ones with higher numerical resolution, in which 
 the mesh numbers in the lateral direction are doubled.
  Middle and right panels are snapshots of the density (left half) 
and the entropy (right half) for 1D (left) and 2D rotating (right) model at 
the epoch when the shock reaches 1000 km. These figures are taken from \citet{suwa}.}
  \label{fig:5}
  \end{center}
\end{figure}

\subsection{Multidimensionality in multi-D radiation hydrodynamic simulations}
Figure \ref{fig:1} depicts the difference between the time evolutions
of 1D (thin gray lines) or 2D (thin orange lines) simulation of the 13 $M_{\odot}$ 
progenitor model.  Until $\sim 100$ ms
after bounce, the shock position of the 2D model (thick red line) is
similar to the 1D model (thick black line). Later on, however, the
shock for 2D does not recede as for 1D, but gradually
expands and reaches 1000 km at about 470 ms after bounce. Comparing
 the position of the gain radius (red dashed line) to the shock
 position for 1D (thick black line) and 2D (thick red line), one
can see that the advection time of the accreting material in the gain
region can be longer in 2D than in 1D.  This longer exposure of cool
matter in the heating region to the irradiation of hot outstreaming
neutrinos from the PNS is essential for the increased efficiency of
the neutrino heating in multi-D models. This can be also depicted in 
Figure \ref{fig:2}. The right-half shows
$\tau_\mathrm{adv}/\tau_\mathrm{heat}$, which is the ratio of the
advection to the neutrino heating timescale\footnote{This quantity is known as a 
 useful quantity to diagnose the success ($\tau_\mathrm{adv}/\tau_\mathrm{heat}\gtrsim 
1$, i.e., the neutrino-heating timescale is shorter than the advection timescale of 
material in the gain region)  or failure 
($\tau_\mathrm{adv}/\tau_\mathrm{heat}\lesssim 1$) of the 
neutrino-driven explosion (e.g., \citet{goshy,jank01,thomp05}).}. For the 2D model (right
panel), it can be shown that the condition of
$\tau_\mathrm{adv}/\tau_\mathrm{heat}\gtrsim1$ is satisfied behind the
aspherical shock, which is deformed predominantly by the SASI ($\ell =2$ mode
 at the snapshot), while
the ratio is shown to be smaller than unity in the whole region behind
the spherical standing accretion shock (left panel:1D). 
Here SASI that develops due to the advection-acoustic cycle in the supernova core
 \citep{fog1,fog2}, is a uni- and bipolar sloshing of the stalled bounce shock 
with pulsational strong expansion and contraction (seen as oscillations in the red 
curves in Figure 1).
Comparing the left-half of each panel, the entropy for the
2D model is shown to be larger than for the 1D model. This is also the
evidence that the neutrino heating works more efficiently in 2D.

 To see clearly the effects of stellar rotation on the postbounce evolution,
 a rapidly rotating model was taken in \citet{suwa}, in which a constant angular 
frequency of $\Omega_0 = 2~{\rm rad}$/s is imposed inside the iron core 
with a dipolar cut off ($\propto r^{-2}$) outside. This 
corresponds to $\beta \sim 0.18 \%$ with $\beta$ 
being the ratio of the rotational to the gravitational energy when the simulation starts. For the rotating model, the dominant mode of the shock
deformation after bounce is almost always the $\ell =2$ mode for the rotating model
 (e.g., Figure \ref{fig:3_added}, although the $\ell = 1$ mode can be as large as the
 $\ell = 2$ mode when the SASI enters the non-linear regime 
($\gtrsim 200$ ms after bounce). In contrast to this rotation-induced 
$\ell = 2$ deformation, the $\ell =1$ mode tends to be larger than the $\ell = 2$ mode 
for the 2D models without rotation in the saturation phase.

The left panel of Figure \ref{fig:5} shows the comparison of the explosion energies for
the 2D models with and without rotation (models with ``-rot'' or "-2D", respectively). 
Although the explosion energies
depend on the numerical resolutions quantitatively, they show a
continuous increase for the rotating models. The
 explosion energies for the models without rotation, on the other
hand, peak around $180$ ms when the neutrino-driven explosion sets in
(see also Figure 1), and show a decrease later on.  The reason for the greater 
explosion energy for models with rotation is due to the bigger mass of
the exploding material.  This is because the north-south symmetric
($\ell =2$) explosion can expel more material than for the unipolar
explosion (compare the middle and right panels in Figure \ref{fig:5}). The explosion
energies when we terminated the simulation are less than $\lesssim
10^{50}$ erg for all the computed models.  For the rotating models, we are
tempted to speculate that the explosion energies could increase later on by a 
linear extrapolation. However, in order to identify the robust 
feature of an explosion,
a longer-term simulation with improved input physics is needed. In combination
 with the assumed rapid rotation, magnetic fields 
 should be also taken into account as in \citet{burr07}, which we 
 are going to study as a follow-up of our rotating 2D models 
(Suwa et al. in preparation).

Extending our 2D modules mentioned above, we are currently running 3D simulations 
 for an 11.2 $M_{\odot}$ star of \citet{woos02} with spectral neutrino transport that 
is solved by the IDSA scheme in a ray-by-ray manner \citep{taki11}. We briefly summarize
 the results in the following.

 \begin{figure}[htbp]
    \centering
    \includegraphics[width=.49\linewidth]{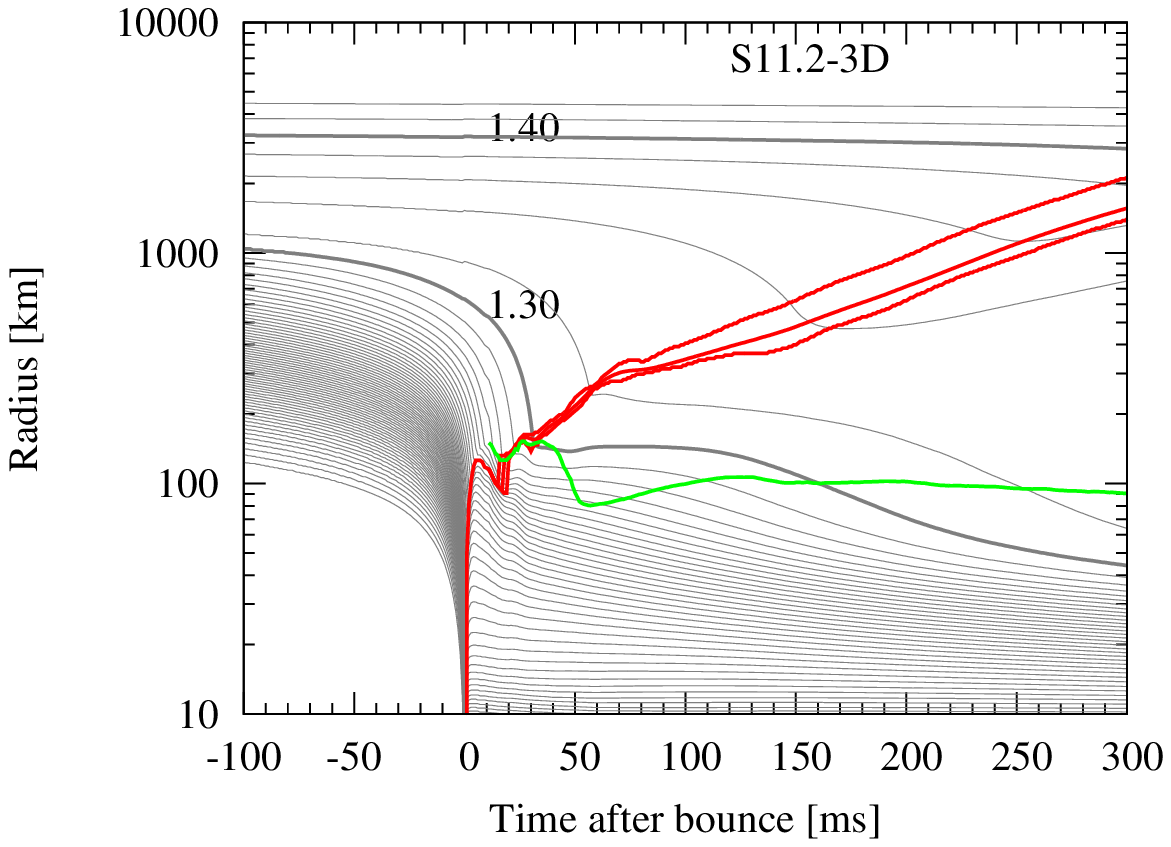}
    \includegraphics[width=.49\linewidth]{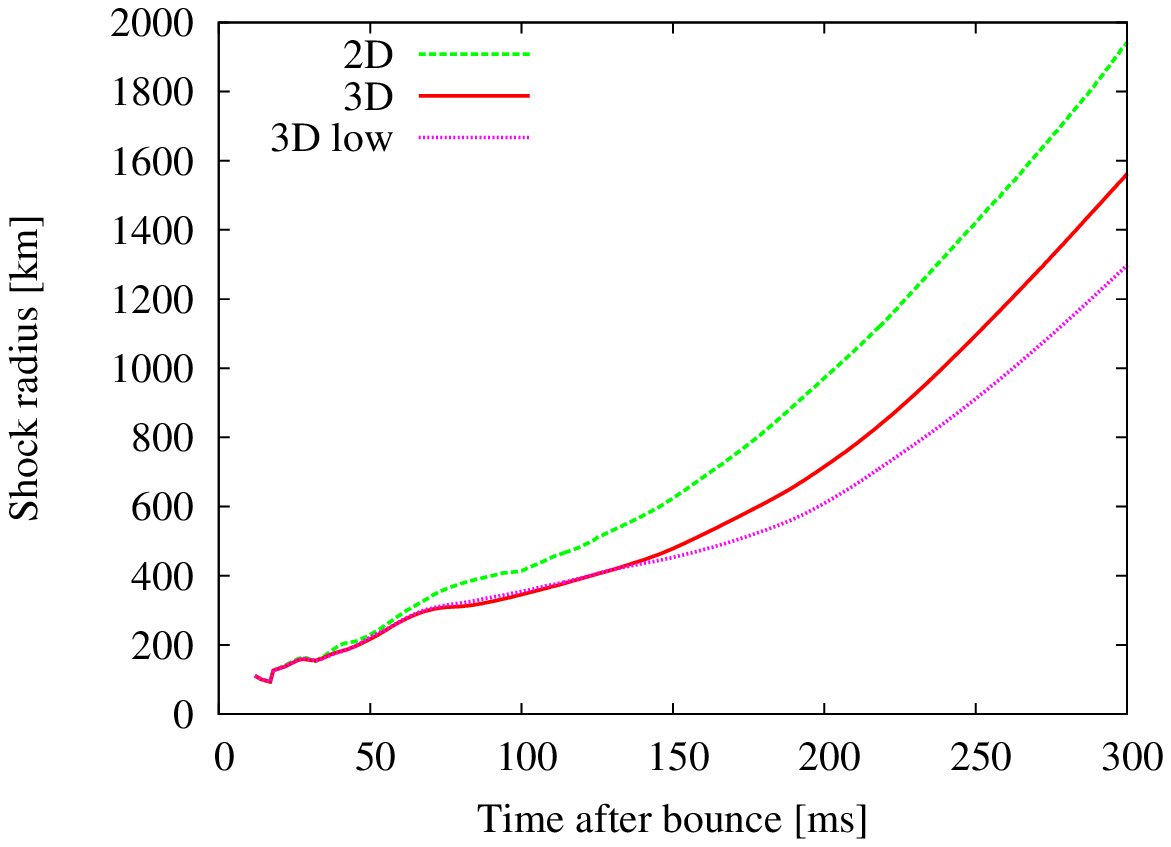}
 \caption{Time evolution of our 3D model \citep{taki11},
 visualized by mass shell trajectories 
in thin gray lines (left panel).
  Thick red lines show the position of shock waves, noting that the maximum (top),
 average (middle), and the minimum (bottom) shock position are shown, respectively. 
 The green line represents the shock position of the 1D model. "1.30" and 
 "1.40" indicates
 the mass in unit of $M_{\odot}$ enclosed inside the mass-shell. Right panel shows 
 the evolution of average shock radii for the 2D (green line), 
 and 3D (red line) models. The "3D low" (pink line) corresponds to the low resolution 3D 
model, in which the mesh numbers are taken to be half of the standard model
(see text). These figures are taken from \citet{taki11}.}
\label{f7}
\end{figure}

 The left panel of Figure \ref{f7} shows mass-shell trajectories 
 for the 3D (red lines) and 1D model (green line), respectively.
 At around 300 ms after bounce, the average shock radius for the 3D model exceeds
 1000 km in radius. On the other hand, an explosion is not obtained 
 for the 1D model, which is in agreement with \citet{buras06}. 
The right panel of Figure \ref{f7} shows 
 a comparison of the average shock radius vs. postbounce time. In the 2D model, 
the shock expands rather 
continuously after bounce. This trend is qualitatively consistent with the 2D result by
 \citet{buras06} (see their Figure 15 for model 
s112\underline{\hspace{0.4em}}128\underline{\hspace{0.4em}}f)\footnote{
 The reason that the average shock of our 2D model expands much faster than theirs would 
 come from the neglected effects in this work including 
 general relativistic effects, inelastic neutrino-electron scattering, and cooling 
by heavy-lepton neutrinos. All of them 
could give a more optimistic condition to produce explosions.
 Apparently these ingredients should be appropriately implemented, which we hope 
 to be practicable in the next-generation 3D simulations.}.

 Comparing the shock evolution between our 2D (green line in 
 the right panel of Figure \ref{f7}) and 3D model (red line), 
the shock is shown to expand much faster for 2D.
 The pink line labeled by "3D low" is for the low resolution 3D 
model, in which the mesh numbers are taken to be half of the standard model.
 Note that the 3D computational grid consists of 300 logarithmically spaced,
radial zones to cover from the center up to 5000 km  and 64 polar ($\theta$) and 32
azimuthal ($\phi$) uniform mesh points, which are used to cover the whole solid angle. 
 The low resolution 3D model has one-half of the mesh numbers in the 
$\phi$ direction ($n_{\phi}$=16), while fixing the mesh numbers in other directions.
 Comparing with our standard 3D model (red line),
  the shock expansion becomes less energetic for the low resolution model
 (later than $\sim 150$ ms). Above results indicate that explosions are easiest 
to obtain in 2D, followed in order by 3D, and 3D (low).
  At first sight, this may look contradicted with the finding by 
 \citet{nordhaus} who pointed out that explosions could be more easily obtained 
 in 3D than in 2D. The reason of the discrepancy is summarized shortly as it follows.

\begin{figure}[htbp]
    \centering
    \includegraphics[width=.45\linewidth]{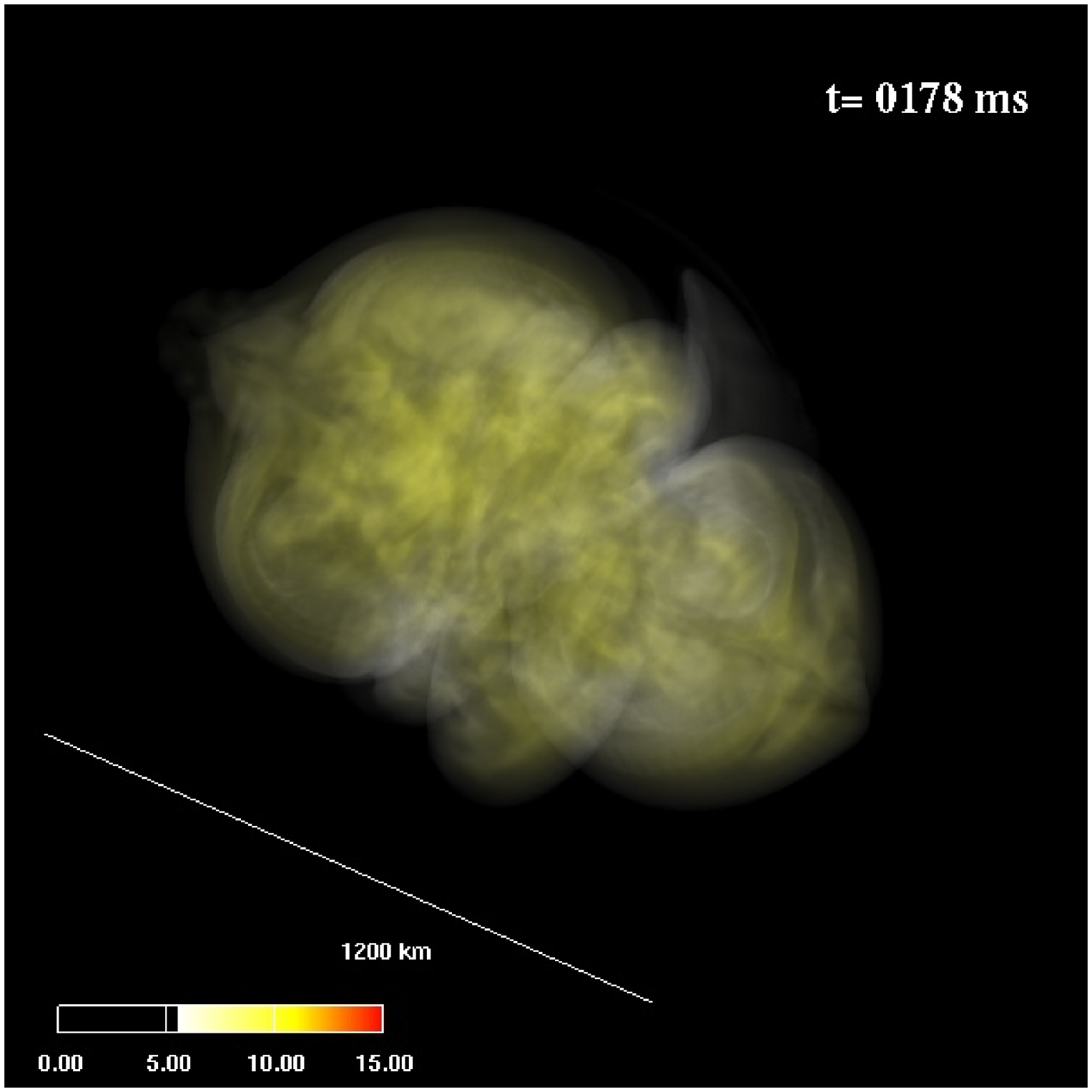}
    \includegraphics[width=.45\linewidth]{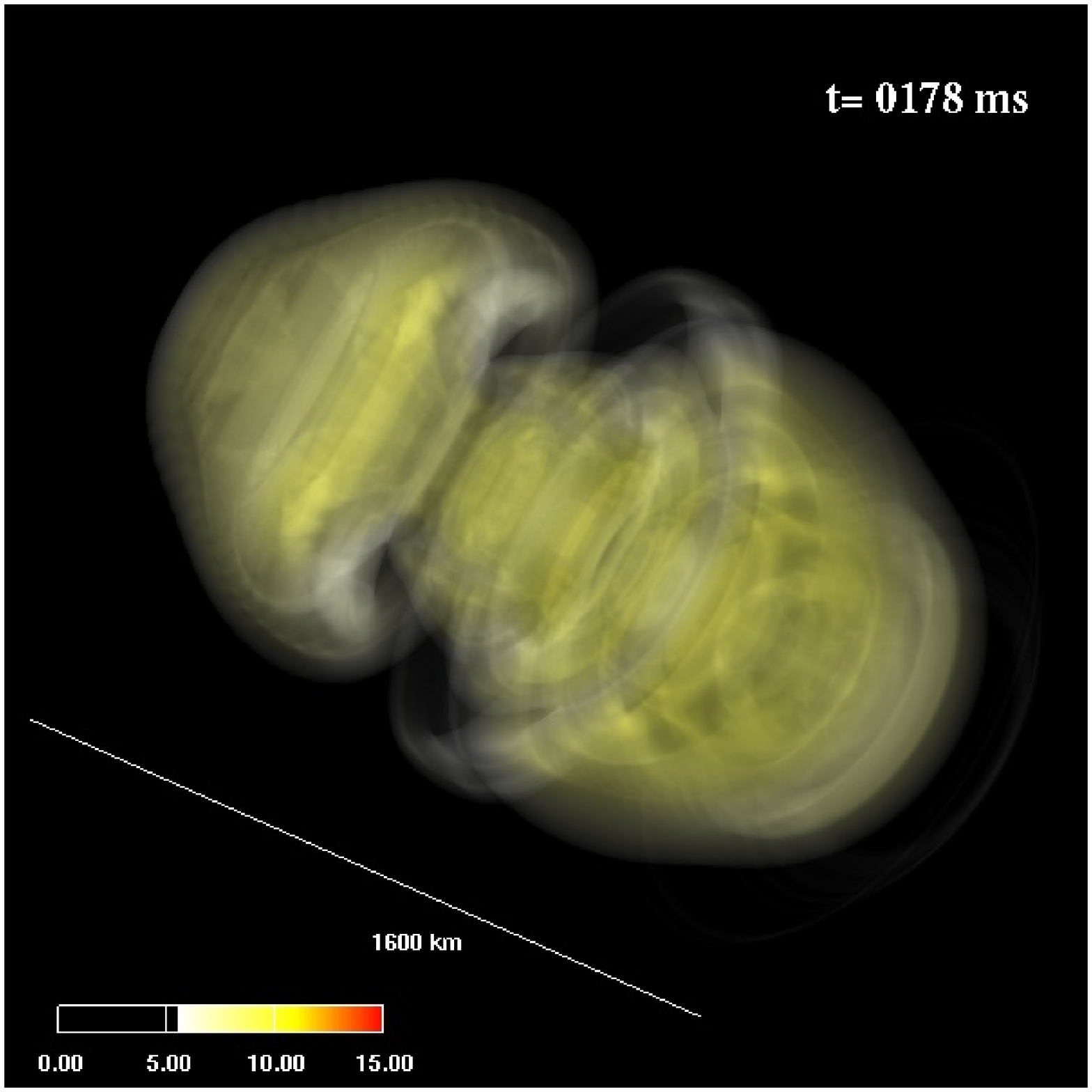}
 \caption{Volume rendering of entropy showing the blast morphology 
 in our 3D (left) and 2D (right) model for the $11.2 M_{\odot}$ progenitor
  of \citet{woos02} (at $t=178$ ms after bounce), respectively. 
The linear scale is indicated in each panel. These figures are taken from \citet{taki11}.}
\label{f5_added}
\end{figure}

\begin{figure}[htbp]
    \centering
    \includegraphics[width=.39\linewidth]{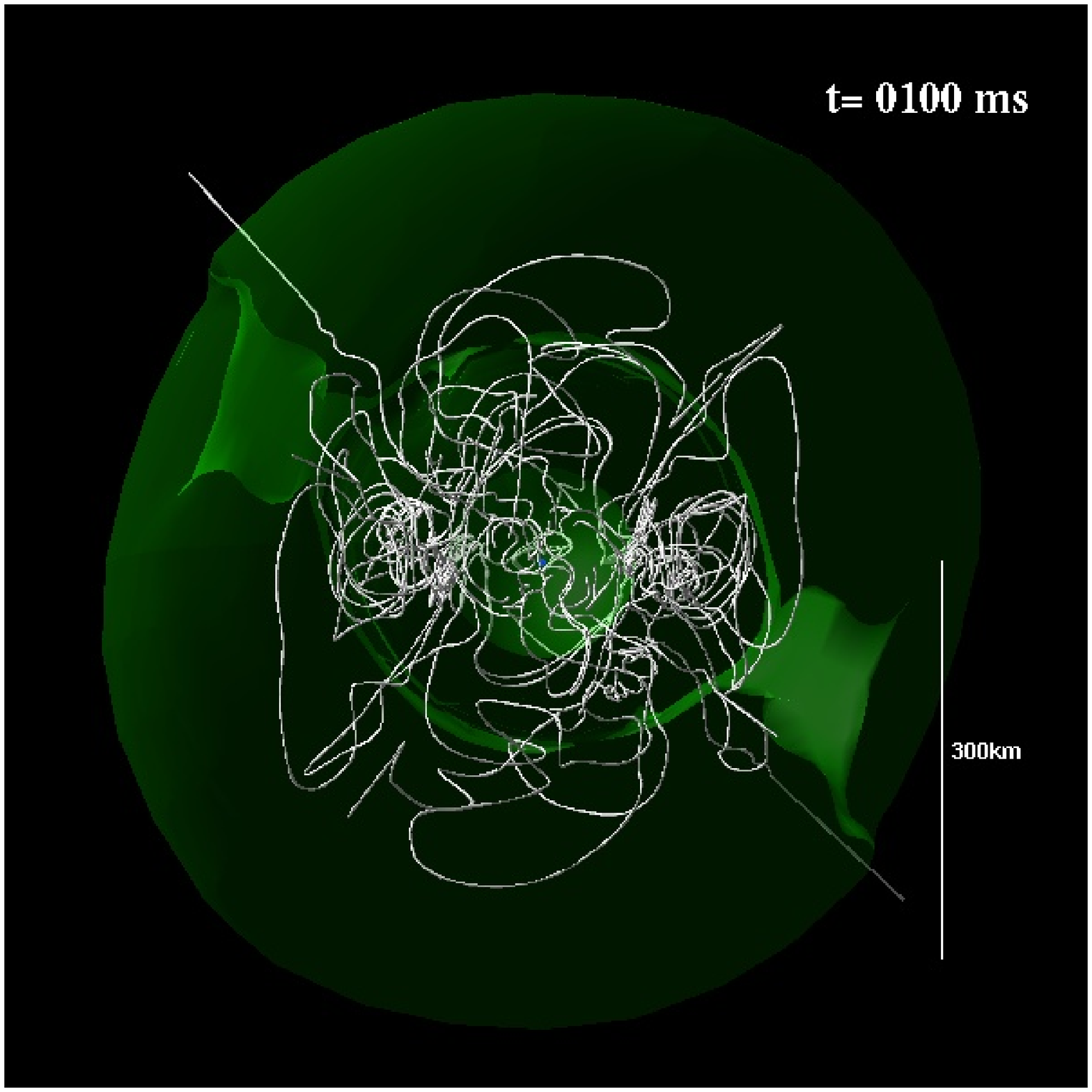}
    \includegraphics[width=.49\linewidth]{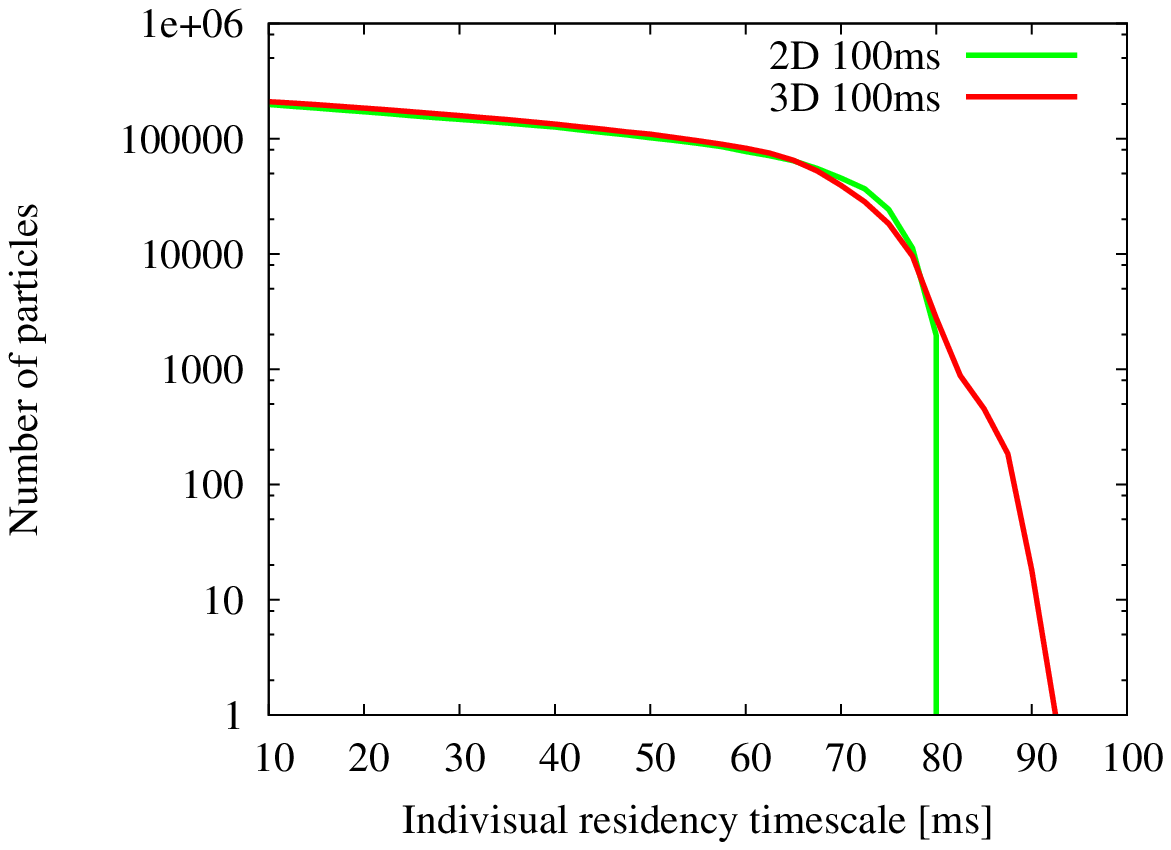}
 \caption{The left panel shows streamlines of selected tracer particles 
advecting through the shock wave to the PNS seen from the polar 
 direction in our 3D model \citep{taki11}. Several surfaces of constant entropy
 marking the position of the shock wave (greenish outside) and the PNS (indicated by the 
central sphere) are shown with the linear scale given in the right bottom edge.
  The right panel shows the number of tracer-particles travelling
 in the gain region as a function of their individual
 residency time between the 2D and 3D model at 100 ms after bounce. 
 If the left panel were for 2D models,  the streamlines would 
 be seen as a superposition of circles with different diameters.
The maximum residency timescale for the 3D model is shown to be longer than that in the 
 corresponding 2D model. These figures are taken from \citet{taki11}. }
\label{to_ref1}
\end{figure}

\begin{figure}[htbp]
    \centering
    \includegraphics[width=.49\linewidth]{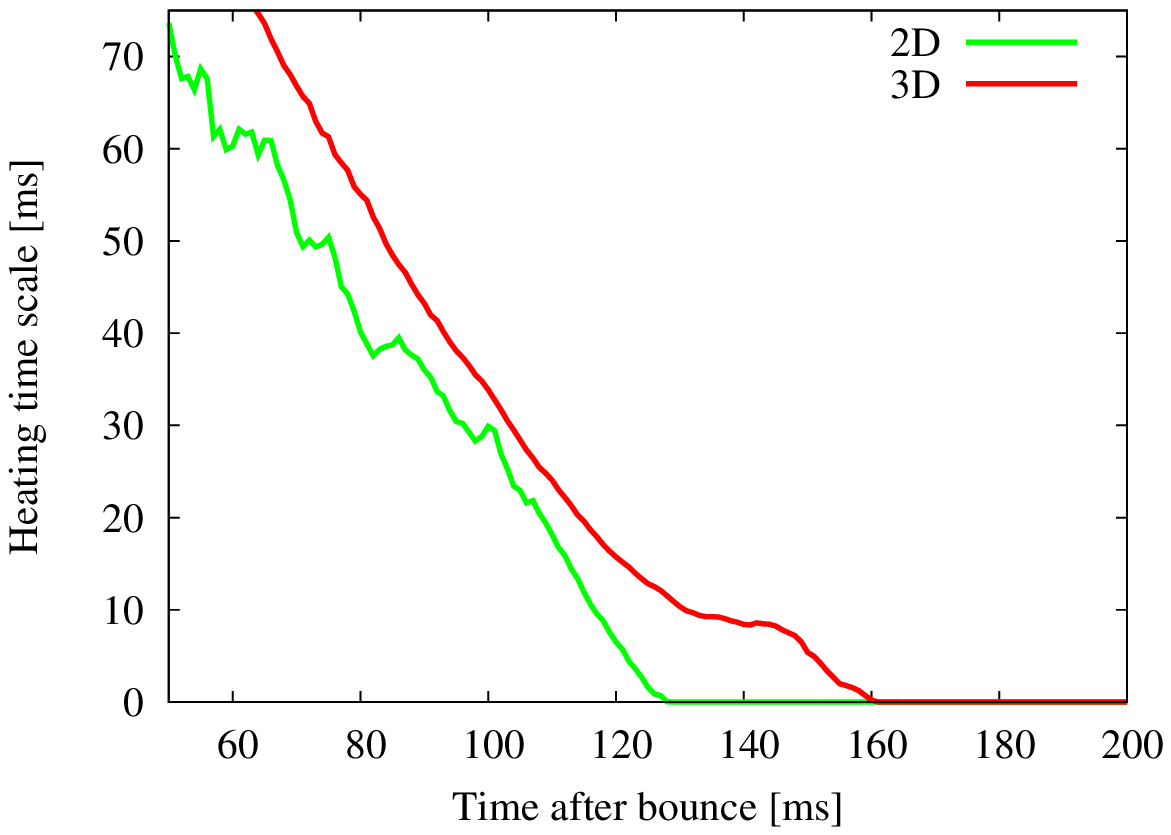}
    \includegraphics[width=.49\linewidth]{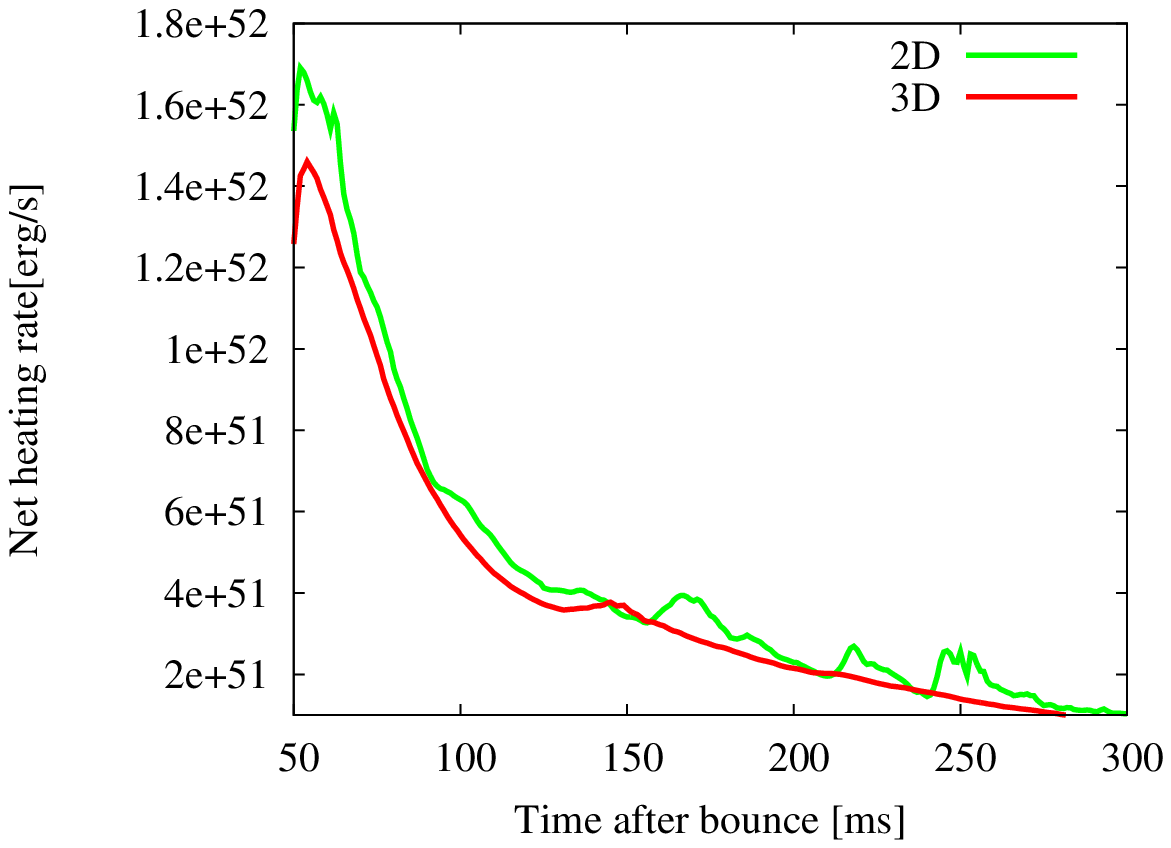}\\
 \caption{Time evolution of neutrino-heating timescale (left) 
and total net rate of neutrino heating (right) in our 2D (green line) 
and 3D model (red line). These figures are taken from \citet{taki11}. }
\label{f11}
\end{figure}

 Figure \ref{f5_added} compares the blast 
morphology for our 3D (left panel) and 2D (right) model\footnote{Note that the 
 polar axis is tilted (about $\pi/4$) both in the left and middle panel.}.
 In the 3D model (left panel), non-axisymmetric structures are clearly seen.
 By performing a tracer-particle analysis, the maximum residency time of material 
in the gain region is shown to be longer for 3D than 2D due to the non-axisymmetric 
flow motions (see Figure \ref{to_ref1}). This 
is one of advantageous aspects of 3D models to obtain the neutrino-driven explosions.
 On the other hand, our detailed analysis 
showed that convective matter motions below the gain radius become much more violent 
in 3D than in 2D, making the neutrino luminosity larger
 for 3D (see \citet{taki11} for more details). Nevertheless the emitted neutrino 
 energies are made smaller due to the enhanced cooling.
 Due to these competing ingredients, the neutrino-heating
 timescale becomes shorter for the 3D model, leading to a smaller 
 net-heating rate compared to the corresponding 2D model (Figure \ref{f11}). 
 Note here that the spectral IDSA scheme, by which 
the feedback between the mass accretion and the neutrino luminosity
 can be treated in a self-consistent manner (not like the light-bulb scheme
 assuming a constant luminosity), 
sounds quite efficient in the first-generation 3D simulations.

As seen from Figure \ref{f7}, an encouraging finding in \citet{taki11} 
was that the shock expansion 
tends to become more energetic for models with finer resolutions.
These results would indicate whether these advantages for driving 3D explosions  
 can or cannot overwhelm the disadvantages is sensitive to the employed numerical 
resolutions\footnote{It is of crucial importance to conduct a convergence test in which a numerical gridding is changed in a systematic way (e.g. \citet{hanke11}).}.
To draw a robust conclusion, 3D simulations 
 with much more higher numerical resolutions and also with more advanced treatment of 
 neutrino transport as well as of gravity is needed, which could be 
 hopefully practicable by utilizing forthcoming petaflops-class supercomputers. 

In addition to the 3D effects, impacts of GR on
 the neutrino-driven mechanism stand out among the biggest open questions  
  in the supernova theory. It should be remembered that using newly derived 
Einstein equations \citep{Misner64}, the consideration of GR was standard in the 
pioneering era of supernova simulations (e.g., \citet{May66}).
One year after \citet{colgate}, \cite{Schwarz67} reported the 
first fully GR simulation of stellar collapse 
to study the supernova mechanism, who implemented a 
gray transport of neutrino diffusion in the 
1D GR hydrodynamics\footnote{Citing from his paper, 
"{\it In this calculation, the neutrino
 luminosity of the core is found to be $10^{54}$ erg/s, or 1/2 a solar rest mass
 per second !! .... This is the mechanism which the supernova explodes".} The 
 neutrino luminosity rarely becomes so high in the modern simulations, but it is 
 surprising that the potential impact of GR on the neutrino-heating mechanism
 was already indicated in the very first GR simulation.}.
  Using GR Boltzmann equations derived by \citet{Lindquist66},
\citet{Wilson71} developed a 1D GR-radiation-hydrodynamic code 
including a more realistic (at the time) description
 of the collisional term than the one in \cite{Schwarz67}. 
 By performing 1D GR hydrodynamic simulations that included 
a leakage scheme for neutrino cooling, hydrodynamical properties 
 up to the prompt shock stagnation were studied in detail 
\citep{VanRiper79,VanRiper81,VanRiper82}. These pioneering studies, albeit 
 using a much simplified neutrino physics than today, did provide a 
 bottom-line of our current understanding of the supernova mechanism
(see \citet{Bruenn01} for a complete list of references for the early GR studies).
 In the middle of the 1980s, 
 \citet{bruenn85} developed a code that coupled 1D GR hydrodynamics to 
 the MGFLD transport of order $(v/c)$ including the so-called standard
 set of neutrino interactions.
Since the late 1990s, 
 the ultimate 1D simulations, in which the GR Boltzmann transport
 is coupled to 1D GR hydrodynamics, have been made feasible by Sumiyoshi-Yamada et al. \citep{Yamada97,Yamada99,Sumiyoshi05,Sumiyoshi07}\footnote{Very recently, they reported
 their success to develop the first multi-angle, multi-energy neutrino transport
 code in 3D \citep{Sumi11}.}
 and by Liebend\"orfer-Mezzacappa-Bruenn et al. 
\citep{Mezzacappa89,Bruenn01,Matthias01,Matthias04} 
(and by their collaborators).

Among them, \citet{Bruenn01} firstly showed that the neutrino luminosity and the 
average neutrino energy of any neutrino flavor during the shock reheating phase
 increase when switching from Newtonian to GR hydrodynamics. 
 They also pointed out that the increase is larger in 
 magnitude compared to the decrease due to redshift effects and gravitational
 time dilation. By employing the current best available weak interactions,
 \citet{Lentz11} reported the update of \citet{Bruenn01} very recently.
  They showed that the omission of observer corrections in the transport equation 
 particularly does harm to drive the neutrino-driven explosions. 
 A disadvantageous trend of GR to produce the neutrino-driven explosions has been
  commonly observed in these 
 full-fledged 1D simulations; the residency time of material
 in the gain region becomes shorter due to the stronger gravitational pull. 
Due to this competition between the gain and loss effects in the end, GR
  works disadvantageously to facilitate the neutrino-driven explosions 
in 1D. In fact, the maximum shock extent in the postbounce phase 
 is shown to be 20\% smaller when switching from Newtonian to GR hydrodynamics 
 (e.g., Figure 2 in \cite{Lentz11}). 

 Among the most up-to-date multi-D models with 
 spectral neutrino transport mentioned earlier (Table 1), 
 the GR effects are at best attempted to be modeled by using a modified gravitational 
 potential that takes into account a 1D, 
 post-Newtonian correction \citep{Buras06a,buras06,marek,bruenn}.
 A possible drawback of this prescription is that a conservation law for the 
 total energy
 cannot be guaranteed by adding an artificial term in the Poisson equation. Since
 the energy reservoir of the supernova engines is the gravitational binding energy,
 any potential inaccuracies in the argument of gravity should be avoided.
There are a number of relativistic simulations of massive stellar collapse 
 in full GR (e.g., 2D \citep{Shibata05a} or 3D 
  \citep{Shibata05b,Ott2007}, and references therein) 
or using the conformally-flatness approximation (CFC) 
(e.g., \citet{dimm02a,Isa2009}).
 Although extensive attempts have been made to include
 microphysics such as by the $Y_e$ formula \citep{Matthias05} or 
 by the neutrino leakage scheme \citep{Sekiguchi10}, 
 the effects of neutrino heating have yet to be included in them, 
which has been a main hindrance
 to study the GR effects on the multi-D neutrino-driven
 mechanism (see, however, \citet{bernhard12,kuroda12}).  

 Putting things together, 
a complete ''realistic'' supernova model should naturally be done
 in full GR(MHD) with multi-D GR
 Boltzmann neutrino transport, in which a microphysical treatment of
 equation of state (EOS) and nuclear-neutrino interactions are
  implemented as realistically as possible. Unfortunately none of the
currently published SN simulations satisfy the "ultimate" requirement (e.g., Table 1).
 In this sense,  all the mentioned studies
 employ some approximations (with different levels of sophistication)
 towards the final goal.

 In the same way, theoretical predictions of the SN multi-messengers
 that one can obtain by analyzing the currently available numerical results, cannot
 unambiguously give us the final answer yet.
 Again we hereby note the feature of this article
 which shows only a snapshot of the moving theoretical terrain.
 Keeping this caveat in mind,
 it is also true that a number of surprising GW features of CCSNe
 have been reported recently both by the 
first-principle simulations (e.g., in Table 1), and also by idealized simulations 
 in which explosions are parametrically initiated mostly
 by the light-bulb scheme. As will be mentioned in 
  the next section, the latter approach
 is also useful to get a better physical understanding of the GW signatures
 obtained in the first-principle simulations\footnote{
 In this sense, these two approaches are complimentary in understanding
 the GW signatures.}. Having shortly summarized a current status of
 CCSN simulations, we are now
 ready to move on to focus on the GW signatures from the next section.


\subsection{Gravitational waves}\label{sec2.1}

 The paper by \citet{ewald82} entitled as
"{\it Gravitational Radiation from Collapsing Rotating Stellar Cores}" unquestionably
 opened our eyes to the importance of making the GW prediction based on realistic 
 SN numerical modeling (see e.g., section 2 in \citet{ott_rev} for a 
 summary of more earlier work which had mainly focused on the
  GW emission in very idealized systems such as in homogeneous spheroids and 
 ellipsoids)\footnote{Needless to say, this kind of approach is still very important to 
extract the physics of the GW emission mechanism.}.
 As one may expect from the title of his paper, rapid rotation, 
if it would exist in the precollapse iron core, leads to
 significant rotational flattening of the collapsing and bouncing core,
 which produces a time-dependent quadrupole (or higher) GW emission.
 Following the first study by M\"uller,  most studies of the past thirty 
have focused on the so-called bounce signals 
(e.g., \citet{ewald82,mm,yama95,zwer97,kotakegw,kota04a,shibaseki,ott,ott_prl,
ott_2007,dimm02,dimmelprl,dimm08,simon,simon1}), and references therein). 

 As summarized by \citet{ott_rev}, a number of 
 important progresses have been recently made to understand features of 
 the bounce signals by extensive 2D GR studies using the CFC
 approximation \citep{dimm02,dimmelprl,dimm08} and also by fully GR 3D
 simulations \citep{ott} both including realistic EOSs and
 a deleptonization effect based on 1D-Boltzmann
 simulations \citep{matthias_dep}.

For the bounce signals having 
a strong and characteristic signature, the iron core must rotate enough rapidly.
However recent stellar evolution 
calculations suggest that rapid rotation assumed in most of the previous studies 
is not canonical for progenitors with neutron star formations
\citep{maeder00,hege05,ott_birth}.
To explain the observed rotation periods of radio pulsars, 
the precollapse rotation periods are estimated to be larger than 
$\sim$ 100 sec 
\citep{ott_birth}. In such a slowly rotating case, the detection of the 
bounce signals becomes very hard even by next-generation laser interferometers 
for a Galactic supernova (e.g., \citet{kota04a}).

Besides the rapid rotation, 
convective matter motions and anisotropic neutrino emission in the postbounce phase
 are expected to be primary GW sources with comparable amplitudes to the bounce signals
 (e.g., \citet{kotake_rev11} for a review). Thus far, various physical ingredients for producing asphericities
 and the resulting GWs in the postbounce phase have been studied, such as the roles 
of pre-collapse density inhomogeneities \citep{burohey,muyan97,fryersingle,fryer04a},
 moderate rotation of the iron core \citep{mueller04}, nonaxisymmetric rotational 
instabilities \citep{rampp,ott_3D}, g-modes \citep{ott_new} and r-modes pulsations 
\citep{nils} of PNSs, and the SASI (\citet{kotake07,kotake_ray,kotake09,marek_gw,murphy})\footnote{see also \citet{ott2011} for the GW signals at 
the black hole formation.}. 
Among them, the most promising GW sources may be convection and 
 SASI, because the degree of the initial inhomogeneities \citep{arnett} and 
the growth of rotational instabilities as well as the r-modes 
 and g-modes pulsations are rather uncertain.

 Based on 2D simulations that parametrize the 
 neutrino heating and cooling by the light-bulb scheme (which we shortly call as 
  {\it parametric SASI simulations} hereafter), we pointed out 
that the GW amplitudes from anisotropic neutrino emission\footnote{
As anisotropic matter motions generate GWs, anisotropic neutrino emission also gives 
rise to GWs, which has been originally pointed out in late 1970's by 
\citet{epstein,turner1979} (see recent progress in \citet{favata}). It is expected as a primary GW source also in 
 gamma-ray bursts \citep{hiramatsu,suwa09} and Pop III stars \citep{suwa07b}.} increase almost monotonically 
with time, and that such signals
 may be visible to next-generation detectors for a Galactic source 
\citep{kotake07,kotake_ray}. By performing such a parametric simulation but without 
 the excision inside the PNS, \citet{murphy} showed 
that the GW signals from matter motions can be a good indicator 
of the explosion geometry.
These features qualitatively agree with the ones obtained by
\citet{yakunin} who reported 
  exploding 2D simulations in which the ray-by-ray MGFLD neutrino transport 
is solved with the hydrodynamics (e.g., Oak-Ridge+ simulations in Table 1).
 \citet{marek_gw} analyzed the GW emission based on 
their long-term 2D ray-by-ray Boltzmann simulations, which seem very close to produce 
 explosions \citep{marek} (e.g., MPA simulations in Table 1).
 They also confirmed that the GWs from neutrinos 
with continuously growing amplitudes (but with the different sign of the 
amplitudes in \citet{kotake07,kotake_ray,yakunin}), 
are dominant over the ones from matter motions. 
They proposed that the third-generation class detectors 
such as the Einstein Telescope are required for detecting the GW signals 
with a good signal-to-noise ratio.

Regarding the GW predictions in {\it 3D} models, 
 \citet{muyan97} coined the first study to analyze the GW signature of 3D non-radial
 matter motion and anisotropic neutrino emission from prompt convection in the outer 
 layers of a PNS during the first 30 ms after bounce. 
 Their first 3D calculations using the light-bulb recipe 
were forced to be performed in a wedge of opening 
angle of $60^{\circ}$.
 Albeit with this limitation (probably coming from
 the computer power at that time), they obtained important findings that because of smaller convective activities inside the cone with slower 
overturn velocities, the GW amplitudes of their 3D models are more than a factor of
 10 smaller
 than those of the corresponding 2D models, and the wave amplitudes from neutrinos
 are a factor of 10 larger than those due to non-radial matter motions.
  With another pioneering (2D) study by Burrows \& Hayes (1996) \cite{burohey},
 it is worth mentioning that
 those early studies had brought new blood into the conventional GW predictions, 
 which illuminated the importance of the theoretical prediction of the 
neutrino GWs.

A series of findings obtained by Fryer et al. in early 2000s \citep{frye02,chris04} 
have illuminated also the importance of the 3D modeling. By running
 their 3D Newtonian Smoothed-Particle-Hydrodynamic (SPH) code coupled to 
 a gray flux-limited neutrino transport scheme, they studied the GW emission
 due to the inhomogeneous core-collapse, core rotation, low-modes convection, 
and anisotropic neutrino emission.
 Although the early shock-revival and the subsequent
 powerful explosions obtained in these 
 SPH simulations have yet to be confirmed by other groups, their approach
 paying particular attention to the multiple interplay between the explosion dynamics, 
the GW signatures, the kick and spins of pulsars, and also the non-spherical explosive 
nucleosynthesis, blazed a new path on which current supernova studies are
 progressing.

We also studied the GW signals from 3D models that mimic neutrino-driven explosions
 aided by the SASI \citep{kotake_ray,kotake09,kotake11}.  In the series of 
 our 3D experimental simulations, the light-bulb scheme was 
used to obtain explosions and the initial conditions were derived from a steady-state 
approximation of the postshock structure and the dynamics only outside an inner boundary
 at 50 km was solved. Based on the results, we show in the following that 
 features of the gravitational waveforms obtained 
 in 3D models are significantly different than those in 2D, 
which tells us a necessity of 3D modeling for a reliable prediction of 
 the GW signals.

\begin{figure}[hbtp]
\epsscale{.7}
\plotone{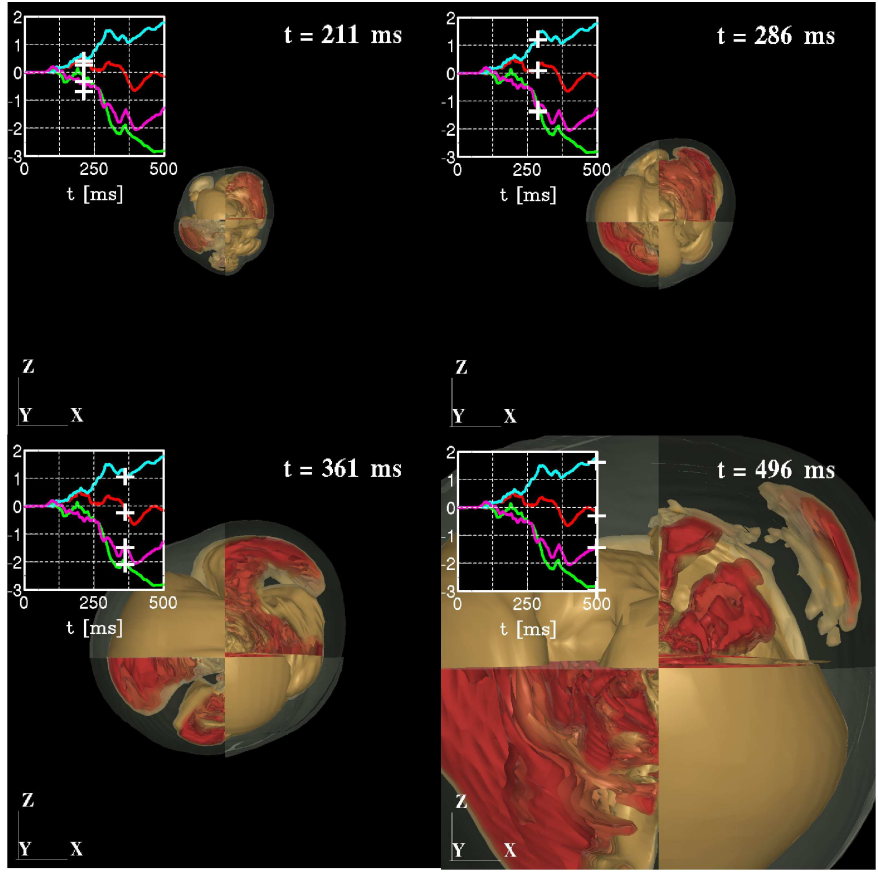}
\caption{Four snapshots of the entropy distributions of a representative 
 3D supernova explosion model (corresponding to model A in \citet{kotake09}).
The second and fourth quadrant of each panel shows 
the surface of the standing shock wave.
In the first and third
quadrant, the profiles of the 
high entropy bubbles (colored by red) inside the section cut by the 
$ZX$ plane are shown. The side
length of each plot is 1000km.
The insets show the gravitational waveforms from anisotropic neutrino emissions, 
with '$+$' on each curves 
representing the time of the snapshot. Note that the colors of the curves are
taken to be the same as the top panel of Figure \ref{fig2_GW}.
 This figure is taken from \citet{kotake09}.}
\label{fig1_GW}
\end{figure}

\begin{figure}[hbtp]
\epsscale{0.5}
\plotone{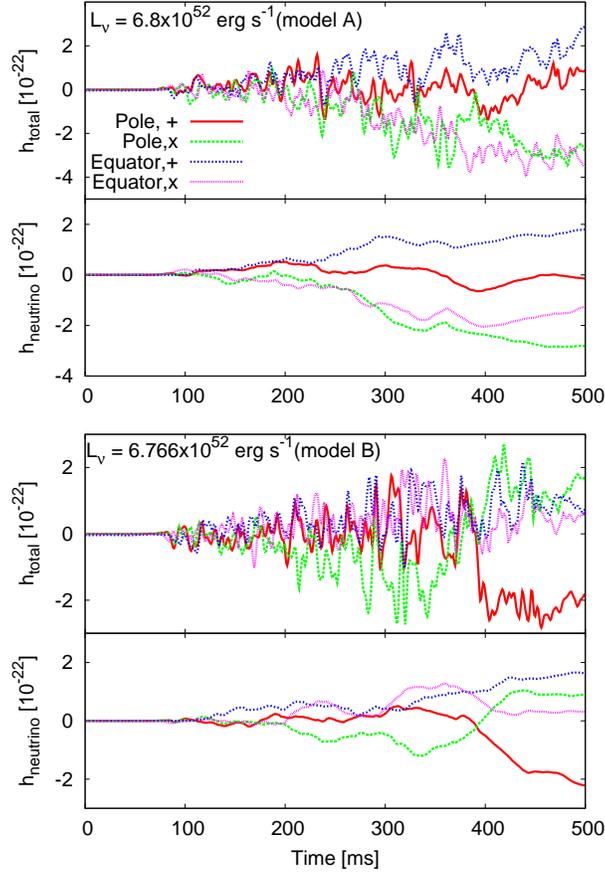}
\caption{Gravitational waveforms from neutrinos (bottom) and from the sum of 
 neutrinos and matter motions (top), seen from the polar
axis and along the equator (indicated by 'Pole' and 'Equator')
with polarization ($+$ or $\times$ modes) for two representative 3D models 
 of A and B (see \citet{kotake09} for details). 
The distance to the SN is assumed to be 10 kpc.}
\label{fig2_GW}
\end{figure}

\subsubsection{Stochastic Nature of Gravitational Waves in 3D simulations}\label{stochastic}

Figure \ref{fig1_GW} shows the evolution of 3D hydrodynamic features
 from the onset of the non-linear regime of SASI (top left) 
until the shock break-out\footnote{This corresponds to the shock emergence
at the outer boundary of the computational domain ($\sim 2000$ km in radius).}(bottom right) with the gravitational waveform from neutrinos 
inserted in each panel. After about $100$ ms, the deformation of the standing 
shock becomes remarkable marking the epoch when the SASI enters the non-linear regime 
(top left of Figure \ref{fig1_GW}).
At the same time, the gravitational amplitudes 
begin to deviate from zero. Comparing the top two panels in Figure \ref{fig2_GW}, 
which shows the total amplitudes (top panel, neutrino + matter) and the neutrino contribution only (bottom),
 it can be seen that the overall structures of the waveforms are 
 predominantly determined by the neutrino-originated GWs with the slower temporal 
 variations ($\gtrsim 30 - 50$ ms), to which the GWs from matter motions 
 with rapid temporal variations ($\lesssim 10$ ms) are superimposed. 

As seen from the top right through bottom left to right panels of Figure
\ref{fig1_GW}, the major axis of the growth of SASI 
is shown to be not aligned with the symmetric axis (:$Z$ axis in the figure)
and the flow inside the standing shock wave is not symmetric with
respect to this major axis (see the first and third quadrant
in Figure \ref{fig1_GW}). This is a generic feature in the computed 3D
models, which is in contrast to the axisymmetric case.
The GW amplitudes from SASI in 2D showed an increasing trend with time  
due to the symmetry axis, along which SASI can develop
preferentially \citep{kotake07,kotake_ray}.
Free from such a restriction, 
 a variety of the waveforms is shown to appear (see waveforms inserted in Figure 5).
Furthermore, the 3D standing shock can also oscillate in all
 directions, which leads to the smaller explosion anisotropy than 2D.
 With these two factors, the maximum amplitudes seen either from the
equator or the pole becomes smaller than 2D.
On the other hand, their sum in terms of the total radiated
energy are found to be 
almost comparable between 2D and 3D models, which is likely to imply the
energy equipartition with respect to the spatial dimensions.

 The (two) pair panels of Figure \ref{fig2_GW} show the gravitational waveforms 
for models with different neutrino luminosity.  The input luminosity for 
the pair panels (models A (top two) and B (bottom two))
 differs only $0.5\%$\footnote{Although any 
 seed perturbation would demonstrate the chaotic behavior, we simply chose
 the value of 0.5 \% as a reference.}. 
Despite the slight difference in the luminosities, 
the waveforms of each polarization 
are shown to exhibit no systematic similarity when seen from the pole or equator.
This also reflects a chaotic nature of the SASI influenced by small
 differences (see also \citet{ewald11}). 

It should be noted that the approximations taken in the simulation, 
such as the excision inside the PNS with its fixed inner boundary and the light bulb
approach with the isotropic luminosity constant with time, 
are the very first step to model the dynamics of the neutrino-heating 
 explosion aided by SASI and study the resulting GWs.
 As already mentioned, the excision of the central regions inside PNSs 
 may hinder the efficient gravitational emission of the oscillating neutron star
\citep{ott_new} and the non-axisymmetric instabilities 
 \citep{ott_3D,simon1} of the PNSs, and the enhanced neutrino emissions inside 
the PNSs \citep{marek_gw}.  Bearing these caveats in mind,
 a piece of encouraging news is that the gravitational waveforms obtained in the 
 2D radiation-hydrodynamic simulations \citep{yakunin}
are similar to the ones obtained in our 2D study using the light-bulb scheme 
 \citep{kotake_ray}. 
 More recently, \citet{ewald11} confirmed the stochastic nature 
by analyzing the GW features obtained in their 3D models \citep{annop}.
 The obtained GW amplitudes as well as the degree of anisotropic neutrino emission
 are almost comparable to our results, which are considerably smaller than 2D models
 \citep{mueller04,marek_gw,murphy,yakunin}.

\begin{figure}[hbtp]
  \begin{center}
    \begin{tabular}{cc}
\resizebox{75mm}{!}{\includegraphics{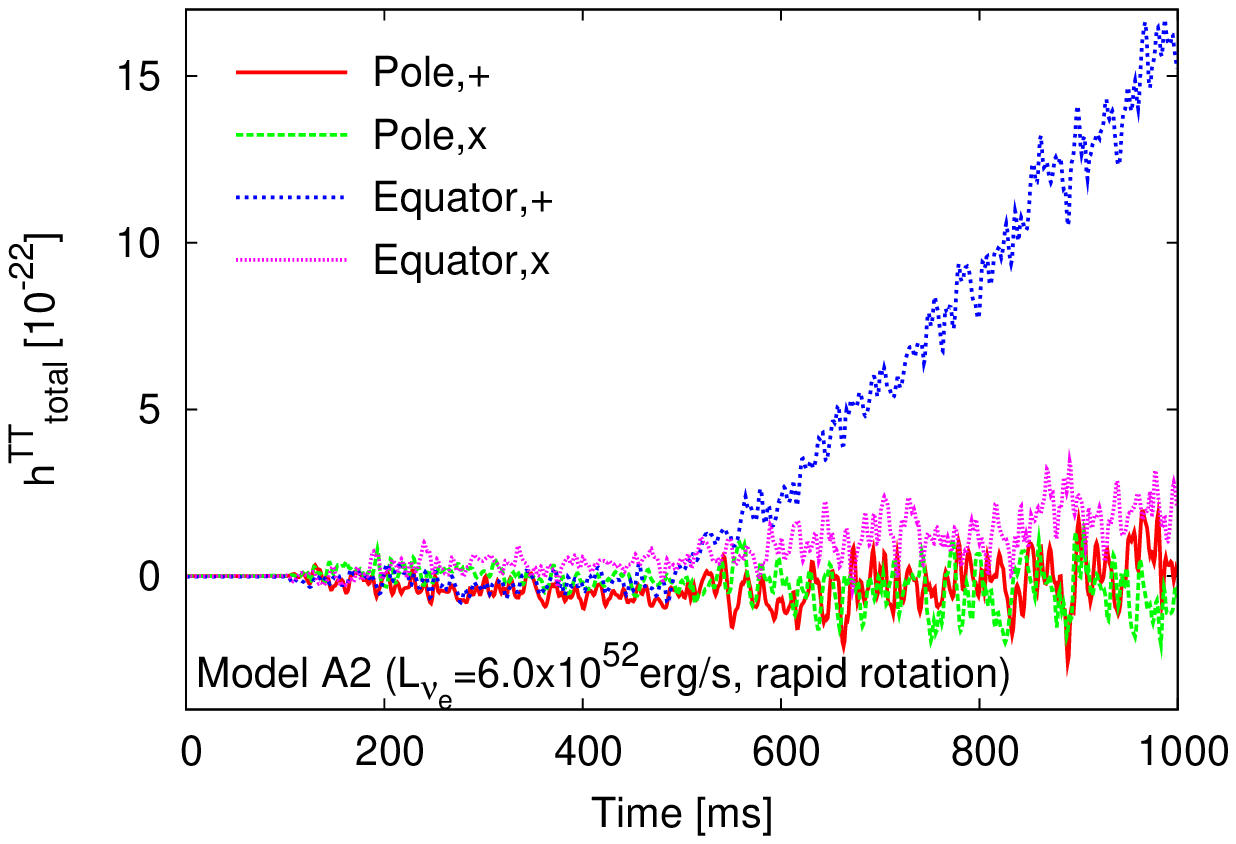}} &
\resizebox{75mm}{!}{\includegraphics{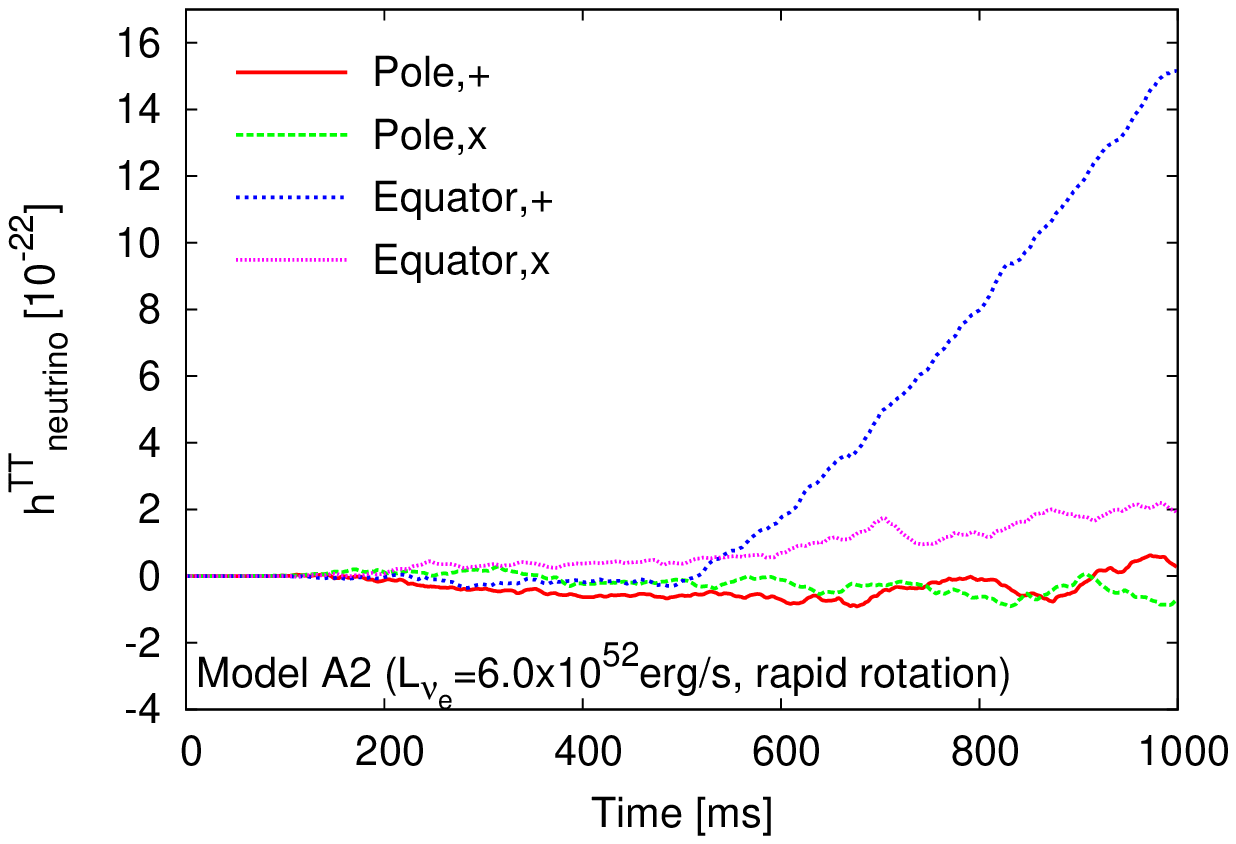}} \\
    \end{tabular}
   \caption{Gravitational waveforms from the sum of neutrinos and matter motions 
(left) and only from neutrinos (right) for a 3D model with rotation (from 
 \citet{kotake11}). 
The time is measured from 
the epoch when the neutrino luminosity is injected from the surface of the 
neutrino sphere. For this 3D model with rotation, the rotational flow is imposed
  to advect to the PNS surface at around $t =400$ ms. The supernova is assumed to be 
located at the distance of 10 kpc.} 
  \label{fig3_GW}
  \end{center}
\end{figure}

\begin{figure}[hbtp]
\begin{center}
 \begin{tabular}{cc}
      \resizebox{80mm}{!}{\includegraphics{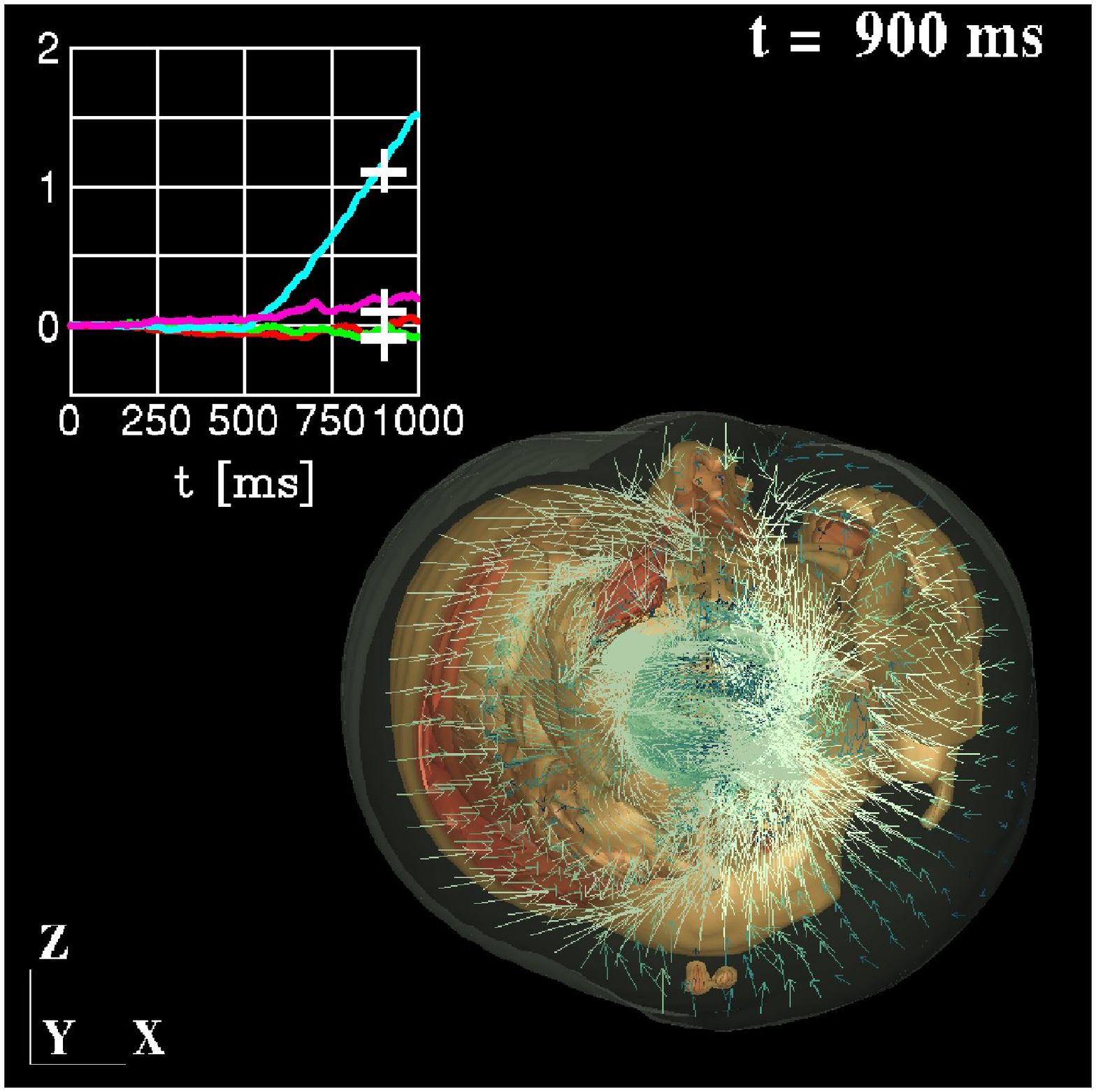}} &
      \resizebox{80mm}{!}{\includegraphics{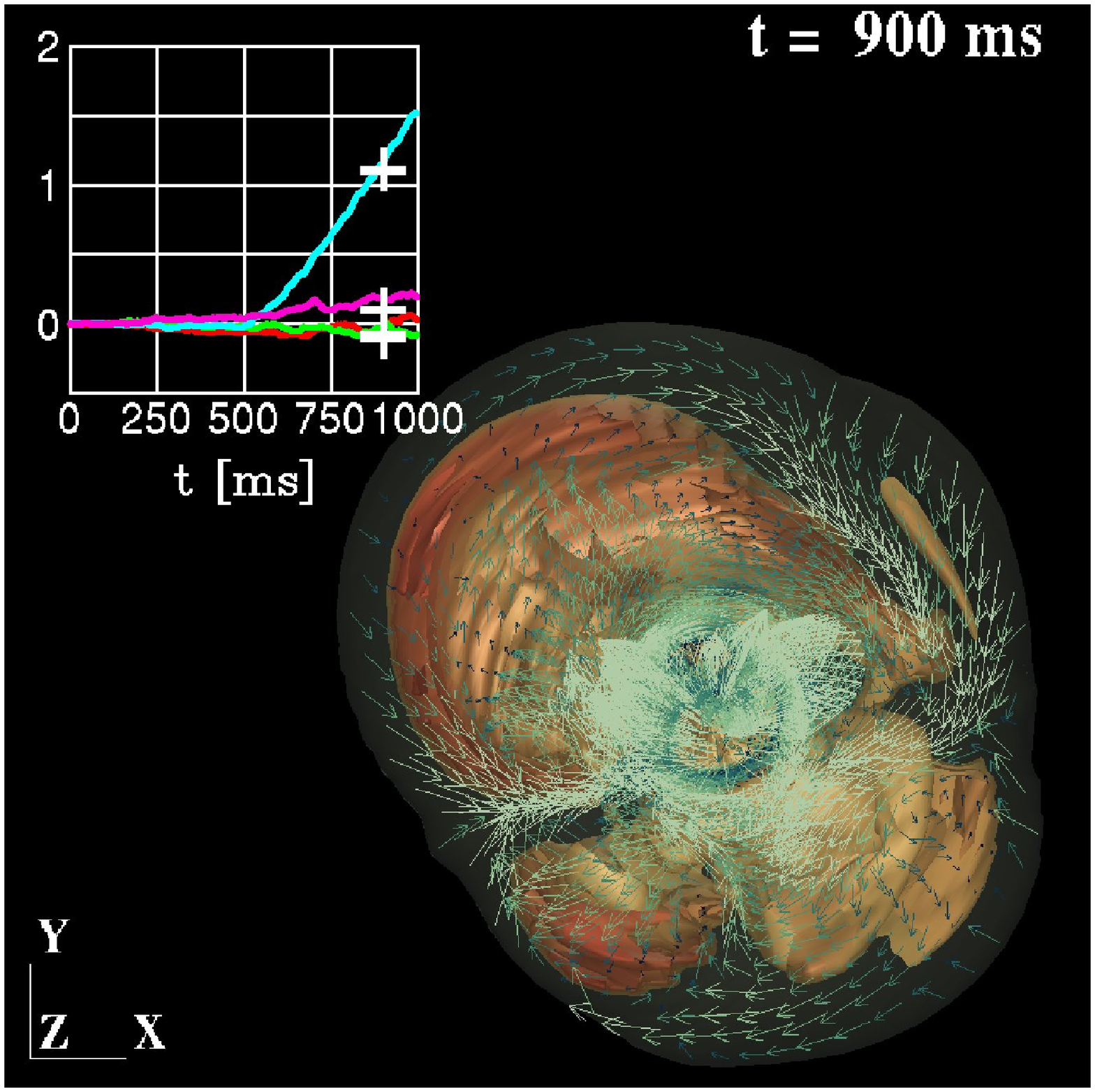}} \\
    \end{tabular}
\end{center}
\caption{Partial cutaway of the entropy isosurfaces and the velocity vectors on the 
 cutting plane for a 3D model that includes rotation. Left and right panels are for the equatorial
 and polar observer, respectively. The insets show the gravitational waveforms with '$+$' on each curves 
representing the time of the snapshot. Note that the colors of the curves are
taken to be the same as the top panel of Figure \ref{fig2_GW}. 
This figure is taken from \citet{kotake11}.}
\label{fig4_GW}
\end{figure}

\subsubsection{Breaking of the Stochasticity Due to Stellar Rotation}\label{break}

More recently, we studied the effects of stellar rotation on the stochastic 
nature of the GWs mentioned above \citep{kotake11}. 
In 3D, the modes of SASI are divided into {\it sloshing} modes 
and {\it spiral} modes (e.g., \citet{iwakami1}).
 Asymmetric $m = 0$ modes so far studied in 2D models 
and axisymmetric $m\neq 0$ modes are classified into the {\it sloshing} modes,
where $m$ stands for the azimuthal index of the spherical harmonics $Y_{l}^{m}$.
In the latter situation, the $\pm m$ modes degenerate
so that the $+m$ modes has the same amplitudes as the $-m$ modes.
If random perturbations or uniformly rotating flows are imposed on 
these axisymmetric flows in the postbounce phase, 
the degeneracy is broken and the rotational modes emerge
 \citep{blo07,iwakami2}.
In this situation, the $+m$ modes has the different amplitudes from $-m$ modes.
Such rotating non-axisymmetric $m\neq0$ modes are called as {\it spiral} modes.
 These non-axisymmetric modes are expected to bring about a 
breakthrough in our supernova theory, because they can help to produce explosions 
 more easily compared to 2D due to its extra degree of freedom 
\citep{nordhaus}, and also because they may have 
a potential to generate pulsar spins (\citet{blo07,rantsiou}, see also \citet{yamasaki,annop}).
 These finding illuminating the importance of stellar rotation motivated us to clarify 
how rotation that gives a special direction to the system (i.e., the 
 spin axis), could affect the stochastic GW features which we observed in 
 the absence of rotation (section \ref{stochastic}).

Figure \ref{fig3_GW} 
shows the gravitational waveform for a typical 3D model 
with rotation (left:total amplitudes,
 right:neutrino only). To construct a model with rotation,
we give a uniform rotation on the flow advecting from the outer boundary 
of the iron core as in \citet{iwakami2}, whose specific angular momentum is 
assumed to agree with recent stellar evolution models \citep{maeder00,hege05}.
 Comparing to Figure \ref{fig2_GW} (for models without rotation),
  one can clearly see a sudden rise in the GW amplitude after around 500 ms
 for the rotating model (blue line in Figure \ref{fig3_GW}), which is 
 the plus mode of the neutrino GW seen from the equator.
 By systematically changing the initial angular 
momentum and the input neutrino luminosities from the 
 PNS, we computed fifteen 3D models in \cite{kotake11} and found 
 that the GW features mentioned 
 above were common among the models.

\begin{figure}[hbtp]
\epsscale{0.6}
\plotone{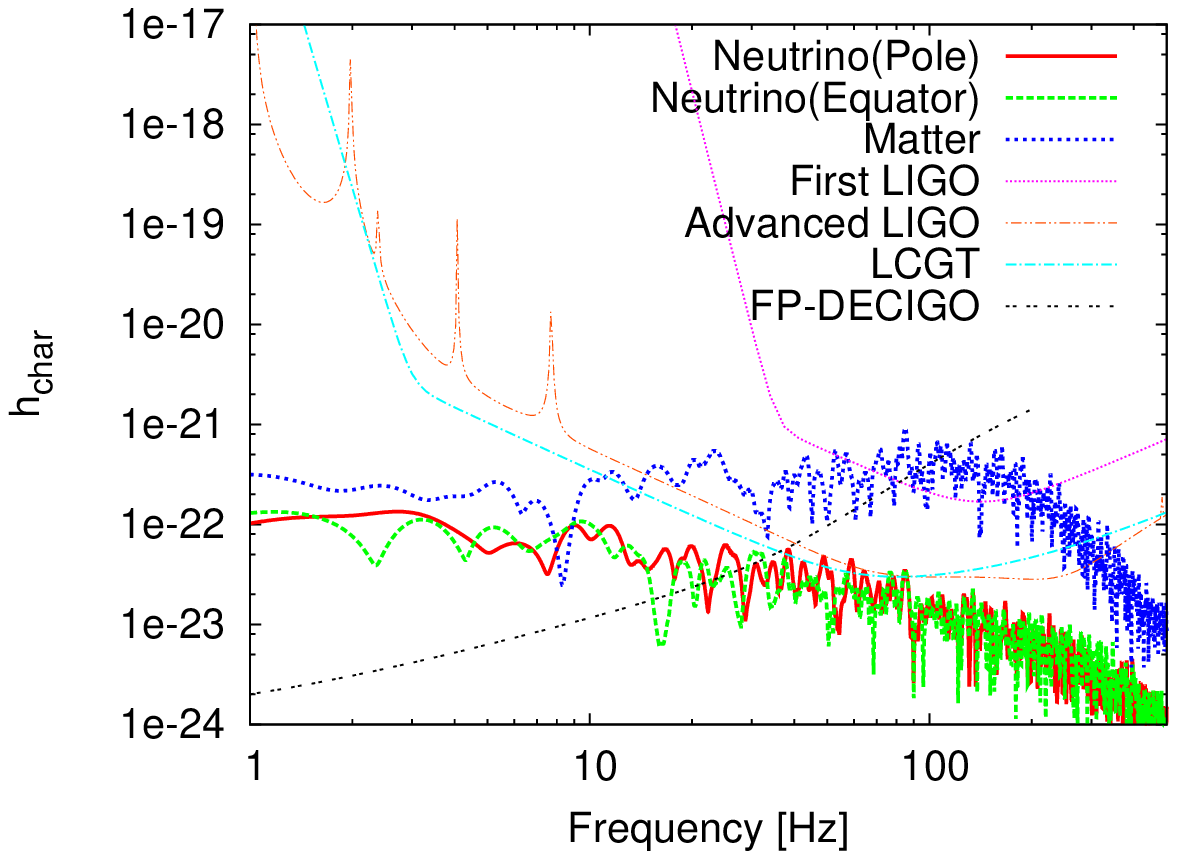}
\caption{Spectral distributions of GWs from matter motions (``Matter'') 
 and neutrino emission (``Neutrino'') seen from the pole or the equator 
for a representative 3D rotating model (e.g., \citet{kotake11}) with 
the expected detection limits of TAMA300 \citep{tamanew}, first LIGO and 
 advanced LIGO \citep{firstligo}, Large-scale
 Cryogenic Gravitational wave Telescope (LCGT) \citep{lcgt} and Fabry-Perot type 
 DECIGO \citep{fpdecigo}.
 It is noted that $h_{\rm char}$ is 
the characteristic gravitational wave strain defined in \citet{flanagan}. 
 The distance to the supernova is assumed to be 10 kpc.
 Note that for the matter signal, the $+$ mode seen from the polar direction is plotted
 (from \citet{kotake11}).}
\label{fig3}
\end{figure}

 Figure \ref{fig4_GW} illustrates a typical snapshot of the flow fields for the 
rotating model (corresponding to the one in Figure \ref{fig3_GW}) when the spiral SASI modes have already entered the non-linear regime,
 seen from the pole (right panel) or from the equator (left panel), respectively.
  From the left panel, one may guess the presence of
 the sloshing modes that happen to develop along the rotational axis ($z$-axis) at this 
 epoch. It should be emphasized that although the dominance of $h^{\rm equ}_{\nu,+}$ 
 observed in the current 3D simulations is similar to the one 
 obtained in previous 2D studies \citep{kotake_ray}, its origin has nothing 
 to do with the coordinate symmetry axis. The preferred direction here is determined by 
 the spin axis. Free from the 2D axis effects, the major axis of the sloshing
 SASI mode changes stochastically with time, and the flow patters behind the
 standing shock simultaneously change in every direction 
 like the non-rotating models.
 As a result, the sloshing modes can make only a small contribution to the GW emission.
  The remaining possibility is that the spiral flows seen in the right panel should be a 
key importance to understand the GW feature mentioned above. In fact, by analyzing 
 the matter distribution on the equatorial plane, we find that the compression of matter 
is more enhanced in the vicinity of the equatorial plane due to the growth of the 
spiral SASI modes, leading to the formation of the spiral flows circulating around the 
spin axis with higher temperatures. As a result, the neutrino emission seen parallel 
 to the spin axis becomes higher than the ones seen from the other direction.
 Remembering that the lateral-angle ($\theta$) dependent function 
 of the GW formulae (e.g., in equation (9) in \cite{kotake_ray}) is positive near 
the north and south polar caps, the dominance of the polar neutrino luminosities 
leads to make the positively growing feature of $h^{\rm equ}_{\nu,+}$ in Figure 
\ref{fig3_GW} (blue line).
 From the spectral analysis of the gravitational waveform 
(Figure \ref{fig4_GW}), it can be readily seen that 
 it is not easy to detect these neutrino-originated GW signatures
 with slower temporal evolution ($\gtrsim O(10)$ms) by ground-based detectors whose 
sensitivity is 
limited mainly by the seismic noises at such lower frequencies. 
However these signals may be detectable by the recently proposed future space 
interferometers like Fabry-Perot type DECIGO (\citet{fpdecigo}, black line in Figure 
\ref{fig4_GW}).
 Contributed by the neutrino GWs in 
the lower frequency domains,
 the total GW spectrum tends to become rather flat over a broad frequency range
 below $\sim 100$ Hz. These GW features obtained in the context of the 
SASI-aided neutrino-driven mechanism are different from the ones expected in the 
other candidate mechanisms, such as the MHD mechanism (e.g., 
\citet{taki_kota}) and
 the acoustic mechanism \citep{ott_new}. Therefore the detection of such signals 
 could be expected to provide an important probe into the explosion mechanism 
(e.g., \citet{ott_rev,ott2009}).

 We like to draw a caution that most of the 3D models cut out the PNS and 
the neutrino transport is approximated by a simple light-bulb
  scheme \citep{kotake11} or by the gray transport scheme \citep{ewald11}. Needless
 to say, these exploratory approaches are but 
the very first step to model the neutrino-heating 
 explosion and to study the resulting GWs.
 As already mentioned, the excision of the central regions inside PNSs 
 truncates the feedback between the mass accretion to the PNS and the resulting 
 neutrino luminosity, which should affect the features of the neutrino GWs.  
 By the cut-out, efficient GW emission of the oscillating neutron star
\citep{ott_new} and non-axisymmetric instabilities 
 \citep{simon,simon1} of the PNSs, and the enhanced neutrino emissions inside 
the PNSs \citep{marek_gw} cannot be treated in principle.
 To elucidate the GW signatures in a more quantitative manner,
 full 3D simulations with spectral neutrino transport 
 are apparently needed (e.g., section 2.1).
 This is unquestionably a vast virgin territory awaited to be 
 explored for the future.

\subsection{Explosive nucleosynthesis}\label{sec2.2}
 In this section, we proceed to discuss possible signatures of supernova 
nucleosynthesis. The study of nucleosynthesis is of primary importance to
 unveil the origins of heavy elements. It could also provide a valuable information 
of the ejecta morphology by observing the aspherical distributions of the synthesized 
elements especially for a nearby CCSN event\footnote{Note that nucleosynthesis 
is not critical for the modeling of the light-curve and spectra for the most frequent 
types of SNe II-P.}.
In the following, we first present a short overview paying particular
 attention to explosive nucleosynthesis, and then discuss 
possible observational signatures that would imprint information of 
multidimensionalities of the supernova engine. 

When in a successful explosion the shock passes through the outer shells, its high
 temperature induces an explosive nucleosynthesis on short timescales 
(e.g., \citet{woos02,woos95,hashimoto95}, and collective references in \citet{jerk11}). 
 The observational determination of the masses of the three main radioactive 
 isotopes $^{56}{\rm Ni}$, $^{57}{\rm Ni}$, and $^{44}{\rm Ti}$ sets one 
 of the main constraints on the explosion dynamics, because the production of 
 these elements is sensitive to the track of density and temperature that the 
expanding material traces (e.g., \citet{woos91}). 
During the shock propagation, iron group elements such as  $^{56}{\rm Ni}$ and its 
 daughter nucleus $^{56}{\rm Co}$ are predominantly produced, which are radioactive
 with a life-time of 8.8 days and 111.5 days, respectively. Most CCSNe enter the so-called nebular phase after the first few months
 when the expanding ejecta becomes optically thin in the continuum.
 In the early nebular phase, $^{56}{\rm Co}$ is the major nuclear power source. 
 As long as the decay particles are trapped by the ejecta, the radiation energy
 supplied by radioactivity is emitted instantaneously, so that the light curve 
 can be described by an exponential decay with time, simply tracing
 the decay of the $^{56}{\rm Co}$ nuclide.
 To explain the bolometric light curve of SN1987A in such a phase,
the $^{56}{\rm Ni}$ mass was determined to be $0.07 M_{\odot}$ \citep{bouchet91}. 

 After several years of explosion, the radioactive output from the ejecta no longer
 balances with the instantaneous input by radioactivity, because the reprocessing
timescale is going to be longer \citep{fransson93}. The bolometric light curve is 
 affected by the delayed release of the ionization energy. 
 After that, a self-consistent modeling is needed, in which one should 
include a detailed 
calculation of the gamma-ray/positron thermalization and a determination of 
the time-dependent temperature, ionization, and excitation 
(e.g., \citet{fransson93,mazzali01,maurer11,jerk11a} and references therein).
 Such a time-dependent modeling by \citet{fransson93} revealed the $^{57}{\rm Ni}$ 
mass of $\sim 3.3 \times 10^{-3} M_{\odot}$ of SN1987A, 
which agrees well with observations (e.g., \citet{varani90,kurfess92} and 
 collective references in \citet{filippenko97}).

By the similar reason to $^{57}{\rm Ni}$ just mentioned above, the determination of 
the $^{44}{\rm Ti}$ is also complicated. The most recent 
 study by \citet{jerk11a} 
gives an estimate of the $^{44}{\rm Ti}$ to be $1.5^{+0.5}_{-0.5} 
\times 10^{-4} M_{\odot}$, which is in good 
 agreement with the eight-year spectrum analysis of SN1987A (e.g., \citet{chugai97},
 see also \citet{lund01}). As shown above, 
the amount of $^{44}{\rm Ti}$ is typically one order-of-magnitude 
smaller than that of $^{57}{\rm Ni}$, however it is crucially important 
for young supernova remnants due to its long life-time ($\sim$ 86 years). 
 It is worth mentioning that NASA will launch the satellite NuSTAR (Nuclear 
 Spectroscopic Telescope Array) to study $^{44}{\rm Ti}$ production in CCSNe. The detector will be able to map out the $^{44}{\rm Ti}$ distribution of the supernova remnant Cassiopeia A and can get velocity distributions of the $^{44}{\rm Ti}$
 in SN 1987A. By comparing detailed modeling of the SN nucleosynthesis in
 the context of 2D and 3D models (e.g., \citet{fryer06,magko}), these are expected to
 both provide direct probes of the explosion asymmetry.
 
Ever since SN1987A, challenges to the classical spherical modeling
 \citep{hashimoto95,woos95,tnh96,rauscher02} have been building also in the 
SN nucleosynthesis (likewise in the explosion theory and GWs mentioned so far).
For many years it has been customary to simulate explosions and the 
 effects of the shock wave on the explosive nucleosynthesis 
by igniting a thermal bomb in the star's interior or by initiating the explosion by a 
strong push with a piston. 2D simulations with manually imparted asymmetries showed
 that bipolar explosion scenarios could account for enhanced $^{44}{\rm Ti}$ synthesis  
along the poles as indicated in SN1987A (e.g., \citet{naga97}).
More recently, 3D effects have been more elaborately studied (\citet{hunger05},
 see also \citet{young06}) as well as the impacts of different explosions by 
employing a number of progenitors 
\citep{jogg10} or by assuming a jet-like explosion~\citep{couch09,tominaga09}, 
which is one of the possible candidates of hypernovae (e.g., \citet{maeda03}).


In addition to the above-mentioned work, nucleosynthesis in a more realistic
simulation that models multi-D neutrino-driven explosions 
has been also extensively studied 
(e.g., \citet{kifo03,kifo06,gawry10} and references therein).
Although a small network has ever been included in the computations, 
these 2D simulations employing a light-bulb scheme \citep{kifo03}
or a more accurate gray transport scheme \citep{scheck06,kifo06}
have made it possible to elucidate nucleosynthesis inside from the iron 
core after the shock-revival up to explosion in a more consistent manner.
\citet{kifo06} demonstrated that the SASI-aided low-mode explosions 
can naturally explain masses and
distributions of the synthesized elements observed in SN1987A.
Their recent 3D results by \citet{hammer10} show that 
the 3D effects change the velocity profiles as well as the
growth of the Rayleigh-Taylor instability, which affects properties of the 
SN ejecta. In simulations with spectral neutrino transport, 
a new nucleosynthesis process, the so-called $\nu$p process, 
is reported to successfully explain some light 
proton-rich($p$-) nuclei including \nuc{Mo}{92,94} and \nuc{Ru}{96,98} (e.g., 
\citet{pruet,froehlich,wanajo} and references therein). It should be noted 
 that self-consistent 
simulations are currently too computationally expensive to follow the dynamics of 
 the expanding shock far outside the central iron core ($\gg $ 1 s after bounce) 
where explosive nucleosynthesis takes place. On the 
 other hand, the light-bulb scheme is not 
 accurate to determine a sensitive balance between neutrino captures proceeding 
via $\nu_e$ or $\bar{\nu}_e$, which is decisive for 
quantifying the $\nu$p processes.
 Therefore the two approaches, namely light-bulb scheme vs. self-consistent neutrino 
transport, may be regarded as playing a complimentary role at present.  

In the next section, we briefly summarize our findings on
 explosive nucleosynthesis in our 2D models that utilize the light-bulb 
scheme to trigger explosions (e.g., \citet{fuji11} for more details).
 By changing neutrino luminosities  
from PNSs systematically ($L_{\nu}$), we discuss how the multidimensionality formed 
  by the SASI and convection could impact on the explosive nucleosynthesis.
 We employ a non-rotating 15$M_\odot$ star with solar metalicity, which has been a
 best-studied supernova progenitor. The abundance pattern of the synthesized 
 elements is estimated by a post-processing procedure, in which 
we have adopted a large nuclear network continuously maintained by the Joint Institute 
for Nuclear Astrophysics (JINA) REACLIB project~\citep{cyburt10}.
 Our readers might wonder why we are returning to 2D results here again after 
  referring to 3D studies in section \ref{sec2.1}. This is 
 simply because our 3D project for the nucleosynthesis is currently in progress, 
 and we like to note again the feature of this review article which shows
 only a snapshot of the moving theoretical terrain.

\subsubsection{Symmetry breaking in explosive nucleosynthesis}\label{exp}

\begin{figure}[htpb]
 \epsscale{0.8}
\plotone{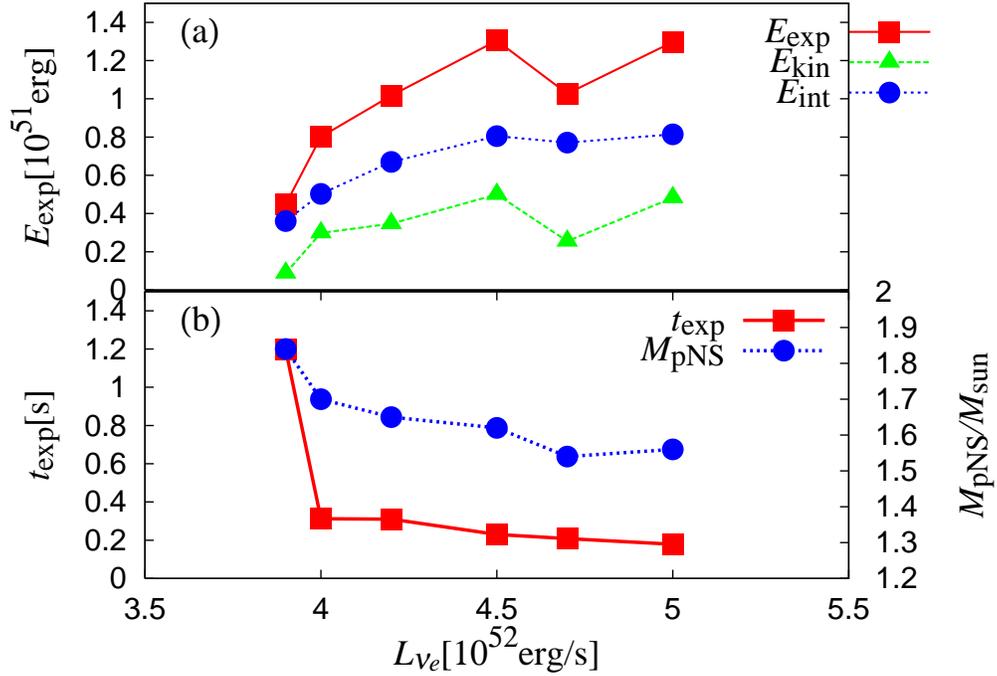}
 \caption{Panel (a) shows explosion energies (red line) vs. 
 input neutrino luminosities ($L_{\nu_e}$), where green and blue line shows 
contributions from kinetic and thermal energy, respectively. Panel (b) 
shows the mass of protoneutron star ($M_{\rm PNS}$) and the explosion timescales 
 ($t_{\rm exp}$) as a function of $L_{\nu_e}$. $M_{\rm PNS}$ is estimated as 
the enclosed mass exceeding $10^{12}{\rm g}~{\rm cm}^{-3}$ and $t_{\rm exp}$ as
 the time scale when 
the mass ejection rate at 100km drops down to $0.1 M_{\odot}/{\rm s}$ 
in our 2D simulations (from \citet{fuji11}).}
 \label{f1_nuc}
\end{figure}

 Figure \ref{f1_nuc} shows how explosion energies (top panel),
 explosion timescales as well as the PNS masses (bottom panel) change  
 with the input neutrino luminosity ($L_{\nu_e}$) for the 2D parametric 
 explosion models (e.g., \citet{fuji11} for more details). As seen, 
 higher $L_{\nu_e}$ makes explosion energy greater, 
the onset of explosion ($t_{\rm exp}$) earlier, the PNS masses ($M_{\rm PNS}$) 
smaller. The PNS 
 masses range in $(1.54-1.62) M_{\odot}$ for models with higher luminosity 
($L_{{\nu_e}} \gtrsim 4.25 \times 10^{52}~{\rm erg}~{\rm s}^{-1}$, panels (b), blue line), which are comparable to the baryonic mass 
(around $1.5 M_{\odot}$) of neutron stars observed in binaries \citep{schwab10}. 
The explosion energies for these high luminosity models
 (panel (a), red line) are also close to the canonical supernova kinetic energy
 of $10^{51}$ erg. To discuss explosive nucleosynthesis in the 
following, we thus choose to focus on the 
model with $L_{\nu_e} = 4.5 \times 10^{52} \rm \, erg \, s^{-1}$ as a reference.

\begin{figure}[htpb]
 \epsscale{1.}
\plotone{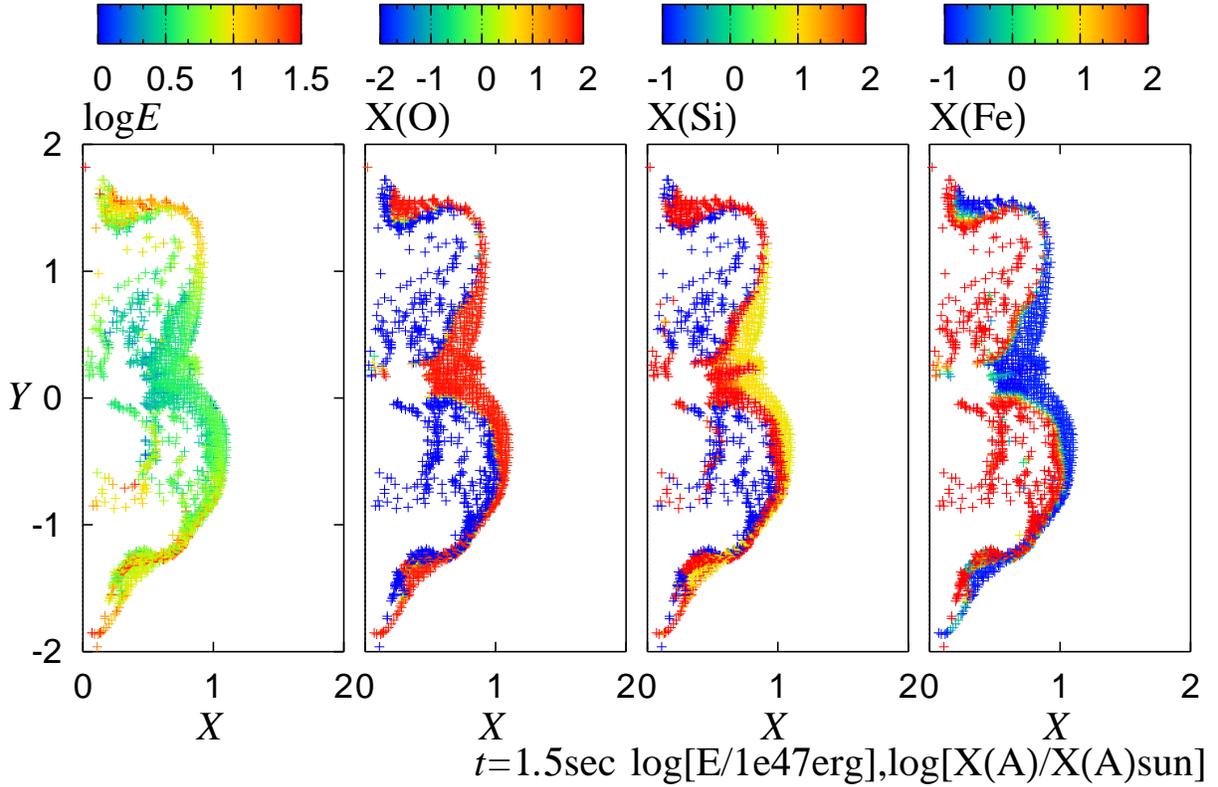} 
 \caption{Distribution of internal energy (left panel) 
and abundances of the ejecta (O, Si, Fe from left to right, normalized
 to solar) 
when the non-spherical shock is propagating in the O-rich layers at around 
$\sim 20000$ km along the pole ($\sim 1.5$ s after bounce). 
For this model, the input neutrino luminosity is set as 
$L_{\nu_e} = 4.5 \times 10^{52} \rm \, erg \, s^{-1}$.
 Note that Fe in the right panel is the sum of \nuc{Fe}{56}, 
\nuc{Ni}{56}, and \nuc{Fe}{54} (from \citet{fuji11}).}
 \label{f2_nuc}
\end{figure}

For the model, Figure \ref{f2_nuc} shows a snapshot of internal energy (left panel)
 and abundances of ejecta (O, Si, Fe) when the SASI-aided low-mode shock is propagating 
in the O-rich layers at around $\sim 20000$ km along the pole. 
 Ejecta with higher internal energies (left panel, colored by red) 
concentrate on the shock front particularly in polar regions 
($\theta < \pi/4, \theta > 3\pi/4$), where both Si and O burning proceed 
to produce abundantly \nuc{Ni}{56} and \nuc{Si}{28}. These aspherical element
 distributions might be responsible for anisotropies of the 
 SN ejecta as observed in SN1987A.
 More recently, \citet{kimura09,uchida09} have analyzed the metal distribution
 of the Cygnus loop and pointed out that the progenitor of the Cygnus loop is a 
 CCSN explosion whose progenitor mass ranges in $\sim 12-15 M_{\odot}$.
 Since the material in this middle-aged supernova remnant has not been completely 
mixed yet, the observed asymmetry is considered to still remain a trace of inhomogeneity 
 produced at the moment of explosion. The asymmetries of the SN ejecta obtained in our 
2D explosion model of a 15 $M_{\odot}$ progenitor have a close correlation with 
  the ones in the Cygnus loop (see \citet{fuji11} for more detailed comparison).
 This might allow
 us to speculate that globally asymmetric explosions induced by the SASI-aided 
neutrino-heating mechanism could account for the origin.

\begin{figure}[htpb]
\begin{center}
 \begin{tabular}{cc}
      \resizebox{82mm}{!}{\includegraphics{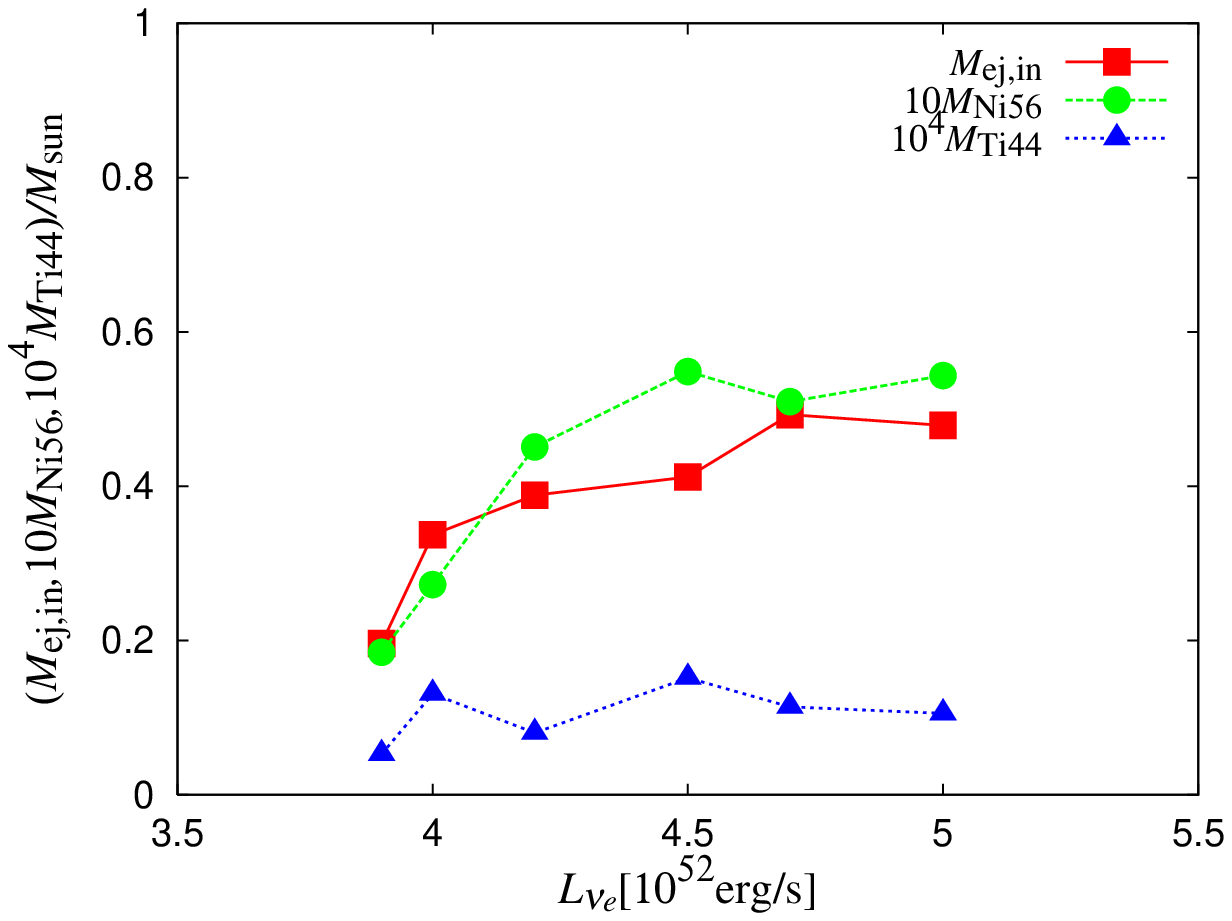}} &
      \resizebox{82mm}{!}{\includegraphics{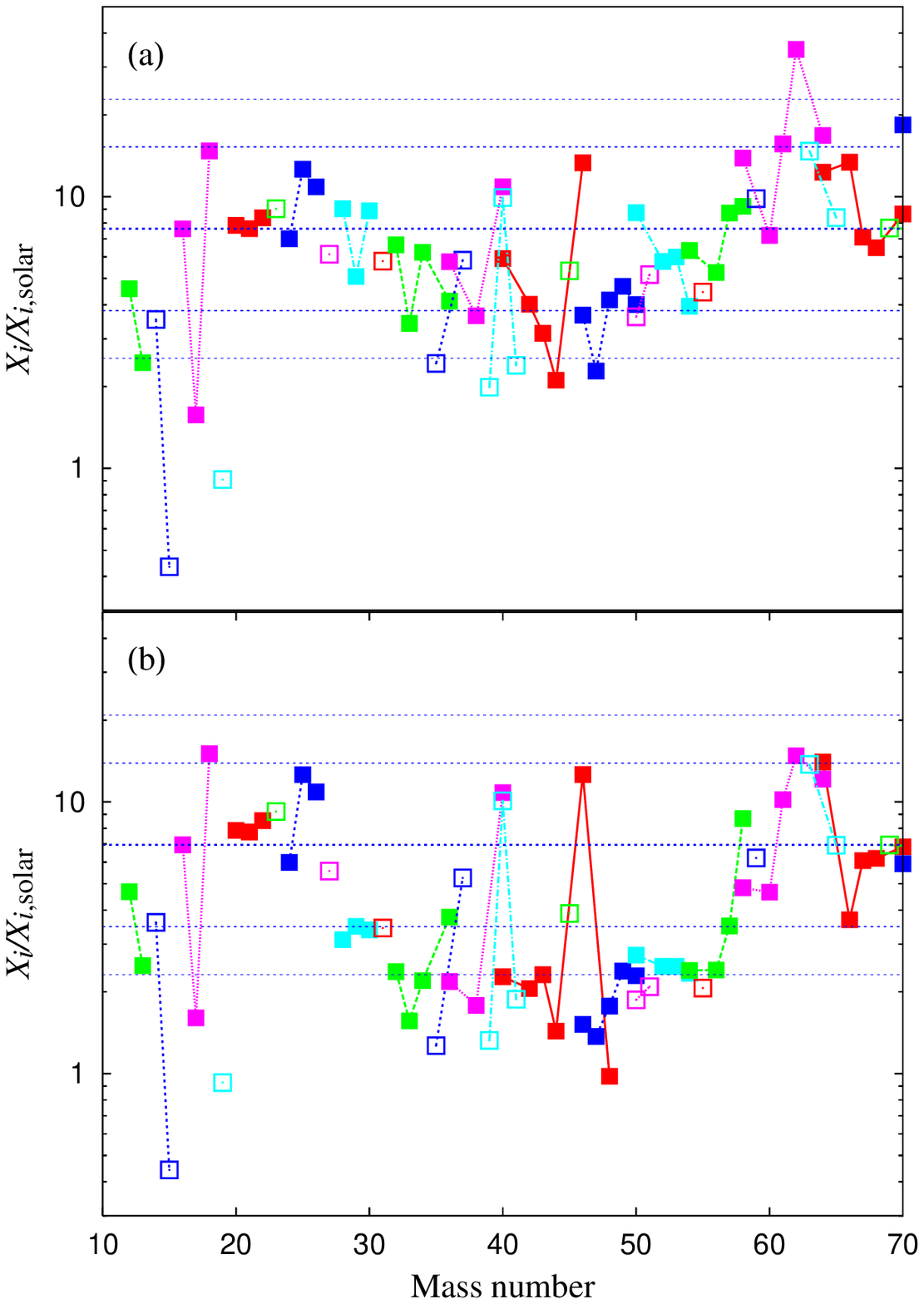}} \\
    \end{tabular}
\end{center}
 \caption{Left panel shows masses of ejecta ($M_{\rm ej,in}$), and 
masses of radioactives (\nuc{Ni}{56} and \nuc{Ti}{44}). Right panel shows
element abundance yields ($X_i/X_{\odot,i}$, normalized to solar) vs. mass number
 for our reference model of  $L_{\nu_e} = 4.5 \times 10^{52} \rm \, erg \, s^{-1}$
 and for (b) $L_{\nu_e} = 3.9 \times 10^{52} \rm \, erg \, s^{-1}$
 Thick horizontal-dashed lines represent a factor equals to that of \nuc{O}{16},
 while two normal and two thin lines denote
 a factor equals to that of \nuc{O}{16} times 2, 1/2, 3, and 1/3, respectively.
 These figures are taken from \citet{fuji11}.}
 \label{f3_nuc}
 \epsscale{1.0}
\end{figure}

Concerning the synthesized amount of \nuc{Ni}{56} and \nuc{Ti}{44}, 
they are $\sim 0.06 M_{\odot}$ 
 and $4\times 10^{-5} M_{\odot}$, respectively for the model 
(left panel in Figure \ref{f3_nuc}). For SN1987A, masses of 
\nuc{Ni}{56} and \nuc{Ti}{44} are deduced to be $\sim 0.07 M_{\odot}$ 
\citep{shige88, woosley88} and $1-2\times 10^{-4} M_{\odot}$ 
\citep[and references therein]{nagataki00}. The obtained mass of
 \nuc{Ti}{44} is, therefore, not enough for SN1987A as well as for
 Cas A ($1.6^{+0.6}_{-0.3} \times 10^{-4} M_{\odot}$) \citep{renaud06}.
 The shortage of \nuc{Ti}{44} is a long-standing problem, which might be 
 solved by 3D effects \citep{hammer10}. For the abundance patterns, our reference model 
(top panel in Figure \ref{f3_nuc}) is shown to
 reproduce a similar trend to the solar system (right panel, top), which is 
 in sharp contrast to the model with lower neutrino luminosity (right panel, bottom).
 
Finally we note that there should are at least two
 barriers to get over in the current-generation
 studies of the CCSN nucleosynthesis. 
As mentioned already, the first one is the need of radiation hydrodynamic 
simulations that follow the dynamics from 
gravitational collapse, shock-revival, shock propagation in the stellar 
 mantle, to stellar explosions in a self-consistent manner.
 The second one is the need of computing
light curves, spectra, and the degree of polarizations in 2D or 3D hydrodynamic models,
 for which one needs to solve the multi-D 
photon transport coupled with multi-D hydrodynamics.
 Unfortunately however, the marriage of the two items,
 albeit ultimately needed for clarifying the photon messengers, may not be 
 done immediately due to their computational expensiveness. For example, one 
 needs to follow the dynamics more than $\sim$ 1 day after the onset of 
gravitational core-collapse for computing the explosive nucleosynthesis, 
because the envelop of a typical supernova progenitor 
 extends up to a radius of $\sim 10^{12}$ cm. It takes furthermore more than 
 $\sim 1$ week before the shock propagation enters to the so-called homologous phase.
 At present, the best available numerical simulation to this end is limited to 2D 
(e.g., \citet{gawry10}).
 For discussing anisotropies in emission-line profiles, one needs to follow the 
 dynamics later than $\sim 1$ year after explosions when the remnant becomes transparent.
 It is a very challenging (but very important) task for the first principle simulations
 to overcome the big gaps regarding the very different timescales. A encouraging news
 is that several groups are pursuing it with the use of advanced numerical 
techniques such as the adaptive-mesh-refinement approach \citep{nordhaus} 
 or Ying-Yang gridding \citep{annop}.
 With an accelerating power (like peta- or exa-scale) of supercomputers, it would 
 be unsurprising that the next (or the next after next!) generation supernova modellers 
 can execute the ultimate {\it ab-initio} simulations, whose findings would be 
 immediately used to make a more precise prediction of the photon messengers.
 
\section{MHD Mechanism}\label{sec3}
 In this section, we move on to focus on the MHD mechanism. After we briefly
 summarize the current status of this topic below, we mention the explosion dynamics 
that proceeds by the field-wrapping mechanism in section \ref{wrap} and 
discuss properties 
of the so-called magnetorotational instability (MRI)
 in sections \ref{mri} and \ref{local} based on our preliminary results. Possible 
 signatures of GWs and neutrinos are discussed 
in section \ref{sec3.1} and \ref{sec3.2}, respectively.

 Numerical simulations of MHD stellar explosions 
have started already in early 1970's shortly after the discovery of pulsars 
 \citep{leblanc,gena,bisno76,muller79,symb84}. 
 However, it is rather only recently that the MHD studies come back to the 
front-end topics in the supernova research followed by a number of 
 extensive MHD simulations (e.g., \citet{arde00,yama04,kota04a,kota04b,kotake05,ober06a,ober06b,burr07,cerd07,suwa07a,suwa07b,komi,taki04,taki09,scheid,martin11} for references therein). 
Main reasons for this activity are observations indicating very asymmetric 
explosions \citep{wang01,wang02}, and the 
interpretation of magnetars \citep{dunc92,latt07}, collapsars and their relevance 
to gamma-ray bursts 
(\citet{dessart08_c,woos06,yoon}) as a possible outcome of the 
magnetorotational core-collapse of massive stars.
 
The MHD mechanism of stellar explosions 
relies on the extraction of rotational free energy of 
collapsing progenitor core via magnetic fields. Hence a high angular momentum of 
the core is preconditioned for facilitating the mechanism \citep{meier}. 
Given (a rapid) rotation of the 
 precollapse core, there are at least two ways to amplify the initial 
magnetic fields to a dynamically important strength, namely by the field wrapping 
by means of differential rotation that naturally develops in the collapsing core, 
and by the MRI (MRI, see \citet{balb98}). Each of
 them we are going to review briefly in the following.

\begin{figure}[hbpt]
 \begin{center}
  \includegraphics[width=1.\linewidth]{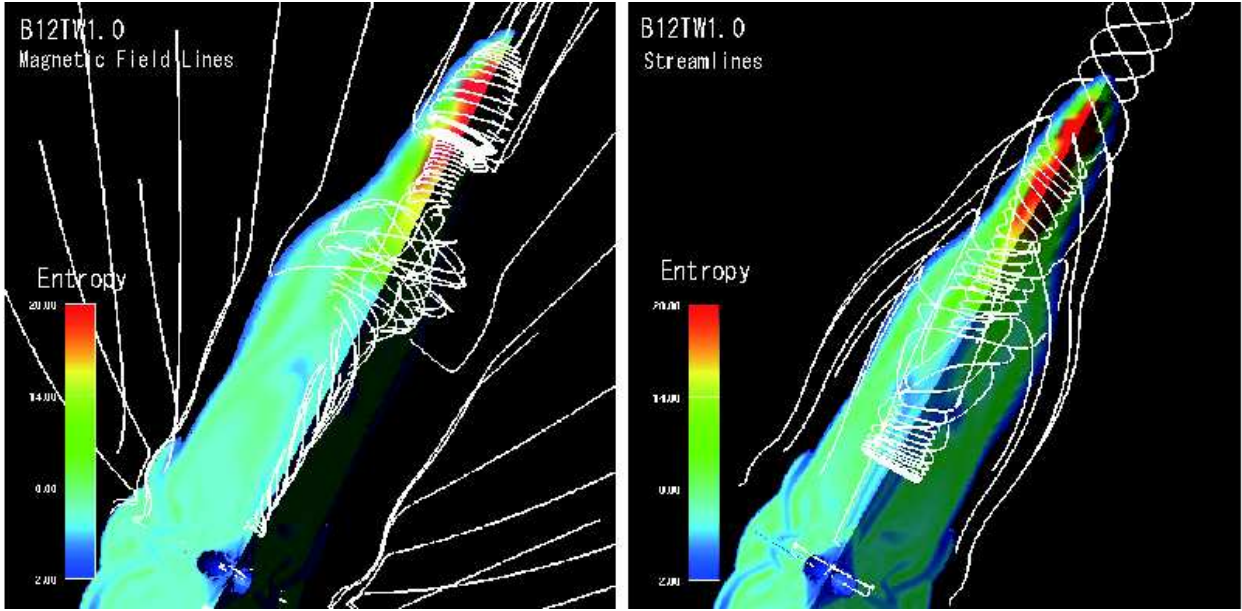}
\caption{
Three dimensional plots of entropy with the magnetic field lines
(left) and the streamlines of the matter (right) during the jet
propagation at $20$ ms and $94$ ms after bounce, respectively (for model B12TW1 taken 
from \citet{taki09}).
The outer edge of the sphere colored by blue represents the radius of 
$7.5 \times 10^7 $ cm.
These panels highlight not only the wound up magnetic field around the
rotational axis (left), but also the fallback of the matter from the jet head of the
downwards to the equator, making a cocoon-like structure behind
the jet (right).
}\label{f1_MHD}
 \end{center}
\end{figure}

\begin{figure}[htbp]
  \begin{center}
    \begin{tabular}{ccc}
      \resizebox{50mm}{!}{\includegraphics{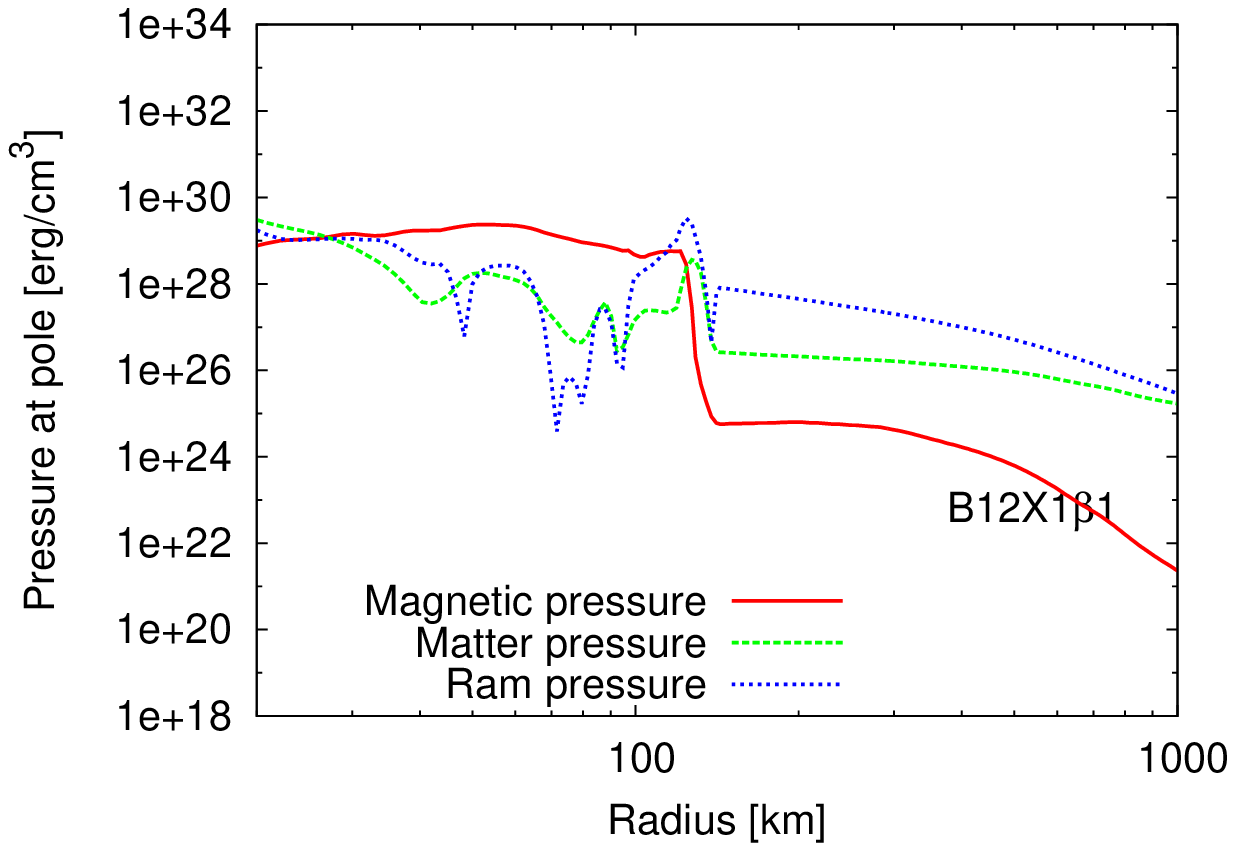}} &
      \resizebox{50mm}{!}{\includegraphics{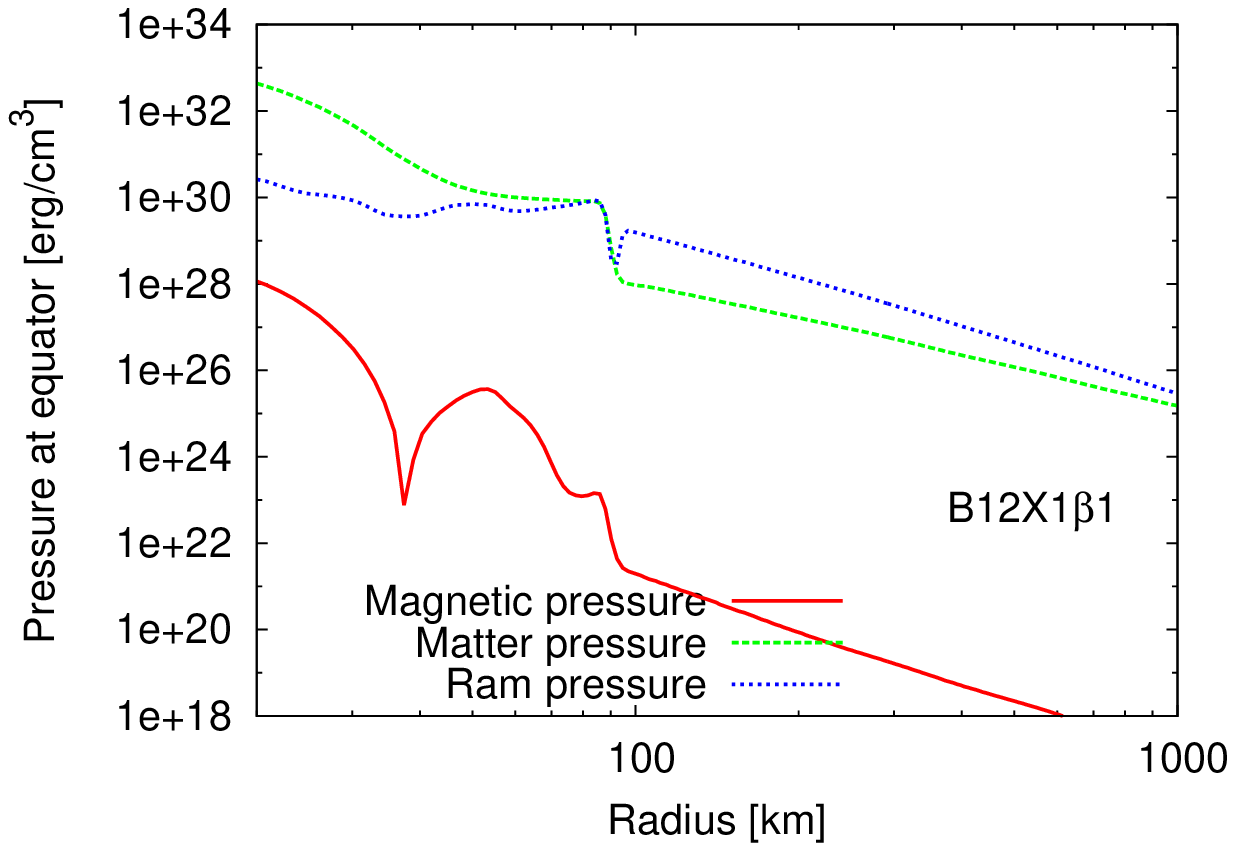}} &
      \resizebox{50mm}{!}{\includegraphics{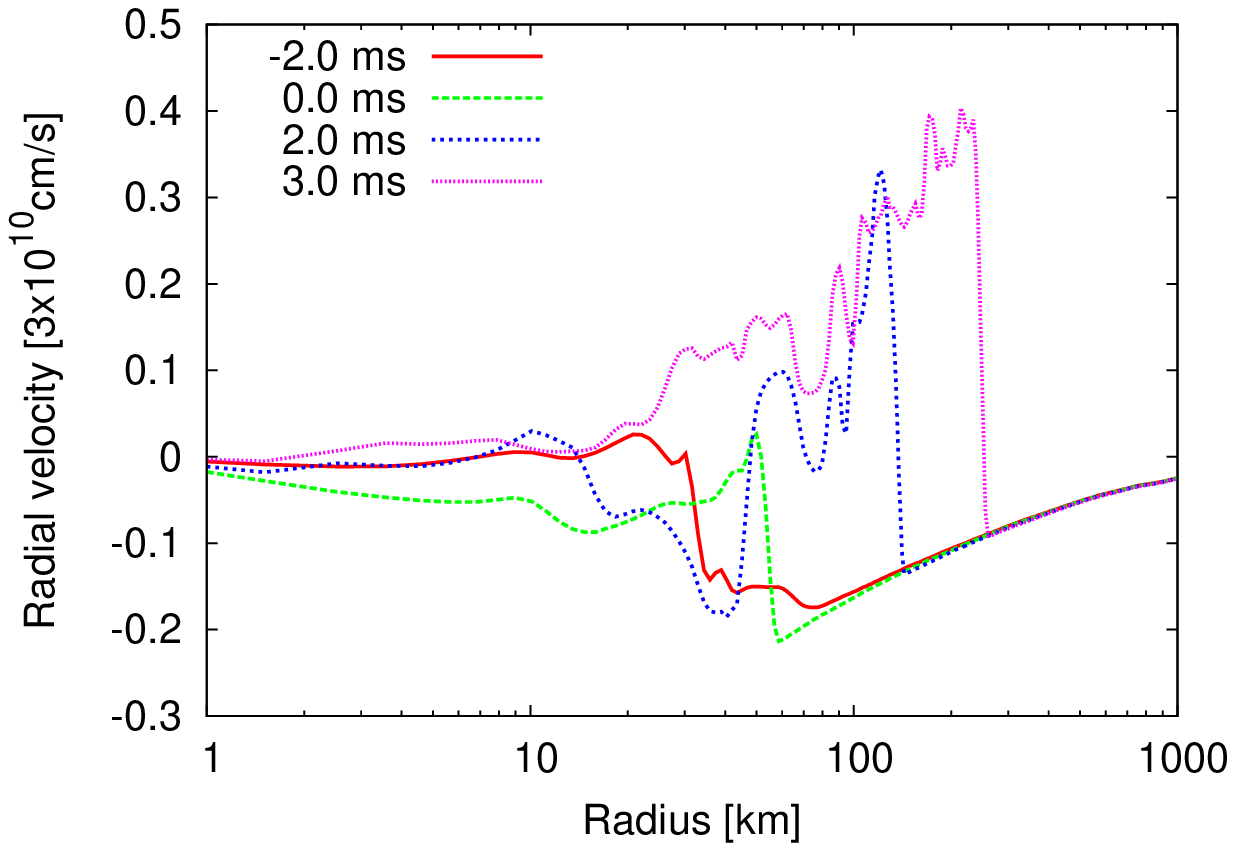}}\\
    \end{tabular}
   \caption{Left and middle panels show magnetic pressure (red line) vs. ram 
pressure (blue line) for a typical MHD model (same as Figure \ref{f1_MHD})
 along the polar axis (left panel)
 or the equatorial plane (middle panel) at 2 ms after the revival of the 
 stalled bounce shock. The matter pressure is 
shown by green line as a reference. The right panel shows a velocity profile along 
 the pole measured from the shock-stall (0.0 ms in the panel).
 For the equator, the magnetic pressure is much less than the ram pressure
 (middle panel), while the magnetic pressure amplified by the field wrapping 
 along the pole becomes as high as the ram pressure of the 
infalling material at the shock front (left panel, note that the shock position can
 be inferred by the discontinuity at around 150 km), 
leading to the MHD-driven shock formation (right panel). These figures 
 are taken from \citet{taki_kota}.}
  \label{f2_MHD}
  \end{center}
\end{figure}

\subsubsection{Field-wrapping mechanism}\label{wrap}

Three dimensional plots of Figure \ref{f1_MHD} are 
useful to see how the field wrapping occurs. For the 2D model
 taken from our 2D special relativistic (SR) MHD simulation \citep{taki09}, 
 the precollapse 
 magnetic field is set to be as high as $10^{12}$ G that is 
uniform and parallel to the rotational 
 axis and the initial $\beta$ parameter (ratio of initial rotational energy to 
the absolute value of the initial gravitational energy) is taken to be 0.1 $\%$ with 
  a uniform rotation imposed in the iron core.
From the left panel, it can be seen that the field lines are strongly
 twisted around the rotational axis. From the induction equation of ideal MHD 
equations, the time evolution of the toroidal fields ($B_{\phi}$) can be expressed
as (e.g., \citet{meier}),
\begin{equation}
\frac{\partial B_{\phi}}{\partial t} \approx  B_X \Bigl( X \frac{\partial \Omega}{\partial
 X} \Bigr),
\end{equation}
where $X$ denotes the distance from the rotational axis and $B_X$ represents the 
 $X$ component of the poloidal fields in the 
 cylindrical coordinates ($X,Z$). Given that a precollapse iron core has 
 strong magnetic field (such as ${B_{X,0}}\sim {10^{12} ~{\rm G}}$ with ${B_{X,0}}$
 representing the initial poloidal fields) and rapid rotation (such as 
$P_{\rm init} \lesssim 4~{\rm s}$ with $P_{\rm init}$ being the precollapse rotation
 period),
the typical amplification of $B_{\phi}$ near core bounce may be estimated as
\begin{equation}
\Delta {B_{\phi}} \approx \frac{t}{P_{\rm rot}} B_{X} \sim 10^{15} G
\Bigl(\frac{t}{100 ~{\rm ms}}\Bigr)\Bigl(\frac{P_{\rm rot}}{1 ~{\rm ms}}\Bigr)
\Bigl(\frac{B_{X,0}}{10^{12} ~{\rm G}}\Bigr), \label{linear}
\end{equation}
 where $t$ represents the typical timescale when the field amplification via the 
 wrapping processes saturates, $P_{\rm rot}$ denotes the rotational period near bounce.
 Equation (2) shows that the toroidal fields grow linearly with time by wrapping the 
poloidal fields ($B_X$), which is the so-called $\Omega$ dynamo 
(e.g., \citet{cerd07}). The $10^{15}$ G class
 magnetic fields are already dynamically important strength to affect the postbounce
 hydrodynamics as will be mentioned below.

The white lines in the right panel in Figure \ref{f1_MHD} show the streamlines of matter.
A fallback of material just outside of the jet head advecting downwards
to the equator (like a cocoon) is seen. In this jet with a cocoon-like structure, 
the magnetic pressure is always dominant over the matter pressure.
 As estimated above, the typical field strength behind the shock is 
$\sim 10^{15}\mathrm{G}$. The critical strength of the toroidal
magnetic field to induce the {\it magnetic} shock revival (i.e. the revival 
 of the stalled bounce shock due to the MHD mechanism) is estimated as follows. 
 The matter in the regions behind the stalled shock is
 pushed inwards by the ram pressure of
 the accreting matter. This ram pressure is estimated as,
\begin{equation}
 P=4 \times
  10^{28}\left(\frac{\rho}{10^{10}\mathrm{g/cm^3}}\right)\left(\frac{\Delta
  v}{2 \times 10^9\mathrm{cm/s}}\right)^2\mathrm{erg/cm^3}, 
\end{equation}
where $\rho$ and $\Delta v$ denotes typical density and radial velocity in
 the vicinity of the stalled shock.
When the toroidal magnetic fields are amplified as large as $\sim
10^{15}\mathrm{G}$ due to the field wrapping behind the shock,
the resulting magnetic pressure, $\frac{B^2}{8\pi}\sim ({10^{15}})^2/{8 \pi}\sim
 10^{29}~{\rm erg}/{\rm cm}^3$, can overwhelm the ram
pressure, leading to the magnetic shock revival (e.g., Figure \ref{f2_MHD}). 
 The onset timescale of the magnetic shock revival is 
 sharply dependent on the precollapse magnetic fields and rotation
 (e.g., \citet{burr07}).
 As the initial field strength is larger
 with more rapid rotation imposed, the interval from the shock stall 
to the MHD-driven shock revival becomes shorter. 
The speed of the jet head is typically mildly relativistic (at most $\sim 0.4 c$),
with $c$ being the speed of light (see also the right panel 
 of Figure \ref{f2_MHD}). At the shock breakout of the iron core, 
the explosion energies generally exceed $10^{51}$ erg for models
 that produce MHD explosions in a shorter delay after the stall of the bounce shock.
These features are in accord with those obtained in more detailed MHD simulations with 
spectral neutrino transport (e.g., \cite{burr07,dessart08_c})\footnote{See, e.g., \citet{taki_kota} for a more detailed comparison.}.

\subsubsection{On the MRI} \label{mri}

We are now in a position to discuss possible impacts of the MRI in supernova cores.
\citet{akiy03} was the first to point out that the interfaces surrounding
 the nascent PNSs quite generally satisfy the 
MRI instability criteria. Therefore any seed magnetic fields can be 
amplified exponentially in the differentially rotating layers, much 
 faster than the linear amplification due to the field wrapping (e.g., Equation 
(\ref{linear})).
 After the MRI enters to the saturated state, the field strength might reach 
 $\sim 10^{15-16}$ G, which is high enough to affect the supernova dynamics.
Not only in the field amplification, the MRI plays 
a crucial role also in operating the MHD turbulence
 (see \citet{hawley95,balb98,masada06}). 
The turbulent viscosity sustained by the MRI can convert a fraction of the shear 
rotational energy to the thermal energy of the system.
 \citet{thomp05} suggested that the additional energy input by the 
 viscous heating can help the neutrino-driven supernova explosion.
Followed by the exponential field amplification and additional heating, 
a natural outcome of the magnetorotational core-collapse is expected to 
 be the formation of 
 energetic bipolar explosions, which might be observed as hypernovae 
(e.g., section \ref{sec2.2}).

 Note again that bipolar explosions obtained in the previous section with 
 the assumption of a strong precollapse magnetic field ($ \gtrsim 10^{12}$G),
  are predominantly driven by the field wrapping processes, not by the MRI.
\citet{shib06} pointed out in their fully 2D GRMHD core-collapse simulations that 
 more than 10 grids are at least needed to capture the fastest-growing mode of the 
MRI (see also \citet{etienne06}).
As well-known, the growth rate of MRI-unstable modes depends on the product 
of the field strength and the wave number of the mode \citep{balb98}.
 When a precollapse rapidly core is strongly magnetized as $10^{11}$ G 
(like in the case of collapsar progenitor), the wavelength is on the order of km 
in the postbounce core. In this case, the exponential field growth of the MRI 
 was successfully captured by high-resolution simulations \citep{shib06}.
 On the other hand, 
the fastest growing modes drop below several meters \citep{MRI} for
 canonical supernova progenitors, which rotates much more slowly with 
 weaker initial fields ($\sim 10^{9}$G) as predicted by recent 
stellar evolution models \citep{maeder00,hege05}.
 At present, it is computationally 
too expensive to resolve those small scales in the global MHD simulations, 
typically more than two or three orders-of-magnitudes smaller than their typical 
finest grid size.
 To reveal the nature of the MRI, local simulations focusing on a small part of the 
MRI-unstable regions are expected to be quite useful as traditionally studied 
 in the context of accretion disks (see \citet{balb98}).

  \citet{MRI} were the first to 
study the growth and saturation level of the MRI in the supernova 
environment\footnote{see, \citet{zhang09} for extensive simulations that 
 focus on the MRI in relativistic outflows in the context of gamma-ray bursts and active galactic nuclei.}. 
To ease a drawback of the local shearing box simulation, they employed 
 the shearing disk boundary conditions by which global radial density 
stratification can be taken into account. By performing 
 such a {\it semi-global} simulation systematically in 2D and 3D
, they derived scaling laws for the termination of the MRI. As estimated in 
 \citet{akiy03}, the MRI was shown to amplify the seed fields exceeding $10^{15}$G. 
 These important findings may open several questions
 that motivate us to join in this effort, such as
  how the nonlinear properties as well as the 
scaling laws could be in the subsequent non-linear state and whether the viscous 
and resistive heating driven by the MRI turbulence could or could not affect 
the supernova mechanism.

 In the following, we briefly summarize the current status of our MRI project,
 in which we perform a local shearing box simulation to study the MRI-driven 
turbulence and their nonlinear properties bearing in mind 
 the application to the supernova core
(Masada, Takiwaki, and Kotake in preparation). 
 The final goal of this project is to determine
 how much the conversion from the rotational energy\footnote{
(that the differential rotation taps)} to the thermal energy occurs
 via the MRI turbulence, which was treated parametrically in \citet{thomp05}. 
As will be shown later, our preliminary results suggest that 
the additional energy supply would affect the neutrino-driven explosion 
in the case of rapidly rotating cores at late postbounce phase.

 To study the nonlinear properties of the MRI,
 we solve viscous MHD equations by a finite-difference code that was 
 originally developed by \citet{sano98}.
The hydrodynamic part is based on the second-order Godunov scheme 
\citep{leer}, which consists of Lagrangian and remap steps. 
The exact Riemann solver is modified to account for the effect of tangential 
magnetic fields. The field evolution is calculated with Consistent MoC-CT 
method \citep{clarke}. The energy equation is solved in the conservative form 
and the viscous terms are consistently calculated in the Lagrangian step. The 
advantages of our scheme are its robustness for strong shocks and the 
satisfaction of the divergence-free constraint of magnetic fields 
\citep{evans,stone}.

Rigorously 
 speaking, 3D simulations are absolutely required for the study of the MRI
 because only by them the disruption
 of channel structures, mode couplings, and the growth of parasitic instabilities
 can be accurately determined (e.g., \citet{pessah09,long}). But in the following,
 we are only able to show our 2D results assuming axisymmetry, which are currently 
 being updated to 3D. Bearing these caveats in mind, we briefly illustrate 
fundamental properties of the MRI\footnote{which is not always familiar 
 with supernova modellers,} and discuss their possible impacts on 
the explosion dynamics.

\begin{figure*}[hbtp]
\begin{center}
\begin{tabular}{cc}
\scalebox{1.0}{{\includegraphics{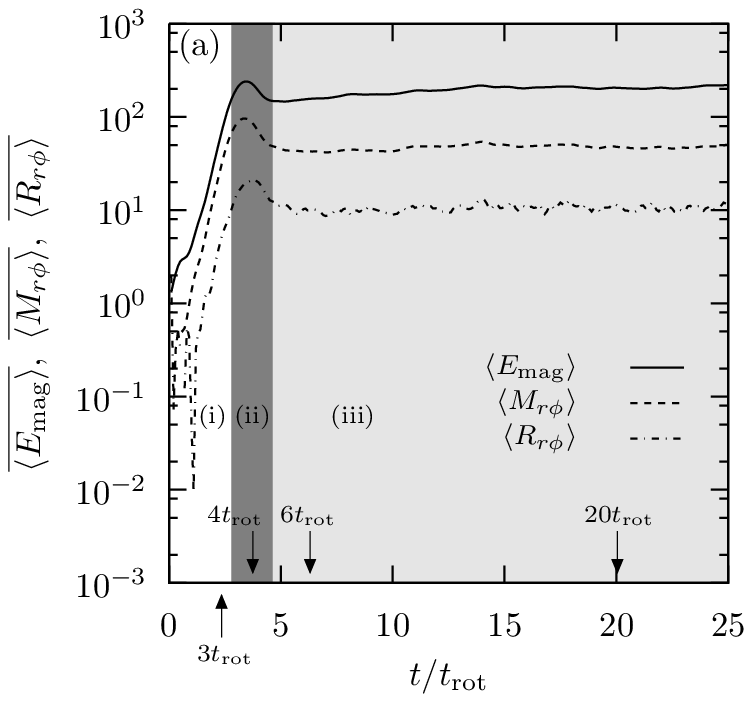}}} &
\scalebox{1.0}{{\includegraphics{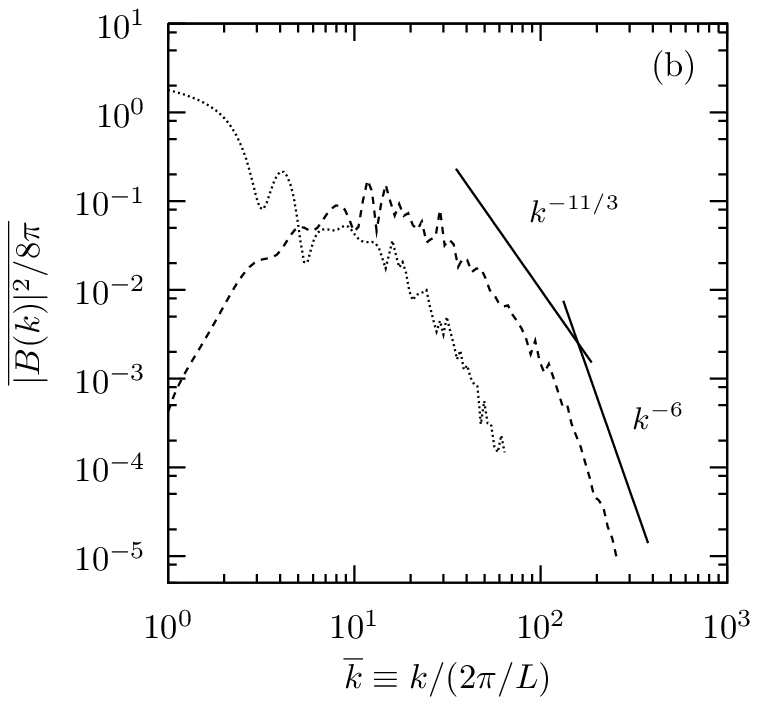}}}
\end{tabular}
\caption{(a) Temporal evolution of the volume averaged magnetic energy, Maxwell and Reynolds stresses 
in the fiducial run are represented by thick, dashed and dash-dotted curves respectively. The vertical axis is 
normalized by the initial magnetic energy $E_{\rm mag,0} \equiv B_0^2/8\pi$, that is 
$\overline{ \langle E_{\rm mag} \rangle} \equiv \langle E_{\rm mag} \rangle /E_{\rm mag,0}$, 
$\overline{ \langle M_{r\phi} \rangle} \equiv \langle M_{r\phi} \rangle /E_{\rm mag,0}$, and 
$\overline{ \langle R_{r\phi} \rangle} \equiv \langle R_{r\phi} \rangle /E_{\rm mag,0}$ where single bracket
denotes the volume average of the physical variables. Horizontal axis is normalized by the rotational period 
$t_{\rm rot} \equiv 2\pi/\Omega$. Three typical evolutionary stages, (i) exponential growth stage, 
(ii) transition stage, and (iii) nonlinear turbulent stage, are denoted by white, dark gray and light gray shaded 
region. 
(b). The amplitude of power spectra of the magnetic energy $|B(k)|^2$ along the $k_x$ and $k_z$ axes. The 
dashed and dotted curves describes the $k_x$ and $k_z$ components respectively. The vertical axis is 
normalized by the initial magnetic energy $E_{\rm mag,0}$. The wave-numbers shown in the horizontal axis 
are normalized by the $2\pi /L$. The upper thick curve demonstrates the Kolmogorov slope for isotropic 
incompressible turbulence $\propto k^{-11/3}$.  The lower slope is proportional to $k^{-6} $ just for the reference. 
In our local simulations, we prepare a 2D shearing box 
 which is threaded by the initial vertical field 
$B_{0} = 1.4 \times 10^{12}$ G with imposing 
initial gas pressure $P_0 = 4\times 10^{28}\ {\rm dyn\ cm^{-2}}$, angular velocity 
$\Omega_0 = 2000\ {\rm sec^{-1}}$ and shear parameter 
$q_0 = 1.25$, respectively. Here the shear parameter represents the degree of 
differential rotation as $q \equiv -d\ln\Omega/d\ln r$ with $r$ being the radial 
coordinate. The other parameters are fixed to mimic the 
 properties in the vicinity of the PNS's surface in the postbounce phase (i.e.
 the density is taken to be $\rho_0 = 10^{12}\ {\rm g\ cm^{-3}}$ and the box size 
 is taken to be $ L_x = 4\ {\rm km}$ (horizontal) $\times$ $L_z = 1\ {\rm km}$ 
(vertical)). 
} 
\label{f1_mri}
\end{center}
\end{figure*}

\begin{figure*}[hbtp]
\begin{center}
\begin{tabular}{cc}
\scalebox{0.7}{{\includegraphics{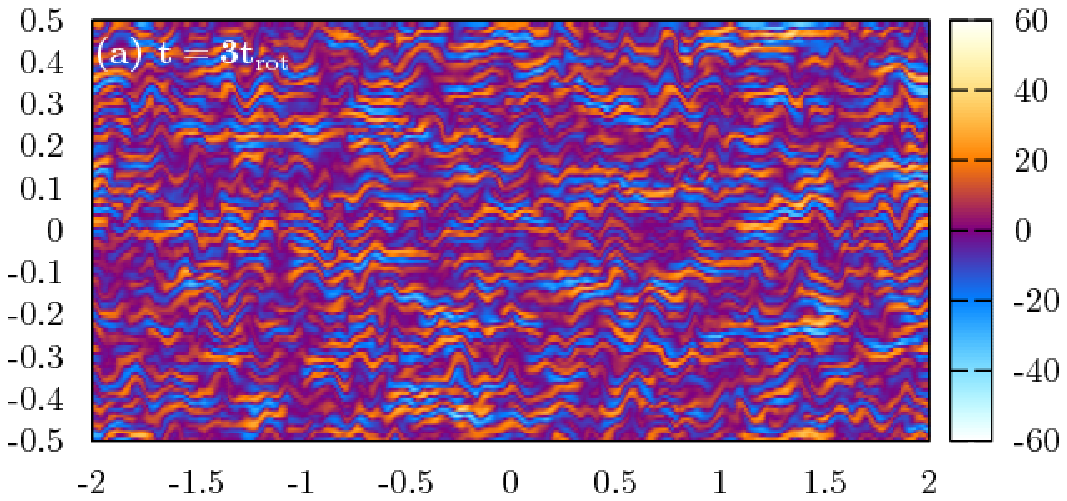}}} &
\scalebox{0.7}{{\includegraphics{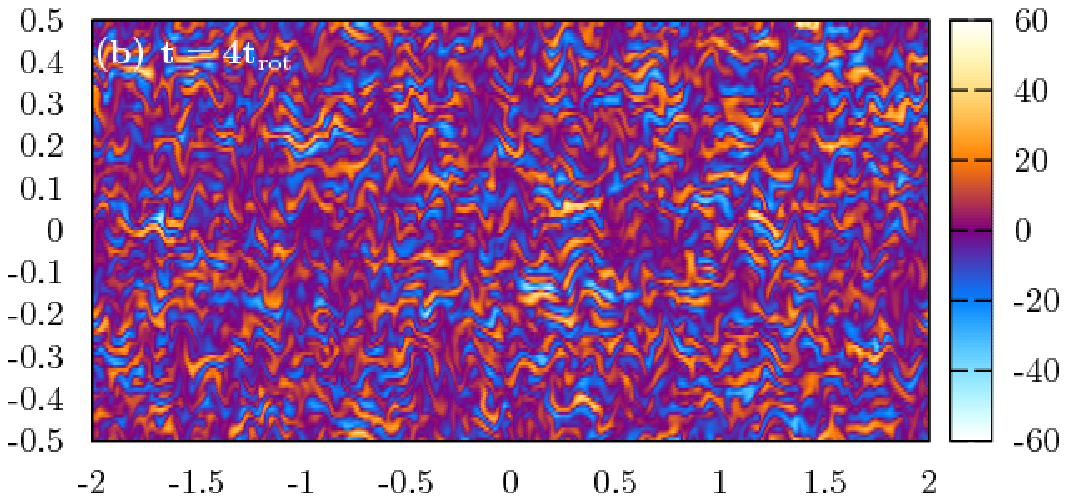}}} \\
\scalebox{0.7}{{\includegraphics{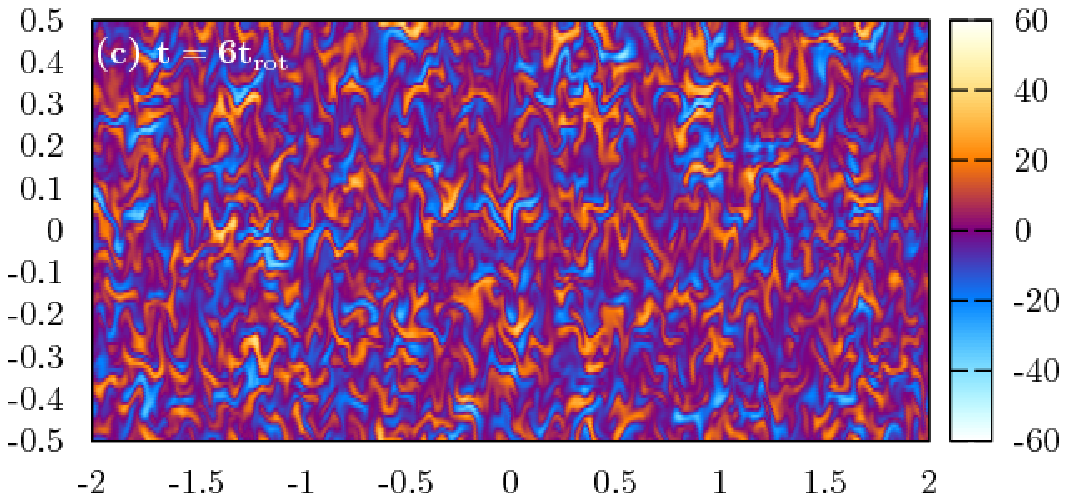}}} &
\scalebox{0.7}{{\includegraphics{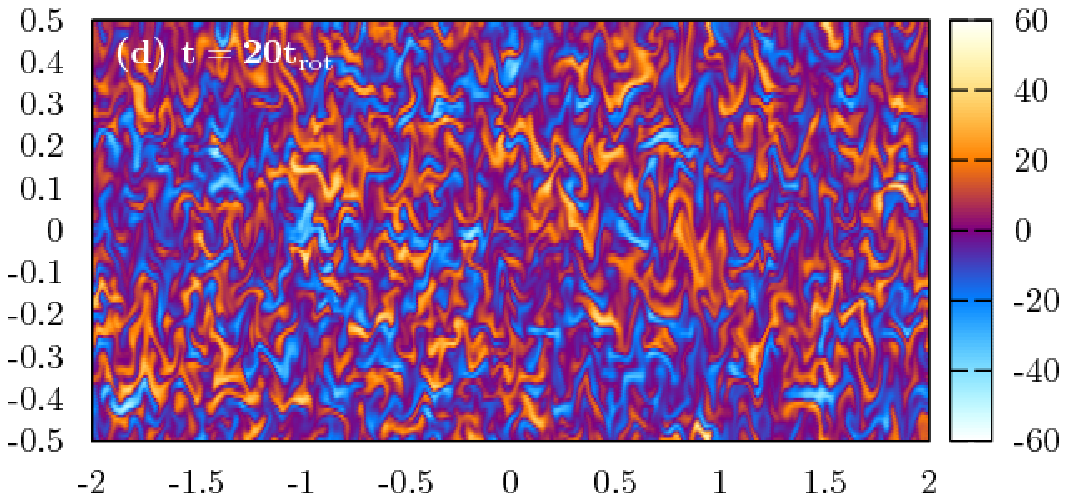}}} \\
\end{tabular}
\caption{Four snapshots representing the temporal evolution of MRI from the linear to the non-linear phase.
  Panels (a), (b), (c) and (d) correspond to the time slice $t = 3t_{\rm rot}$, $4t_{\rm rot}$, $6t_{\rm rot}$
and $20t_{\rm rot}$ respectively. The color bar indicates the size of the toroidal 
magnetic field normalized by the strength of the initial vertical magnetic field 
($B_0 = 1.4 \times 10^{12} $ G). Note that the vertical and horizontal are 
normalized by the typical spatial scale of the local portion in the proto-neutron stars which is selected as $L = 1$km 
in the fiducial run. Three typical evolutionary stages are observed, (i) exponential growth stage, 
(ii) transition stage, and (iii) nonlinear turbulent stage (compare with Figure 
\ref{f1_mri}). } 
\label{f2_mri}
\end{center}
\end{figure*}

\begin{figure*}[hbtp]
\begin{center}
\begin{tabular}{cc}
\scalebox{0.8}{{\includegraphics{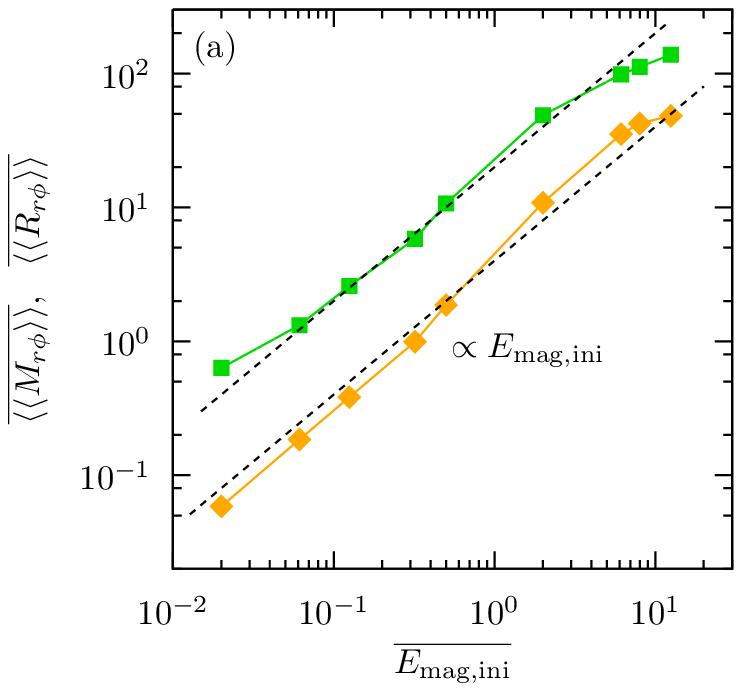}}} &
\scalebox{0.8}{{\includegraphics{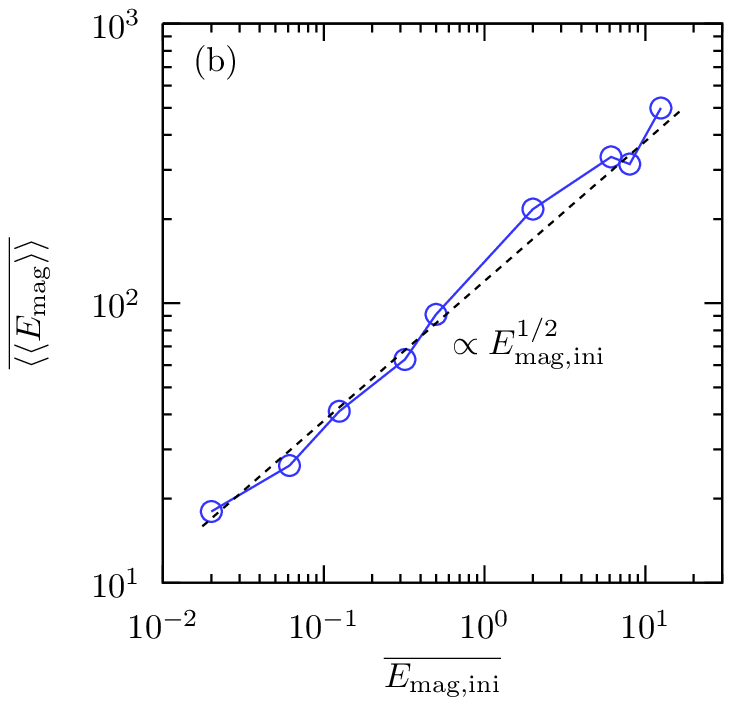}}} 
\end{tabular}
\caption{The dependence of the time- and volume-averaged 
 (a) Maxwell (green squares) and Reynolds stresses 
(orange diamonds) and (b) magnetic energy 
(blue circles) and in the turbulent state 
on the strength of the initial magnetic energy $E_{\rm ini}$. 
Both the vertical and horizontal axes are normalized by the initial 
magnetic energy of the fiducial model $E_{\rm mag,0}$. The initial 
field strength is varied from $\sim 10^{11}$ to $\sim 6 \times 10^{12}$ G. The dashed 
lines denote reference slopes proportional to (a) $E_{\rm mag}^{1/2}$ and 
(b) $E_{\rm mag}$ respectively. We take the time-average of volume-averaged quantities 
during $90t_{\rm rot} < t < 100t_{\rm rot}$.} 
\label{f3_mri}
\end{center}
\end{figure*}

\subsubsection{Local simulations in axisymmetry} \label{local}

As it is well known, there are three typical 
evolutionary stages observed in our local simulations. They are 
(i) exponential growth stage, (ii) transition stage, and (iii) 
nonlinear turbulent stage \citep{hawley92,hawley95}.
 Each stage is denoted by white, dark gray and light gray shaded regions in Figure 
\ref{f1_mri}. 

In the stage (i), the channel structure of the magnetic field 
exponentially evolves and inversely cascades to the larger spatial scale 
with small structures merging as the magnetic field is amplified. 
This is because the channel modes of the MRI are a exact solution 
 for the nonlinear MHD equations \citep{goodman}. 
The temporal evolution of channel structures for the toroidal 
magnetic field and their inversely cascading nature are shown in panels (a) and 
(b) of Figure \ref{f2_mri}. Then in the stage~(ii), the channel structure of 
the magnetic field is disrupted via the parasitic instability and magnetic 
reconnection at the transition stage \citep{goodman,MRI}. 
The channel disruption induces a drastic phase-shift from the coherent structure 
to the turbulent tangled structure of the field as is illustrated by panel (c)
 of Figure \ref{f2_mri}. The magnetic energy stored in the amplified magnetic field 
is then converted to the thermal energy of the system. 
Finally in the stage (iii), the nonlinear stage emerges after the channel 
 structures are disrupted to be a turbulence state driven by the MRI (see 
panel (d) in Figure \ref{f2_mri}). Figure~\ref{f1_mri}a 
shows that, at this stage, the turbulent Maxwell stress 
($M_{r\phi} =  - B_rB_\phi /4\pi$) dominates over the Reynolds stress ($R_{r\phi} = 
\rho v_r \delta v_\phi$). 
 Therefore the turbulent Maxwell stress is the main player to convert
 the free energy stored in the differential rotation 
into the thermal energy of the system. 

Figure \ref{f1_mri}b shows the power spectra 
of the magnetic energy in the nonlinear turbulent stage. The logarithmic slope of the 
spectrum roughly coincides with $-11/3$ 
in the regime $10 \lesssim \bar{k} \lesssim 100$ while 
it steepens at higher wave-numbers where numerical effects become important
 (like $k^{-6}$). This indicates that the turbulent state in the MRI can be 
  represented by the Kolmogorov spectrum for a homogeneous, incompressible 
turbulence that scales as $k^{-11/3}$. 
 These features would be consistent with the 3D properties of the MRI-driven 
turbulence \citep{hawley95}. 
It has been pointed out in the previous axisymmetric local shearing box simulation 
 that the channel structures do neither disrupt nor transit to the turbulent state 
without a relatively strong resistive effect \citep{sano,MRI}. We however adopted
 the simulation box with a large aspect ratio ($L_x/L_z =4$) 
and quasi-isothermal equation of state ($\gamma \simeq 1$), both of which could
 facilitate to disrupt the channel structures \citep{bodo} as in 3D simulations. 

Figure \ref{f3_mri} shows the time- and volume-averaged 
 Maxwell (green squares) and Reynolds stresses (orange diamonds, panel (a)), and  
magnetic energy (blue circles, panel (b))
 in the turbulent stage as a function of the initial magnetic energy. 
 It is shown that the Maxwell stress depends on the initial magnetic flux and 
 it provides the 
 following scaling relation,
\begin{equation}
\langle\langle M_{r\phi} \rangle\rangle 
\propto E_{\rm mag, ini},
\end{equation}
with $E_{\rm mag,ini} \equiv B_{\rm ini}^2/8\pi$. 
 This suggests that the magnetic flux initially penetrating the system 
controls the nonlinear transport properties of the MRI-driven 
turbulence. This is qualitatively consistent with the previous results 
in the accretion disk \citep{hawley95,sano98,sanosan04,pessah} 
while the power-law index obtained here is slightly larger.
  Note that the magnetic energy at the nonlinear turbulent stage has a linear 
dependence on the initial magnetic flux (because $\langle\langle E_{\rm mag} 
\rangle\rangle \propto E_{\rm mag,init}^{1/2} \propto B_{\rm ini}$ 
(Figure \ref{f3_mri}(b))). This is also qualitatively consistent with 
Hawley et al. (1995). In this way, we investigated parameter dependence
 of the rotational shear ($q$ parameter) as well as the initial pressure ($P$)
 on the saturated value of the Maxwell stress and deduced
 a scaling relation that yields 
$M_{r\phi} \propto B^2 P^{1/4} [q/(2-q)]^{1/4}\Omega^{1/2}$. Here the shear parameter 
represents the degree of 
differential rotation as $q \equiv -d\ln\Omega/d\ln r$ with $r$ being the radial 
coordinate.

Using the scaling relation obtained in our local simulations, the
 heating rate maintained by the MRI turbulence can be estimated as,
\begin{eqnarray}
\epsilon_{\rm MRI}  && =   M_{\rm r \phi} q \Omega \nonumber \\
			       && =  \left[ f(B_{\rm pb}/B_0)^2(P_{\rm pb}/P_0)^{1/4} s_{\rm pb}^{1/4}(\Omega_{\rm pb}/\Omega_0)^{1/2} \right]\ q\Omega \nonumber \\
			       && \simeq  10^{30}{\rm erg\ cm^{-3}}\ {\rm sec}^{-1} \times \\
			       &&f_{30}\ q_{{\rm pb} , 1}\; s_{{\rm pb},1}^{1/4}\left( \frac{B_{\rm pb}}{10B_0}\right)^2
			       \left( \frac{P_{\rm pb}}{100P_0} \right)^{1/4}
			       \left(\frac{\Omega_{\rm pb}}{\Omega_0} \right)^{3/2} \;, \nonumber
\label{f}
\end{eqnarray}
where $f_{30}$ represents a typical amplification factor of the stress 
normalized by $30$ (e.g., Figure \ref{f3_mri}), $q_{{\rm pb},1}$ is the shear rate 
normalized by unity and $s_{{\rm pb},1}$ is the ratio of the vorticity to the 
shear normalized by unity. Since we are interested in the growth of the MRI
in the postbounce phase, the original magnetic flux, gas pressure angular velocity, and 
the shear rate $B_0 = 1.4 \times 10^{12} G$, $P_{0} = 4 \times 10^{28}{\rm erg}~
{\rm cm}^{-3}$, $\Omega_{0} = 2000~{\rm rad}~{\rm s}^{-1}$, and $q = 1.25$ 
 are replaced by $B_{\rm pb}$, $P_{\rm pb}$, $\Omega_{\rm pb}$ and $q_{\rm pb}$ with some
 amplification factors.

 A typical volume in which the MRI-driven turbulence is active, may be estimated as
 $V_{\rm MRI}  = 4\pi R^2 h\simeq  10^{21}\ 
{\rm cm^3}R_7^2h_6 $, where $R_7 = R/10^7$ cm is typical radius of the 
 PNS normalized by $10^7$ cm, and $h_6 = h/10^{6}$ cm is the 
typical radial width of the MRI-active layer normalized by $10^6$ cm. Then the 
energy releasing rate $L_{\rm MRI} \equiv \epsilon_{\rm MRI} V_{\rm MRI}$ becomes,
\begin{eqnarray}
L_{\rm MRI} & \sim & 10^{51} R_7^2h_6\ {\rm erg\ sec^{-1}} \times  \\
&& f_{30}\ q_{{\rm pb},1} s_{{\rm pb},1}^{1/4}\left( \frac{B_{\rm pb}}{10B_0}\right)^2
			       \left( \frac{P_{\rm pb}}{100P_0} \right)^{1/4}
			       \left(\frac{\Omega_{\rm pb}}{\Omega_0} \right)^{3/2}\;.  \nonumber 
\end{eqnarray}
 Typical timescales of neutrino-driven explosions observed in recent
 supernova simulations are $\gtrsim 400$ ms after bounce (e.g., \citet{marek,suwa}). 
So the energy deposition could be as high as 
  $10^{50}~{\rm erg}$ if the core rotates rather rapidly as taken 
in the above estimation.
 If this is the case, the turbulent heating due to the MRI is expected  
 to play an important role for assisting the neutrino-driven explosion.

It should be noted that 
 nonaxisymmetric modes of the 
parasitic instability could disrupt the channel structures 
more effectively than in 2D \citep{pessah09,long}.
To obtain more accurate estimate of the amplification factor of $f$ 
(equation (\ref{f})), 3D simulations are unquestionably required, 
 which we are currently undertaking. 

\subsection{Gravitational waves}\label{sec3.1}

As already mentioned in section \ref{sec2.1}, rapid rotation, necessary for 
producing strong bounce GW signals, is likely to obtain $\sim$ 1\% of massive star 
population (e.g., \citet{woos_blom}). This can be really the case, albeit minor, for
 progenitors of rapidly rotating metal-poor stars,
 which experience the so-called chemically homogeneous evolution \citep{woos06,yoon}. 
 In such a case, the MHD mechanism could work to produce 
  energetic explosions, which 
 are receiving great attention
 recently as a possible relevance to magnetars and collapsars
 (e.g., \citet{macf99,macf01}, see \citet{dessart08_c,harikae_a,harikae_b} 
for collective references), 
 which are presumably linked to the formation of long-duration gamma-ray bursts (GRBs)
 (e.g., \citet{mesz06} for review).

Among the previous studies focusing on the bounce signals (e.g., references in 
 section \ref{sec2.1}),
 only a small portion of papers
has been spent on determining the GW signals in the MHD mechanism
 \citep{yama04,kota04a,ober06b,cerd07,shib06,scheid}.  
 This may be because the MHD effects on the dynamics as well as
 their influence over the GW signals can be visible 
 only for cores with precollapse magnetic fields over $B_{0} \gtrsim 10^{12}$ G 
\citep{ober06b,kota04a}. 
Considering that the typical magnetic-field strength of GRB 
progenitors is at most $\sim 10^{11-12}$ G \citep{woos06}, 
this is already an extreme situation. 
In a more extremely case of
 $B_0 \sim 10^{13}$ G, a secularly growing feature in the waveforms
was observed \citep{ober06a,shib06,scheid}. In the following, we summarize 
 the GW signatures based on our 2D SRMHD simulations \citep{taki_kota} and discuss how 
 a peculiarity of the MHD mechanism could be imprinted in the GW signals.

\begin{figure}[htbp]
    \centering
    \includegraphics[width=.49\linewidth]{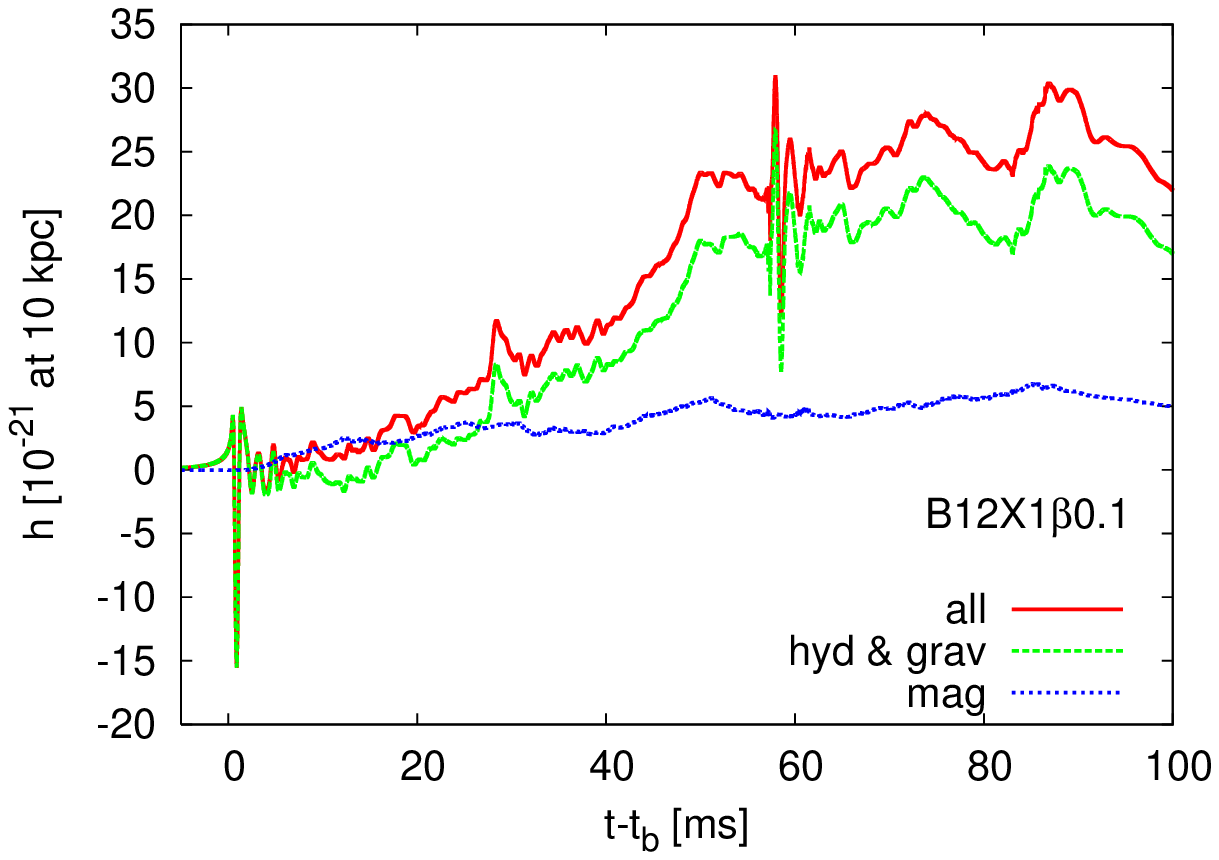}
    \includegraphics[width=.49\linewidth]{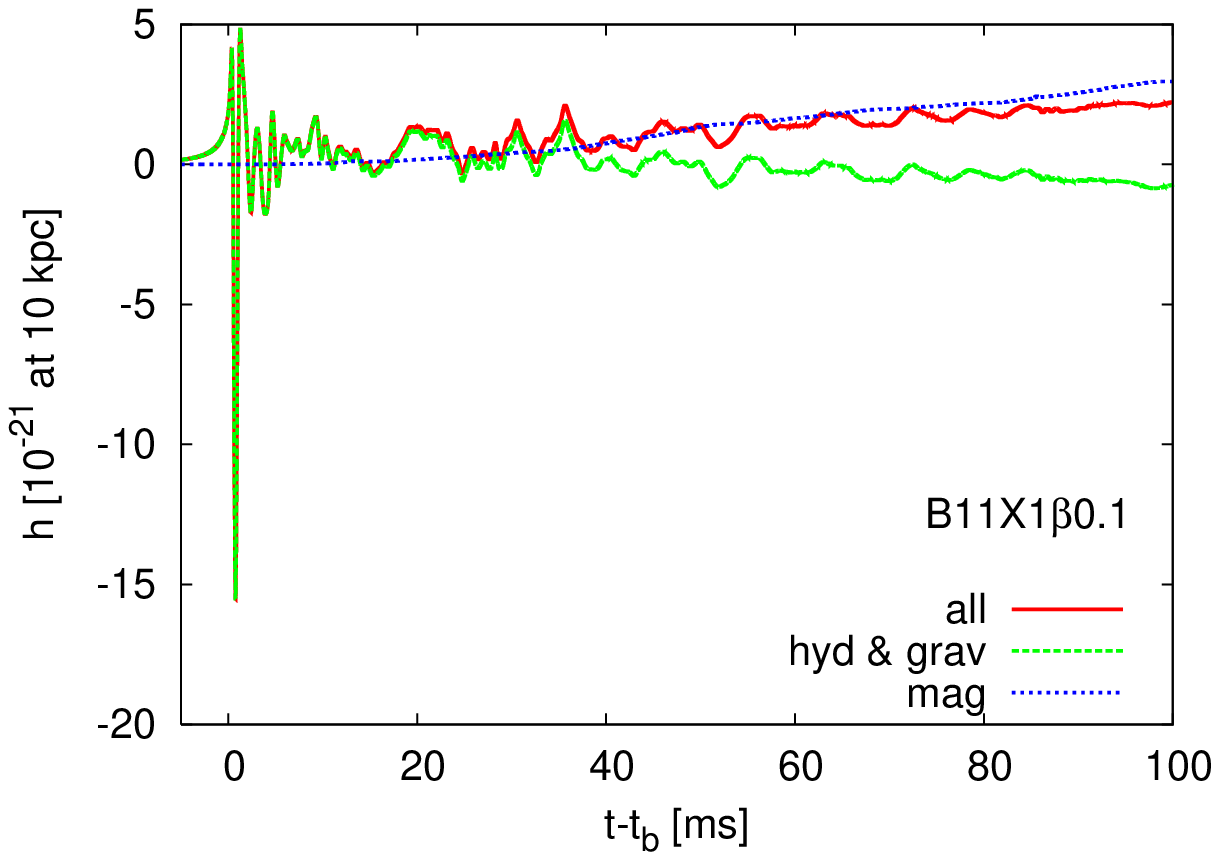}\\
    \caption{Gravitational waveforms with an increasing trend (left) or not
  (right panel). Initial rotation parameter is both set to be  $\beta = 0.1$ \%,
 while the precollapse magnetic field is taken as $10^{12}$ G (left panel) and 
 $10^{11}$ G (right panel), respectively (from \citet{taki_kota}).
 The total wave amplitudes are shown by the red line, 
while the contribution from the magnetic fields and from the sum of 
hydrodynamic and gravitational parts
 are shown by blue and green lines, respectively (e.g.,  equation (\ref{mag}) and 
 equations (\ref{quad},\ref{grav})). Note that the bounce GW signals 
(($t - t_b \lesssim 20$ ms) are not affected by the magnetic fields significantly, 
and they are categorized into the so-called type I or II waveforms.}
    \label{f1_gwMHD}
\end{figure}

 The left panel of Figure \ref{f1_gwMHD} shows the gravitational
 waveform with the increasing trend, which 
 is obtained for a model with strong precollapse 
magnetic field ($B_{0} = 10^{12}$G) also with rapid rotation initially imposed
 ($\beta$ parameter = 0.1 \%). Such a feature cannot be 
 observed for a weakly magnetized model ($B_{0} = 10^{11}$G. right panel).
To understand the origin of the increasing trend, it is straightforward to 
 look into the quadrupole GW formula, which can be expressed as 
\begin{equation}
h_{ij}^{TT}({\mbox {\boldmath  $X$}},t) 
\stackrel{\ell = 2, m = 0}{=} \frac{1}{R}A^{E2}_{20}\left(t - 
\frac{R}{c}\right)T^{E2,20}_{ij}(\theta,\phi),
\label{htt}
\end{equation}
 where $T^{E2,20}_{ij}(\theta,\phi)$ is 
\begin{equation}
T^{E2,20}_{ij}(\theta,\phi) = \frac{1}{8}\sqrt{\frac{15}{\pi}}\sin^2 \theta,
\label{Tij}
\end{equation}
(e.g., \cite{thorne}).
 ${A_{20}^{\rm{E} 2}}$ can be expressed as 
\begin{equation}
 {A_{20}^{\rm{E} 2}} =  {A_{20}^{\rm{E} 2}}_{\rm (hyd)} + 
 {A_{20}^{\rm{E} 2}}_{\rm (mag)} + {A_{20}^{\rm{E} 2}}_{\rm (grav)},
\label{A20}
\end{equation}
On the right hand side, the first term is related to anisotropic 
 kinetic energies which we call as hydrodynamic part, 
\begin{eqnarray}
 {A_{20}^{\rm{E} 2}}_{\rm(hyd)} &=& \frac{G}{c^4} \frac{32 \pi^{3/2}}{\sqrt{15 }} 
\int_{0}^{1}d\mu 
\int_{0}^{\infty}  r^2  \,dr  {f_{20}^{\rm{E} 2}}_{\rm(hyd)}, \\
{f_{20}^{\rm{E} 2}}_{\rm(hyd)} 
&=& \rho_{*}W^2( {v_r}^2 ( 3 \mu^2 -1) + {v_{\theta}}^2 ( 2 - 3 \mu^2)
 - {v_{\phi}}^{2} - 6 v_{r} v_{\theta} \,\mu \sqrt{1-\mu^2}),
\label{quad}
\end{eqnarray}
 the second term is related to anisotropy in gravitational potentials 
which we call as the gravitational part,
\begin{eqnarray}
 {A_{20}^{\rm{E} 2}}_{\rm(grav)} &=& \frac{G}{c^4} \frac{32 \pi^{3/2}}{\sqrt{15 }} 
\int_{0}^{1}d\mu  \int_{0}^{\infty}  r^2  \,dr
 {f_{20}^{\rm{E} 2}}_{\rm(grav)} ,\\
{f_{20}^{\rm{E} 2}}_{\rm(grav)}
&=&
 \left[\rho h (W^2+(v_k/c)^2) + 
 \frac{2}{c^2}\left( p+\frac{\left|b\right|^2}{2}\right)
-\frac{1}{c^2}\left((b^{0})^2+(b_{k})^2 \right)\right]\nonumber \\
& & \times \left[- r \partial_{r} \Phi (3 \mu^2 -1) + 3 \partial_{\theta} \Phi \,\mu
\sqrt{1-\mu^2}\right],
\label{grav}
\end{eqnarray}
and finally the third term is related to anisotropy in magnetic energies that 
 we refer to as magnetic part,
\begin{eqnarray}
 {A_{20}^{\rm{E} 2}}_{\rm(mag)} &=& - \frac{G}{c^4} \frac{32 \pi^{3/2}}{\sqrt{15 }} 
\int_{0}^{1}d\mu  \int_{0}^{\infty}  r^2  \,dr {f_{20}^{\rm{E} 2}}_{\rm(mag)}, \\ 
 {f_{20}^{\rm{E} 2}}_{\rm(mag)}
&=&
 [{b_r}^2 ( 3 \mu^2 -1) + {b_{\theta}}^2 ( 2 - 3 \mu^2)
 - {b_{\phi}}^{2} - 6  b_{r} b_{\theta} \mu \sqrt{1-\mu^2}];
\label{mag}
\end{eqnarray}

\begin{figure}[htbp]
    \centering
    \includegraphics[width=.44\linewidth]{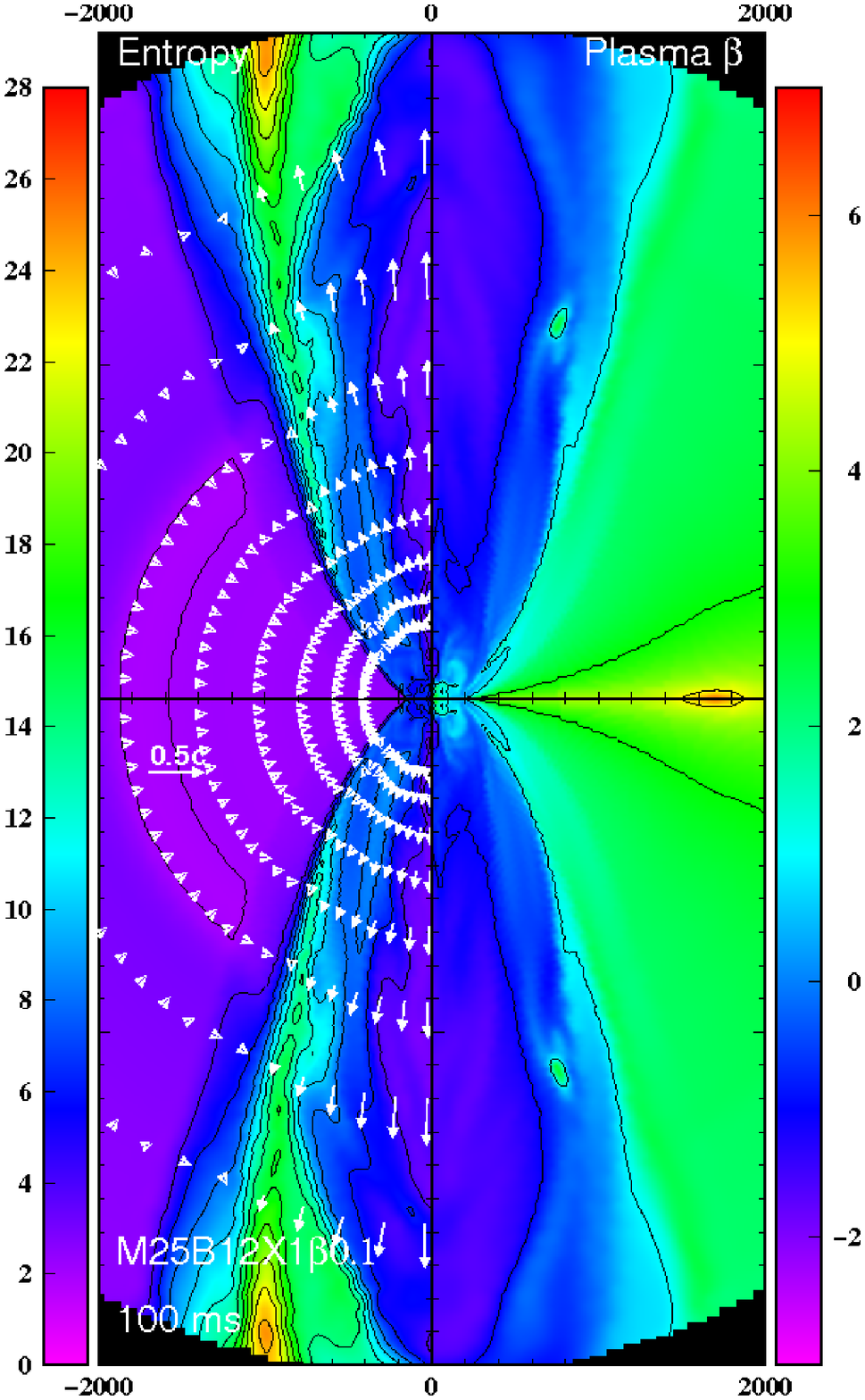}
    \includegraphics[width=.44\linewidth]{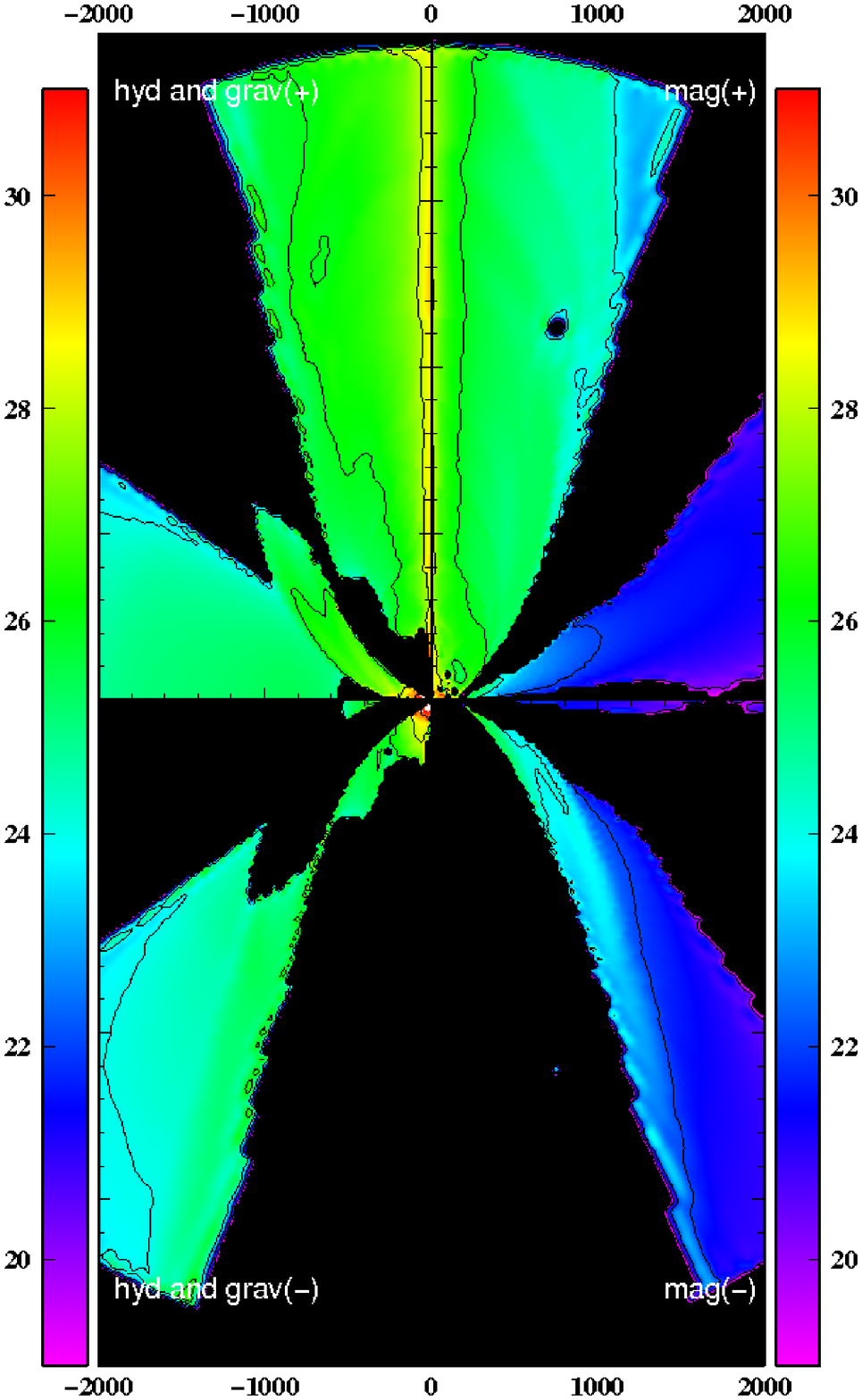}
    \caption{
 Left panel shows distributions of entropy [$k_B$/baryon] (left) and
 logarithm of plasma $\beta$ 
(right) for model B12X1$\beta$0.1 at 100ms after bounce. 
The white arrows in the left-hand side show the velocity fields, 
which are normalized by the scale in the middle left edge ($0.5 c$).
Right panel shows the sum of the hydrodynamic and gravitational parts 
(indicated by ``hyd and grav'' in the left-hand side) and the magnetic part 
 (indicated by "mag" in the right-hand side), respectively. The top and bottom 
panels represent the positive and negative contribution (indicated by (+) or (-))
 to ${A_{20}^{\rm{E} 2}}$,
 respectively (see text for more detail).  The side length of each plot is 
4000(km)x8000(km). This figure is taken from \citet{taki_kota}.}
 \label{fig:inc_cont}
\end{figure}

The right panel of Figure \ref{fig:inc_cont} shows contributions
 to the total GW amplitudes (equation (6)) for the strongly magnetized model
 (left panel of Figure \ref{f1_gwMHD}),
 in which the left-hand-side panels are for the sum of the hydrodynamic and 
gravitational part,
 namely $\log\left(\pm\left[{f_{20}^{\rm{E} 2}}_{\rm (hyd)} + {f_{20}^{\rm{E} 2}}_{\rm (grav)}\right]\right)$ 
(left top($+$)/bottom($-$)(equations (\ref{quad},\ref{grav})), and the right-hand-side 
panels are for the magnetic part, namely 
$\log\left(\pm{f_{20}^{\rm{E} 2}}_{\rm (mag)}\right)$ 
(right top($+$)/bottom($-$)) (e.g., equation (\ref{mag})).
 By comparing the top two panels in Figure 19,
 it can be seen that the positive contribution is  
 overlapped with the regions where the MHD outflows exist.
The major positive contribution is from the kinetic term of the MHD outflows 
 with large radial velocities (e.g., $ + \rho_{*}W^2 {v_r}^2$  in equation (\ref{quad})).
 The magnetic part also contributes to the positive trend (see top right-half
  in the right panel (labeled by mag(+))). This comes 
from the toroidal magnetic fields (e.g., $ + {b_{\phi}}^{2}$ in equation
 (\ref{mag})), which dominantly contribute to drive MHD explosions. 

\begin{figure}[htbp]
    \centering
    \includegraphics[width=.49\linewidth]{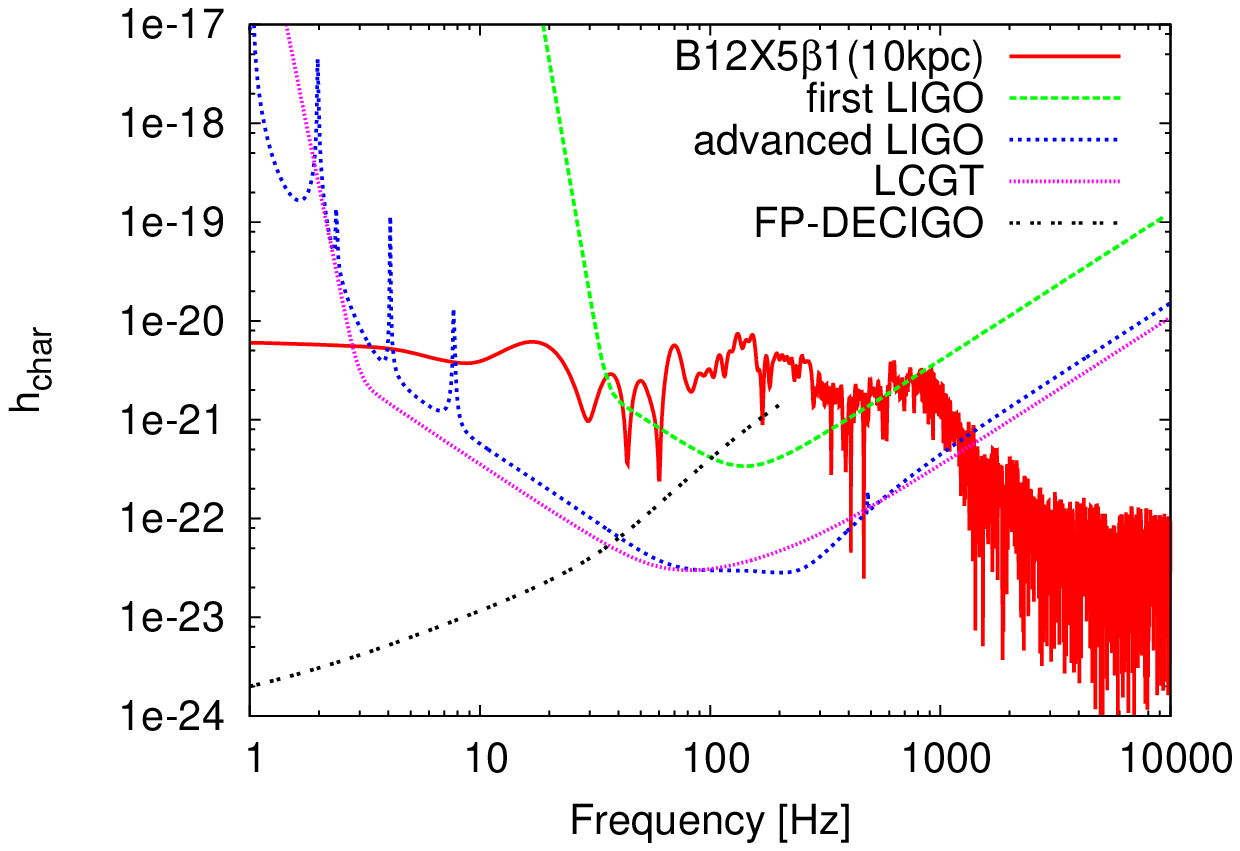}
    \includegraphics[width=.49\linewidth]{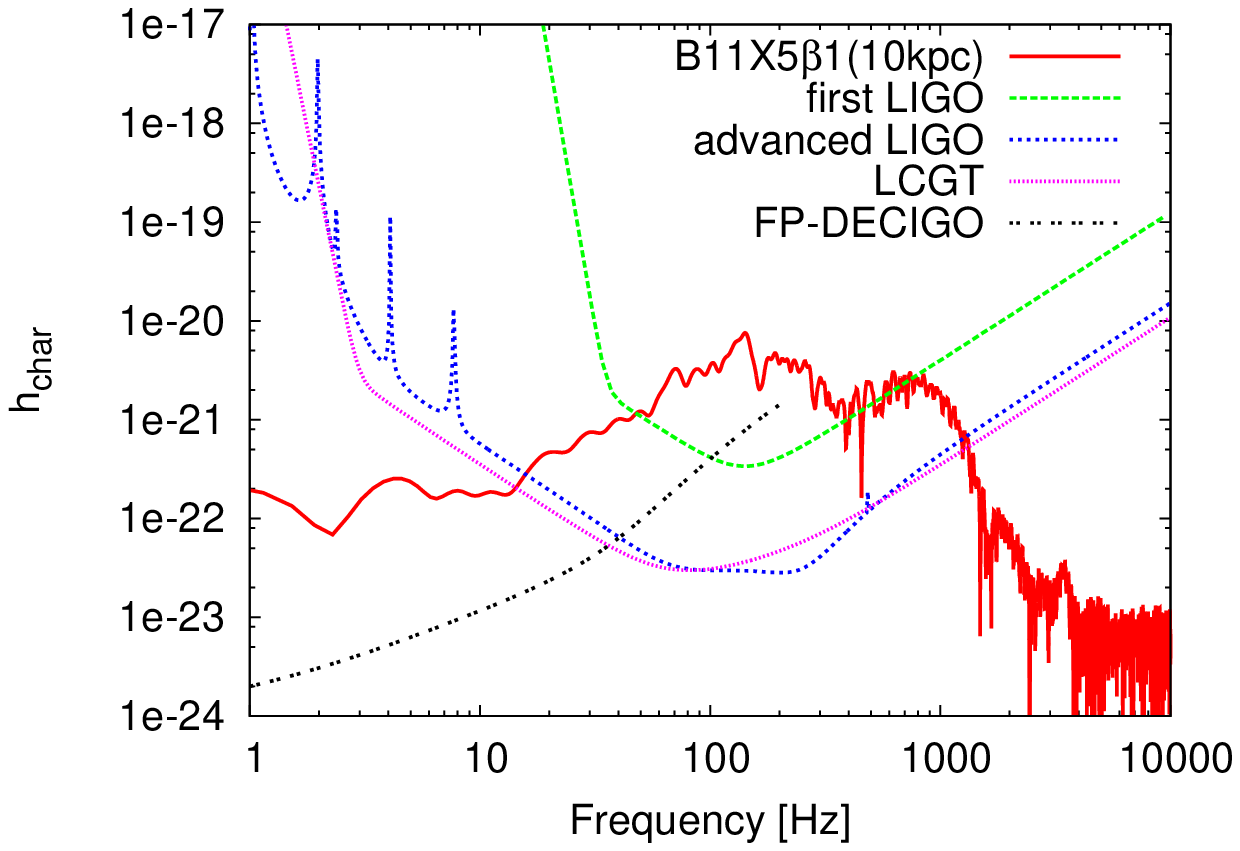}\\
    \caption{
 Gravitational-wave spectrum for representative models with the 
 expected detection limits of the first LIGO \citep{firstligonew}, 
the advanced LIGO \citep{advancedligo}, Large-scale
 Cryogenic Gravitational wave Telescope (LCGT) \citep{lcgt}, and Fabry-Perot type 
 DECIGO \citep{fpdecigo,kudoh}.
It is noted that $h_{\rm char}$ is 
the characteristic gravitational wave strain defined in \citet{flanagan}. 
The supernova is assumed to be located at the distance of 10 kpc.
 This figure is taken from \citet{taki_kota}.}
    \label{fig:spectrum}
\end{figure}

 Figure \ref{fig:spectrum} shows the GW spectra for a pair models of B$12X5\beta1$ 
and  B$11X5\beta1$ like in Figure 
\ref{f1_gwMHD} that with or without the increasing trend. Regardless of 
 the presence (left panel) or absence (right panel) of the increasing trend,
  the peak amplitudes in the spectra are around $1$ kHz. This is because they come 
 from the GWs near at bounce when the MHD effects are minor. 
 On the other hand, the spectra for lower frequency domains (below $\sim 100$ Hz) 
are much larger for the model with the increasing 
 trend (left panel) than without (right panel).
 This reflects a slower temporal variation of the secular drift 
inherent to the increase-type waveforms (e.g., Figure \ref{f1_gwMHD}).
It is true that the GWs in the low frequency domains mentioned above are
 relatively difficult to detect due to seismic noises, but
 a recently proposed future space interferometers like 
Fabry-Perot type DECIGO is designed to be sensitive in the frequency regimes
\citep{fpdecigo,kudoh} (e.g., the black line in Figure \ref{fig:spectrum}). 
 Our results suggest that 
these low-frequency signals, if observed, could be one 
 important messenger of the increase-type waveforms that are likely to be 
associated with MHD explosions exceeding $10^{51}$ erg. 


\clearpage

\subsection{Neutrino Signals}\label{sec3.2}
As of now, SN1987A remains the only astrophysical neutrino source outside 
 of our solar system.  Even though the detected events were just two dozen, these events have 
been studied extensively (yielding $\sim$ 500 papers) and have allowed us to 
have a confidence that our basic picture of the supernova 
 physics is correct (e.g., \citet{Sato-and-Suzuki}, see
 \citet{raffelt12,raffelt_review} for a recent review).
 For a next nearby event, it is almost certain that the flagship detectors
 like Super-Kamiokande and IceCube are able to detect a SN 
"neutrino light curve" with high statistics. For the detectors, 
the horizon to the sources now extends out to about 100 kpc and thus covers our 
galaxy and its satellites.
 A future megaton-class detector reaches as far as Andromeda galaxy 
 at a distance of 780 kpc (e.g., Hyper-Kamiokande, Memphys, and LBNE) 
and large-scale scintillator (e.g., HALO \citep{engel} that is an 
 upgrade of SNO) or liquid-Argon detectors
 (GLACIER \citep{autie} that is an upgrade of ICARUS) 
 will also play an important role (see \citet{scholberg}) 
for a complete list and details).

Before core bounce, neutrinos of different flavors are initially 
 tapped in their relative neutrino spheres with increasing their Fermi energies 
as the central density becomes higher. After bounce, the shock dissociates 
 nuclei into free nucleon, which drastically increases the number of protons,
 leading to a sharp increase in electron capture and production of the prompt
 $\nu_e$ burst. At this {\it neutronization} epoch 
(in the first 10-20 ms postbounce), the largest difference in the light-curves
 among $\nu_e$ and the rest of the neutrino species can be seen (e.g., Figure 1 
 in \citet{Kach}). Since neutrino oscillations take place only when 
 there is a difference in the neutrino fluxes among different species
 (see section 3 \citet{kota06} for more details on the neutrino oscillations),
 the oscillation effects in the neutronization epoch would be strongest. Unfortunately, 
however, they are very hard to measure because the currently-running detectors are 
 primarily sensitive to $\bar{\nu}_e$ signals that are produced by 
the inverse beta process $\bar{\nu}_e + p \rightarrow e^+ + n $. 
 It should be noted that the detection the $\nu_e$ bursts is important 
 for studying not only neutrino oscillations but also for determining the direction 
and distance to the source \citep{Kach}. In a megaton-class \u{C}herenkov
 detector in which gadolinium is added to catch neutrons \citep{gazooks}, 
and large liquid-Argon detectors (like ICARUS and GLACIER cited above)
 are expected to be
powerful $\nu_e$ detectors.

During the subsequent accretion phase (approximately before 1 s postbounce),
 the bounce shock stagnates and matter falls in through the stalled shock
 to the center, releasing gravitational energy that powers neutrino emission.
For relatively well-studied progenitors in the mass range of 
 10.8 to 25 $M_{\odot}$ (e.g., \citet{kitaura,thom03,Kach,huedel,fischer10} 
and references  therein), 
one generally finds $L_{\nu_e} \sim
 L_{\bar{\nu}_e} > L_{\nu_x}$ and $\langle E_{\nu_x} \rangle > 
\langle E_{\bar{\nu}_e} \rangle \gtrsim \langle E_{\nu_e} \rangle$ with $L_{\nu_i}$ and
 $\langle E_{\nu_i} \rangle$ representing luminosity and average neutrino energy for 
 the neutrino species of $i= {\nu_e}, \bar{\nu}_e,~\nu_x$(;$\nu_{\mu}, \nu_{\tau}$ and
 their anti-particles). The accretion phase is followed by the PNS\footnote{(proto-neutron  star)} cooling 
 in which the accretion stops and the PNS settles to become a neutron star after a 
successful revival of the stalled bounce shock into explosions. Since the explosion 
mechanism is still under debate (see discussion in section \ref{sec2}), it is at present 
 difficult to tell the accurate turn-over time. 
But it should be less than 1-2 s postbounce, otherwise the central PNS would collapse 
 into a black hole \citep{oconnor}.
 In the PNS cooling phase, the luminosities of all neutrino species
 become similar $L_{\nu_e} \sim  L_{\bar{\nu}_e} \sim L_{\nu_x}$ (with the 
 energy hierarchy of $\langle E_{\nu_x} \rangle > 
\langle E_{\bar{\nu}_e} \rangle > \langle E_{\nu_e} \rangle$)
 and decrease monotonically with time. Due to the similarity in the 
luminosities and spectra among $\bar{\nu}_e$ and $\nu_x$ and also due to the 
 darkening with time, it is much harder to 
 see flavor oscillation effects by the currently running detectors.

  Neutrinos streaming out of the iron core 
  interact with matter via the Mikheyev-Smirnov-Wolfenstein (MSW)
 effect (e.g., \citet{mikheev1986}) firstly in propagating
 through progenitor envelope and then through the Earth 
before reaching detectors.
 Such effects have been extensively investigated so far 
from various points of view, 
 with a focus such as on the progenitor 
dependence of the early neutrino burst (e.g., \citet{takahashi5, Kach})
and on the Earth matter effects 
(e.g., \cite{lunardini, dighe04}).
The conversion efficiency via the MSW effect depends 
 sharply on density, equivalently electron fraction gradients, thus 
 sensitive to discontinuities produced by 
the passage of supernova shocks.
 Such shock effects 
 have been also extensively studied (e.g., \citet{schirato02,lunardini03,tomas,
das07,lunardini08}, see 
 \citet{kneller} for a recent review).
  The shock passage to the so-called high-resonance regions may be observed
 as a sudden decrease in $\nu_e$ events in the case of normal mass hierarchy 
(or $\bar{\nu}_e$ in the case of inverted mass hierarchy). Such features monitoring
 time evolution of the density profile like a tomography, thus could provide a 
powerful test of the mixing angle and the mass hierarchy 
(e.g., \citet{dighe00,takahashi1,tomas,fogli05}). 

 It is rather recently that the importance of collective flavor 
 oscillations was widely recognized.
Neutrinos streaming out of the neutrino spheres are so dense
 that they provide a large matter effect for each other. 
 The collective effect usually takes place between the neutrino sphere and 
 the mentioned MSW region.
 Regardless of the inherent non-linearity and
 the presence of multi-angle effects, the final outcome for the emergent 
 neutrino flux seems to be converging after a series of 
 extensive study over the past years at least for the 1D models.
 Although the collective effects
 will not be significant to assist the neutrino-driven explosions
 (e.g., \citet{chakra,dasgup2}, see however 
\citet{suwa11,dasgup}), they are predicted to emerge as a distinct observable feature in their energy spectra
\cite[see][for reviews of the rapidly growing research
field and collective references therein]{raff07,duan08,duan10,dasg10}.

 An important lesson from SN1987A is that for explaining the duration of the 
 events, there was not other energy-loss channel 
 but for the ordinary neutrinos in the context of 
the Standard Model of particle physics (e.g., \citet{raffelttext,raffelt_review}).
 The next SN event
  could provide an opportunity to study also a nontrivial property of neutrinos, 
 such as the magnetic dipole moments.
 The resonant spin-flavor conversion has been also studied both analytically 
 (e.g., \cite{schech81,lim1988, akhmedov1988} 
and references therein) and numerically 
(e.g., \cite{ando03}), 
which can transform some of the prompt ${\nu}_e$ burst into $\bar{\nu}_e$
 in highly magnetized supernova envelopes, leading to a huge $\bar{\nu}_e$ burst.
 The mentioned important ingredients related to the flavor conversions due 
 to the MSW effects, collective effects, and the electromagnetic effects 
in the supernova environment have been studied often one by one in each study 
without putting all the 
effects together, possibly in order to highlight the new ingredient. In this sense,
 all the studies mentioned above should be regarded as complimentary towards the precise 
 predictions of SN neutrinos. 
 
 Here it should be mentioned that 
most of those rich phenomenology of SN neutrinos 
have been based on 1D spherically symmetric simulations 
(e.g., \citet{totani98,lunardini03,takahashi5,tomas,Kach,luna08,duan08,huedel,cherry11,
Srdjan}
 and references therein). 
Apart from a simple parametrization to mimic anisotropy and Stochasticity of the shock
 (e.g., \cite{friedland,choubey,reid11}),
 there have been not so many studies focusing how global asymmetry of 
 the explosion dynamics obtained in recent supernova simulations (e.g., section \ref{sec2} and \ref{sec3}) would affect the neutrino oscillations 
 (see, however, \cite{kneller,das08_multi}). 
This is mainly due to the lack of multi-D supernova models, 
which are very computationally expensive to continue the simulations 
till the shock waves propagate outward until they affect neutrino transformations.
 It is only recently that several studies along this line have been reported 
\citep{lund,brandt},
 in which the collective or MSW effects are rendered to be treated in an approximate
 manner. By analyzing the 2D results of a 15 $M_{\odot}$ model by \citet{marek} who 
included one of the best available neutrino transfer approximations (e.g., Table 1), 
 \citet{lund} pointed out that fast time variations caused by convection 
 and SASI lead to significant modulations around a few hundred Hz, which can be 
 visible in IceCube or future megaton-class detectors for the galactic SN source.
 Based on the 2D results of 20 $M_{\odot}$ models by \citet{ott_multi} who
 reported the first 2D multi-angle transport simulations, 
 \citet{brandt} also obtained the similar results. From their rapidly rotating model,
  they also pointed out that a rapid rotation of 
precollapse SN cores imprints strong asymmetries in the neutrino flux \citep{kota03a} 
 as well as in its light curves \citep{jamu,brandt}.

In the following, we briefly summarize our findings \citep{kawa} in which we studied
 exploratory the neutrino oscillations in the context of 
the MHD mechanism.
As mentioned in section \ref{sec3}, it is recently possible for special relativistic 
 simulations to follow the dynamics of MHD explosions continuously, starting 
from the onset of gravitational collapse, through
 core-bounce, the magnetic shock-revival, till the shock-propagation to the stellar 
 surface. Based on our models \citep{taki09}, we calculated numerically 
the neutrino flavor conversion in the highly non-spherical envelope 
through the MSW effect. The neutrino transport was simply 
treated by a leakage scheme
 (e.g., \citet{taki09}). The emergent neutrino spectra was assumed to take a Fermi-Dirac 
type distribution function and the neutrino temperature was estimated to take the 
average matter temperature on the neutrino sphere.
 As explained section 3.0.2,
the density profile along the polar and equatorial direction is very different due to
 the strong explosion anisotropy inherent to the MHD explosions.
 In the following, we pay attention to the anisotropic shock effect
 on the MSW effect. As will be discussed, we could observe 
 a sharp dip in the neutrino event only seen from a polar direction, albeit
 depending on the mass hierarchy and the 
  the mixing angle of $\theta_{13}$. An advantage of the MHD models is that 
  the shock revival can occur much faster after bounce compared to the other proposed
 mechanism. Therefore the neutrino luminosity could remain higher than those in the 
other mechanisms, which could potentially enhance the detectability 
due to the early shock arrival to the resonance region.
For simplicity, the effects of neutrino self-interactions are 
 treated very phenomenologically and the resonant spin-flavor is not considered.
 Even though far from comprehensive in this respect, 
we presented the first discussion how the magneto-driven explosion 
anisotropy has impacts on the emergent neutrino spectra 
and the resulting event number observed by the 
SK for a future Galactic supernova (e.g., \citet{kawa}).

\begin{figure}[htpb]
\plotone{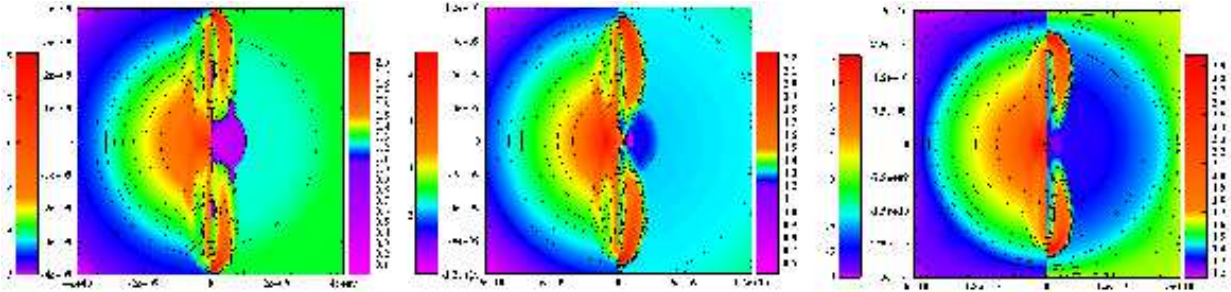} 
 \caption{Snapshots showing MHD explosions of 
  CCSNe at 0.8, 2.0 and 4.3 sec after core bounce from
  left to right (e.g., \citet{taki09}). In each figure, contour of the logarithmic density
  [$\mathrm{g/cm^3}$] (left) and entropy per baryon [$k_{\mathrm{B}}$]
  (right) are shown. The unit of the horizontal and the vertical axis is
  in [cm]. Note that the  difference of the length scale for each panel
  and that the outermost radius of the  progenitor is 
  $\sim 3 \times 10^{10}$ [cm].}
 \label{f1_nu}
 \epsscale{1.0}
\end{figure}

\subsubsection{Possible neutrino signatures from MHD explosions}\label{mag_new}

One prominent feature of the MHD models is a high degree of the explosion asphericity. 
Figure \ref{f1_nu} shows
 several snapshots featuring typical hydrodynamics of the model, 
from near core-bounce (0.8 s, left), during the 
 shock-propagation (2.0 s, middle), till near the shock break-out from the star (4.3 s, 
right), in which time is measured from the epoch of core bounce.
It can be seen that the strong shock propagates outwards with time along 
 the rotational axis.
On the other hand, the density profile hardly changes in the equatorial
direction, which is a generic feature of the MHD explosions.

As well known, the flavor conversion through the pure-matter MSW effect occurs 
in the resonance layer, where the density is 
\begin{eqnarray}
\rho _{\mathrm{res}} \sim  1.4 \times 10^3 \mathrm{g/cm}^3
 \left( \frac{\Delta m^2}{10^{-3} \mathrm{eV}^2}\right)
 \left( \frac{10 \mathrm{MeV}}{E_{\nu}} \right)
 \left( \frac{0.5}{Y_e} \right)
 \cos 2 \theta \label{eq:rho-res}
\end{eqnarray}
where $\Delta m^2$ is the mass squared difference, 
$E_{\nu}$ is the neutrino energy, ${Y_e}$ is the number of electrons per baryon,
and $\theta$ is the mixing angle.
 Since the inner supernova core is too dense to allow MSW resonance conversion, 
we focus on two resonance points in the outer supernova envelope.
One that occurs at higher density is called the H-resonance, and the other, 
 which occurs at lower density, is called the L-resonance. 
$\Delta m^2 $ and $\theta $ correspond to $\Delta m^2_{13}$ and
$\theta_{13}$  at the H-resonance and to $\Delta m^2_{12}$ and
$\theta_{12}$ at the L-resonance.

\begin{figure}[hbtp]
\epsscale{1}
\begin{center}
\plottwo{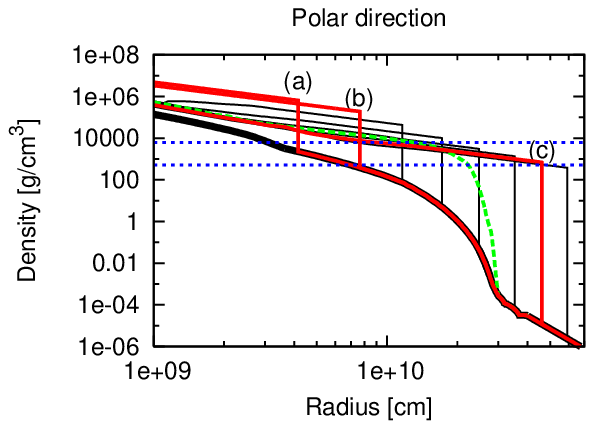}{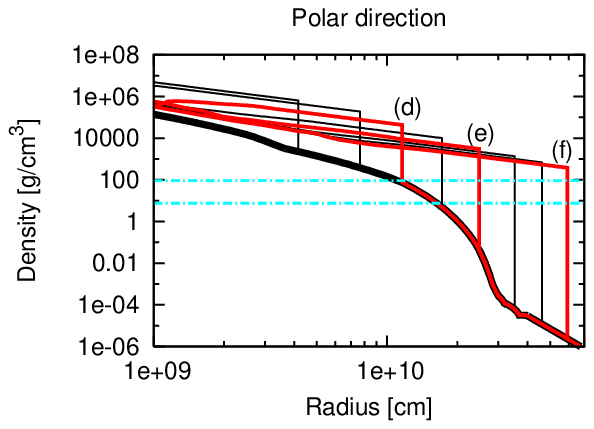}
\caption{The density profile in the polar direction every 0.4 s as a
  function of radius. (a), (b) and (c) correspond to 0.4, 0.8 and 2.8
  s, respectively.   
  The horizontal blue lines show approximately density
  of the H-resonance for different neutrino energies which are 5 (above) and 60
  MeV (below), respectively. The green line is an original density
  profile for 2.0 s. Same as the left panel, but for 1.2 (d), 2.0(e), and 3.2 (f)
  s, respectively. Note that a sharpness of the shock along the polar direction 
is modified from the one in the MHD simulations (green line showing
 an original density profile for 2.0 s), which 
is made to be inevitably blunt due to the employed numerical scheme using 
 an artificial viscosity to capture shocks (taken from \citet{kawa}). 
}\label{f2_nu}
\end{center}
\end{figure}

Figure \ref{f2_nu} are evolutions of the density  
profiles in the polar direction every 0.4 s
as a function of radius.
Above and below horizontal lines show approximately density
of the resonance for different neutrino energies which are 5 and 60
MeV, respectively. 
Blue lines (left panel) 
show the range of the density of the H-resonance, and sky-blue lines (right panel)
 show that of the L-resonance. 
Along the polar axis, the shock wave reaches to the H-resonance,
$\sim O(10^3$)g/cm$^3$ at $\sim 0.5 $ s, and the L-resonance, $\sim O(1)$g/cm$^3$
 at $\sim 1.2$ s. It should be noted that 
those timescales are very early in comparison with the ones predicted in the 
neutrino-driven explosion models, typically $\sim $ 5 s and $\sim$ 
15 s for the H- and L- resonances,
 respectively (e.g., \cite{fogli05,tomas}). 
This arises from the fact that the MHD explosion 
is triggered promptly after core bounce, which is in sharp contrast to the 
neutrino-driven {\it delayed} explosion models (\cite{fogli05,tomas}). 
The progenitor of the MHD 
models, possibly linked to long-duration gamma-ray bursts, is more compact 
due to a chemically homogeneous evolution \citep{yoon,woos06},
 which is also the reason for the early shock-arrival to the resonance regions.


\begin{figure}[htbp]
    \centering
    \includegraphics[width=.49\linewidth]{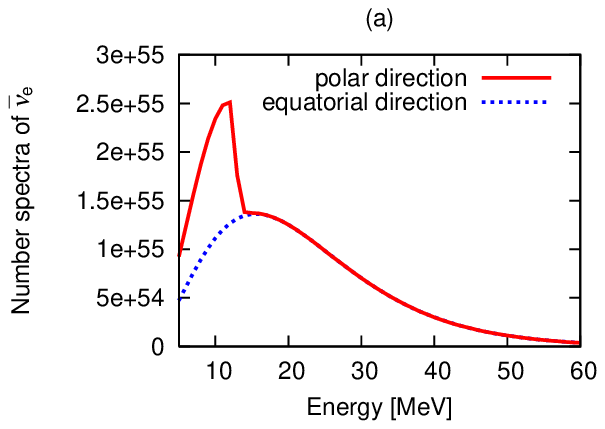}
    \includegraphics[width=.49\linewidth]{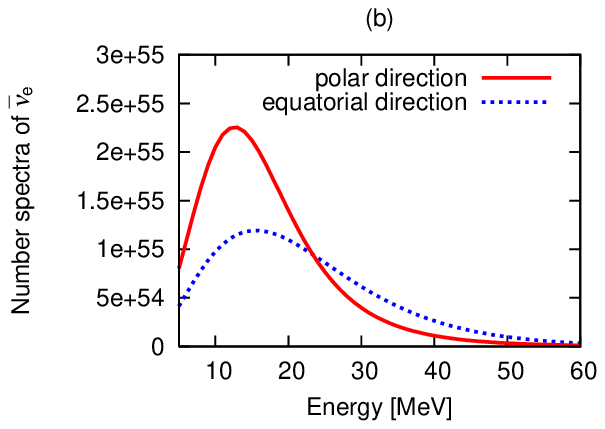}
    \includegraphics[width=.49\linewidth]{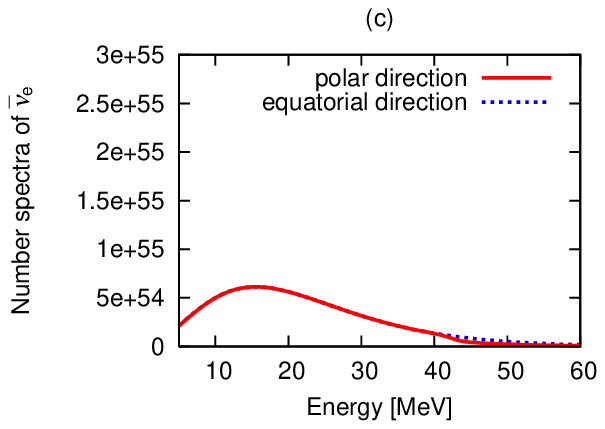}
    \caption{\label{fig:spctreb}$\bar{\nu}_e$ spectra at the surface of the star
  in the case of (a), (b) and (c).  
  We assume the inverted mass hierarchy and $\sin ^2 2
 \theta_{13}=10^{-3}$. The solid lines (red lines) and the dotted
  lines (blue lines) are the spectra of polar direction and equatorial
 direction, respectively.  Note in this figure that the neutrino luminosity and 
 spectra are modeled to be the same as those obtained in a 1D full-scale 
 numerical simulation by the Lawrence Livermore group \citep{wils85}.
 These figures are taken from \citet{kawa}.}
    \label{f3_nu}
\end{figure}

Figure \ref{f3_nu} is the energy spectra of $\bar{\nu}_e$ 
for (a), (b) and (c) in the case of inverted mass hierarchy.
The solid lines (red lines) and the  dotted lines (blue lines) are
the spectra of the polar direction and the equatorial direction,
respectively.
At the panel (a), an enhancement of the low-energy side of the polar direction
 is seen, which is as a result of the supernova shock reaching to the H-resonance
 region. The survival probability of $\bar{\nu}_e$ can remain non-zero
 when the steep decline of the density at the shock front changes the
resonance into non-adiabatic (see \citet{kawa} for more details).
As the shock propagates from the high density region to low energy region, 
the shock effect transits from low-energy side to high-energy side in 
spectra, because the density at the resonance point, $\rho_{\mathrm{res}}$, is 
proportional to $E_{\nu}^{-1}$. 
In fact, the energy spectrum of the polar direction becomes softer 
in the high-energy side than for the equatorial direction 
 as shown in Figure \ref{f3_nu} (b). 

\begin{figure}[hbtp]
\epsscale{0.8}
\begin{center}
\plottwo{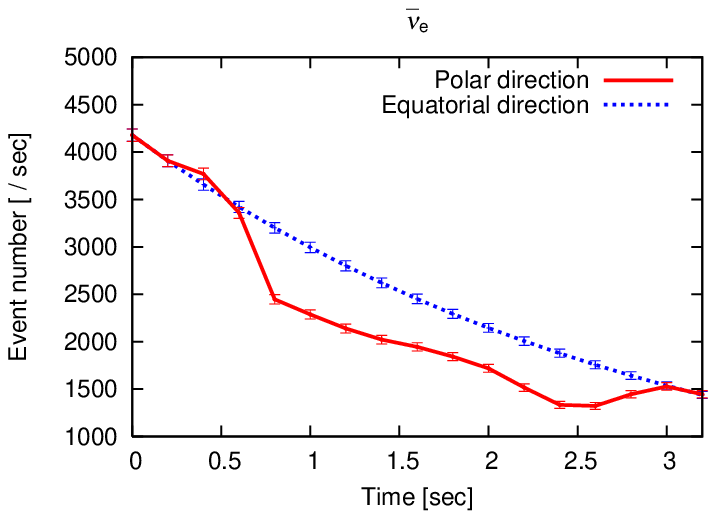}{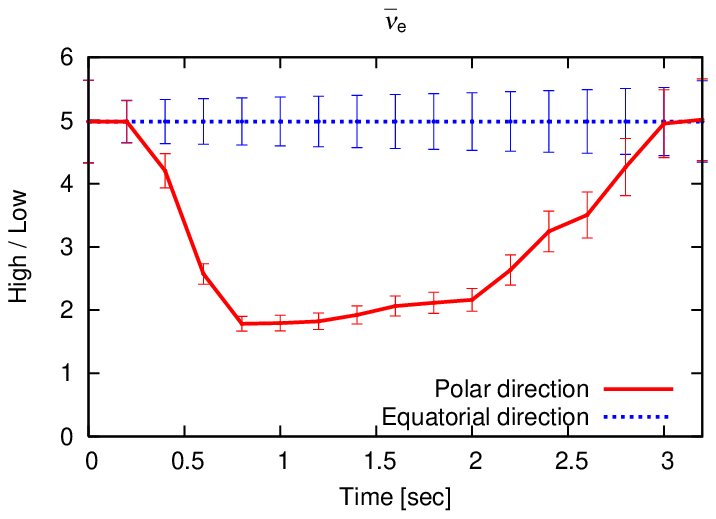}
\caption{Same as Figure \ref{f3_nu} but for 
the expected event number of $\bar{\nu}_e$ (left panel) in the
  SK as a function of the time.
  The solid line (red line) and the the dotted line (blue line) is 
 for the polar and the equatorial direction, respectively.
  The supernova is assumed to be located at the distance of 10 kpc. 
The error bars represent the 1$\sigma$ statistical errors only.
 Right panel is for the ratio of high- to low-energy events which 
 can be a useful quantity to characterize the change (dip in the 
 right panel) of neutrino signals by the shock effect 
(see \citet{kawa} for details).}\label{f4_nu}
\end{center}
\end{figure}

 Figure \ref{f4_nu} shows the time evolution of the event number in
  SK, where the solid line (red line) and the dotted line (blue line) are 
 for the polar and the equatorial direction, respectively. 
 The shock effect is clearly seen.
The event number of the polar direction shows a steep decrease,
 marking the shock passage to the H-resonance layer.
 The change of the event number by the shock passage 
 is about 36\% of the event number without the shock.
 Since the expected events are $\sim 2500$ at the sudden decrease 
(Figure \ref{f4_nu}), it seems to be quite possible to identify such a feature
 by the SK class detectors. 
 Such a large number imprinting the shock effect, 
 possibly up to two-orders-of magnitudes larger than the ones 
 predicted in the neutrino-driven explosions (e.g., Fig 11 in \cite{tomas}),
 is thanks to the mentioned early shock-arrival to the resonance layer, 
 peculiar for the MHD explosions. 

Finally we exploratory discuss possible 
 impacts of the spectral swap between $\bar{\nu}_e$ and $\nu_x$, which is one 
of a probable outcome of the collective neutrino oscillations \citep{fogli07}.
In our model, the original neutrino flux of $\bar{\nu}_e$ is larger than that 
 of $\nu_x$ for the low-energy side, which is vice versa for the high-energy side.
Due to the spectral swap, the above inequality sign reverses (see
 \citet{kawa} for more detail). As a result,
 the event number in the polar direction is expected to increase
by the influence of the shock.
To draw a robust conclusion to the self-interaction effects, one needs to 
 perform a more sophisticated analysis such as the multi-angle approach 
 (e.g., \cite{duan10}), which is a major undertaking.
\clearpage
\section{Summary and Discussion}

\begin{table}[htpb]
\plotone{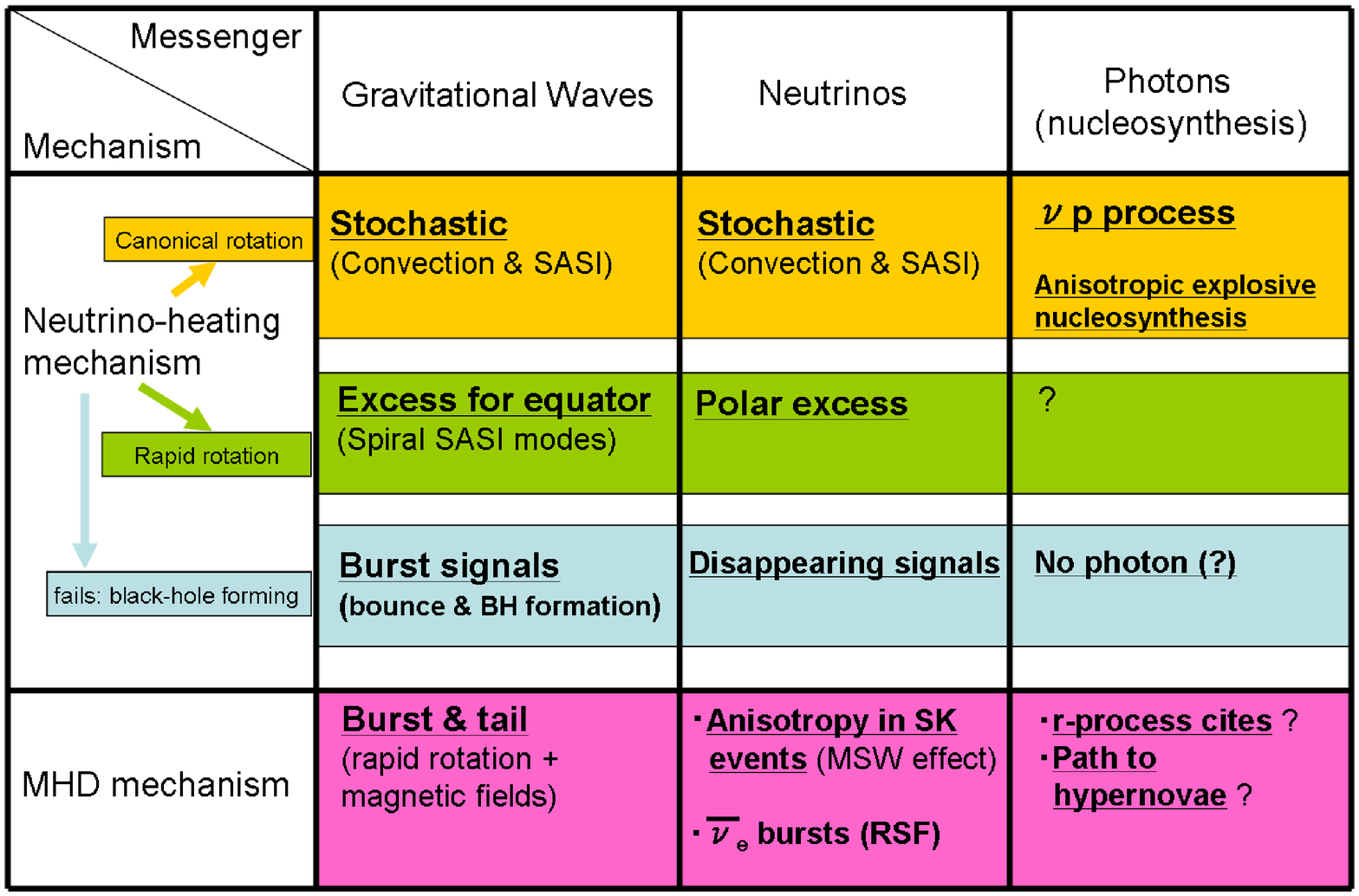} 
 \caption{A summary illustrating the candidate supernova mechanisms (vertical direction
 in the table) and their expected features of the emergent 
multimessengers (horizontal direction).
 Regarding the photon messengers, possible signatures associated with nucleosynthesis 
 are only partly presented (see section 1 for a number of important findings regarding 
 the light-curve and spectra modeling).}
 \label{table1}
 \epsscale{0.8}
\end{table}

The aim of writing this article was to provide an overview of what we
currently know about the explosion mechanism, neutrinos, 
gravitational waves, and explosive nucleosynthesis in CCSNe 
 to bridge theory and observation through multimessenger astronomy.
 Not to inflate the volume, we had to limit 
 ourselves to focus primarily on the findings based on our numerical studies, 
which are thus 
limited to GWs (section \ref{sec2.1}) and explosive nucleosynthesis 
(section \ref{sec2.2})
 in the neutrino-heating mechanism (section \ref{sec2}),
 and to GWs (section \ref{sec3.1}) and neutrino 
signals (section \ref{sec3.2})
 in the MHD mechanism (section \ref{sec3}).
 Apparently our current presentation is weak on the photon side mainly because 
 there is a big gap to connect the outcomes obtained in
 the multi-D neutrino-radiation-hydrodynamic
 simulations covering only the very centre of the star,
 to the light-curve and spectra modeling that requires 
 multi-D photon-radiation-hydrodynamic modeling (far)
 after the shock-breakout of the star. The gap is not only the matter of 
 physical scale, but also the matter of numerical difficulty.
 Note that table \ref{table1} is not intended to show an overview covering 
 everything concerning the SN multi-messengers, which is far beyond the scope 
 of this review.
 We boldly present the incomplete table here, hoping that the shortage 
  might give momentum to theorists for fixing the problems and for gaining 
  much more comprehensive multimessenger perspectives for the future.

In the neutrino-heating mechanism, 
stellar rotation holds a key importance to 
 characterize the SN multimessengers. The whole story may be summarized as follows. 
If a precollapse iron core has a ``canonical'' rotation rate as predicted by recent 
stellar evolution calculations \citep{maeder00,hege05}, 
collapse-dynamics before bounce proceeds spherically
 and the postbounce structures interior to the PNS are essentially spherical. Typically 
 later than $\sim 100$ ms after bounce, the bounce shock turns to a standing accretion 
shock, and the activities of convective overturns and SASI 
become vigorous with time 
 behind the standing shock. Since anisotropies of the neutrino flux and matter motions 
 in the postbounce phase are 
 governed by the non-linear hydrodynamics, GWs emitted in this epoch 
 also change stochastically with time (indicated by "Stochastic" in the table, 
 e.g., section \ref{sec3.1} for more details). 
For detecting these signals for a Galactic CCSN event with a good 
signal-to-noise ratio, we need next generation detectors 
such as the advanced LIGO, LCGT, and the Fabry-Perot(FP)-type DECIGO.
 Neutrino signals at the non-linear epoch also change stochastically with time,
 which is expected to be detected by currently running 
 detector such as by IceCube and SK \citep{brandt,marek_gw,lund} for the 
 galactic event.
 It should be also mentioned that 
detailed neutrino transport including the most up-to-date 
interaction rates should be accurately implemented in the multi-D simulations to make 
 a reliable prediction of the neutrino signals (e.g., \citet{huedel,fischer10}).
 When material behind the stalled shock successfully absorbs enough neutrino energy 
to be gravitationally unbound from the iron core, 
 the $\nu$p process is expected to successfully explain some light 
proton-rich nuclei in the neutrino-driven wind phase
 \citep{pruet,froehlich,wanajo}. When the revived supernova 
shock (successfully) attains the kinetic energy as energetic as $10^{51}$ erg, 
explosive nucleosynthesis that proceeds in the SASI-aided low-mode explosions
  is expected to partly explain the solar abundance yields as well as the observed
 non-uniform morphology of the synthesized elements (e.g., section \ref{exp}).

 One important notice here is that explosion energies 
 obtained in the state-of-the-art multi-D simulations are typically underpowered 
by one or two orders-of-magnitudes to explain the canonical supernova kinetic energy 
($\sim 10^{51}$ erg, e.g., Table 1).
Moreover, the softer EOS, such as of the
\citet{latt91} (LS) EOS with an incompressibility at nuclear
densities, $K$, of 180 MeV, is employed in those simulations (e.g., in multi-D 
simulations of the MPA, Oak Ridge+, and Tokyo+ groups in Table 1).  On top
of a striking evidence that favors a stiffer EOS based on the
nuclear experimental data ($K=240 \pm 20$ MeV, \citet{shlo06}), 
the soft EOS may not account
for the recently observed massive neutron star of $\sim 2 M_{\odot}$
\citep{demo10} 
\citep[see the maximum mass for the LS180 EOS in][]{oconnor,kiuc08}. 
With a stiffer EOS, the explosion energy may be
even lower as inferred from \citet{marek} who did not obtain the
neutrino-driven explosion for their 
model with $K=263$ MeV\footnote{On the other hand,
 they obtained 2D explosions for Shen EOS ($K=281$MeV) (H-T. Janka, private 
communication).}.  What is
then missing furthermore ?  We may get the answer by going to 3D
simulations (section 2.1)
 or by taking into account new ingredients,
such as exotic physics in the core of the protoneutron star
\citep{sage09,tobi11}, viscous heating by the magnetorotational instability
 (section \ref{mri}), or energy dissipation via Alfv\'en waves
 \citep{suzu08}. In addition to the 3D effects,
 GR is expected to help the onset of multi-D neutrino-driven explosions
 \citep{bernhard11,kuroda12}\footnote{e.g.,
 \citet{bernhard12} in 2D simulations but with detailed neutrino transport or
 \citet{kuroda12} in 3D simulations
 but with a very approximate neutrino transport}.
 At present, the most up-to-date neutrino transport 
 code can treat the multi-energy and multi-angle 
transport in 2D \citep{ott_multi} and even in 3D simulations \citep{Sumi11} but 
 only limited to the Newtonian simulations. 
 Extrapolating a current growth rate of supercomputing power, 
the dream simulation (i.e. 6D simulations with full Boltzmann neutrino transport 
 in full GR simulations) is likely to be practicable not in the distant future 
(in our life-time!) presumably by using the exascale platforms.

In addition to these numerical advancements,
 the physical understanding of the supernova theory via 
 analytical studies is also in a steady progress.
What determines the saturation levels of 
 SASI ?  A careful analysis on the parasitic instabilities has been 
 reported to answer this question \citep{jerome}. What determines mode couplings between 
 small-scale convection eddies and large-scale SASI modes ?  To apply the theory
 of turbulence \citep{murphy_meakin} should put a milestone to address this question.
 Besides the two representative EOSs of Lattimer-Swesty and Shen, new sets of 
 EOSs have been recently reported \citep{hempel,furusawa,gshen}. Some of them
  will enable SN modellers to go beyond a single nucleus approximation,
 which is an important improvement for an accurate description of 
 neutrino-nucleus interaction.

Rotation, albeit depending on its strength, can give a special direction 
(i.e., spin axis) in the supernova cores. As discussed in section \ref{break},
 stochastic nature of the GWs becomes weak in the presence of rotation as a 
 result of the spiral SASI\footnote{It should be noted
 that the feature can be seen even in a slowly
  rotating progenitor that agrees with recent stellar evolution calculation 
\citep{maeder00,hege05} (see section \ref{break} for more details).}(indicated by "excess for 
 equator'' in Table 2). 
 Although it is not easy to detect these low-frequency
   GWs from anisotropic neutrino emission 
 by currently running laser interferometers,  
   a recently proposed space-based interferometers (like FP-type DECIGO)
 would permit detection for a galactic event. 
 Contributed by the neutrino GWs in the lower frequency domain,
 the total GW spectrum is expected to become rather flat over a broad frequency range
 below $\sim 1$ kHz. This GW feature obtained in the context of the 
SASI-aided neutrino-driven mechanism is different from the one  in the 
other candidate supernova mechanisms, such as the MHD mechanism (section \ref{sec3.1})
 and the acoustic mechanism \citep{ott_new}, thus could 
provide an important probe into the explosion mechanism. 
 Rapid rotation induces a polar 
excess of neutrino emission with a harder spectrum than on the equator
 \citep{jamu,kota03a,brandt} (indicated by "polar excess" in the table). This 
 should be also the case of the MHD mechanism, which deserves
 further investigation for a quantitative discussion (symbolized 
 by Polar excess (?) in the table).
These (rotation-induced) neutrino signatures, if observed, will carry an important 
 information 
about the angular momentum 
profiles hidden deep inside massive stellar cores.
 Concerning the $\nu$p process and explosive synthesis in rapidly rotating cores
(as well as in the MHD explosions),
 there still remains a vast virgin territory for further investigation
 (symbolized by "$\nu$p process (?)" in the table).

 If the neutrino-heating mechanism (or some other mechanisms)
 fails to blow up massive stars,
 central PNSs collapse to BHs. Recent GR simulations by \citet{ott2011}  
  show that the significant GW emission is associated at the moment of
 the BH formation, which can be 
a promising target of the advanced LIGO for a galactic source. 
As pointed out by \citet{sumi}, the disappearing 
 neutrino signals that mark the epoch of BH formations also can be a target 
 of SK and IceCube \citep{icecube11}.
 When quarks and hyperons appear in the postbounce core,
  a neutrino burst produced by the sudden EOS softening and 
 by the subsequent rebounce \citep{sage09} is likely to be detected by IceCube 
for a galactic event \citep{icecube11}. The interval between core-bounce
 and the BH formation 
  depends on the details of exotic physics in the super-dense PNS cores \citep{naka1,naka2}. All of these observational signatures should provide an important probe 
 into the so-called dense QCD region in the QCD phase diagram (e.g., \citet{hatsuda} 
 for recent review) to which lattice 
 calculations are hardly accessible at present. 
  Concerning photons, no optical outbursts are expected
 in the BH-forming SNe (indicated by "no photon (?)" in Table 2), if not for 
rapid rotation and strong magnetic fields prior to core-collapse.

 If a precollapse core rotates enough rapidly (typically initial rotation period
 less than 4 s) with strong magnetic fields (higher than $\sim 10^{11}$ G), 
  the MHD mechanism can produce bipolar explosions along the rotational axis 
 predominantly driven by the field wrapping processes (section
 \ref{wrap}). If the MRI can be sufficiently resolved in global simulations, 
the MRI would exponentially amplify the initial magnetic fields to a dynamically important strength 
 within several rotational periods (section \ref{mri} and \ref{local}). 
The next-generation supercomputing resources are 
needed again to see the outcome. If such simulations would be executable, 
the MHD outflows will be produced even for more weakly magnetized 
cores than currently predicted. 
 The GW signals in the MHD explosions are characterized by a 
burst-like bounce signal plus a secularly growing tail (section \ref{sec3.1}, 
indicated by ``burst and tail'' in the table). 
 Similar to the neutrino GWs in the neutrino-driven mechanism, 
  a future detector (like FP-type DECIGO) is again necessary for detecting
 the low-frequency tail component. As discussed in section
 \ref{mag_new}, the pole to equator anisotropy of the shock propagation 
in the MHD explosions affect
 the neutrino signals through the MSW effects.
  The anisotropic shock passage to the H-resonance regions leads to 
 a sudden decrease in the SK events ($\sim 2500$ events for a galactic source),
 which is quite possible to identify by the SK-class detectors (indicated by 
 "anisotropy in SK events" in Table 2). Since the MHD mechanism has such distinct
 signatures, a planned joint analysis of neutrino and GW 
data \citep{leonor} would be potentially very powerful to 
 tell the MHD mechanism from the other candidate mechanisms.
 If the resonant spin-flavor conversion (indicated as RSF in Table 2) 
  occurs in the highly magnetized core,
 the neutronization burst of $\nu_e$ converts to 
 that of $\bar{\nu}_e$, which is thus expected to be a 
probe into the magnetic moment of neutrinos (e.g., \cite{ando03}).
 Finally the MHD explosions are considered to be a possible $r$-process cite
\citep{nishi06,fujimoto}\footnote{See 
collective references in \citet{wanajo} for other plausible $r$-process cites.}.
 This is because the mass ejection
  from the iron core occurs in much shorter timescales compared to 
the delayed neutrino-heating mechanism, so that the ejecta can stay 
 neutron rich before it becomes proton-rich via neutrino capture reactions. 
 Most of these possibilities regarding the MHD mechanism have been proposed so far 
in simulations with 
a crude treatment of neutrino transport, which is now rather easily
 re-examined by the state-of-the-art multi-D simulations (if the code is MHD).

Finally, what is about the story if the MHD mechanism fails ? (indicated by "path 
 to hypernovae in Table 2).  The rapidly rotating PNS collapses into a BH, probably 
 leading to the formation of accretion disk around the BH. 
Neutrinos emitted from the accretion disk heat matter in the polar funnel region
 to launch outflows 
 or strong magnetic fields in the
cores of the order of $10^{15}$ G also play an active role both in driving 
the magneto-driven jets and in extracting a significant amount of
energy from the BH (e.g.,
\cite{whee00,thom04,uzde07a} and see references therein). 
 This picture, often referred as collapsar (e.g., \citet{macf99,macf01}), has been
 the working hypothesis as a central engine of long-duration GRBs for 
these 10 years (see references in \citet{dai98,thom04,uzde07a,bucci,metzger}
 for other candidate mechanism including magnetar models).
 However it is still controversial whether the generation of 
 the relativistic outflows predominantly proceeds via the neutrino-heating mechanism
 or the MHD mechanism. In contrast to a number of findings illustrated in Table 2,  
 much little things are known about the BH-forming supernovae and also about
 the collapsar.
 This is mainly because the requirement for making a {\it realistic}
 numerical modeling is very computationally expensive, which (at least)
 necessitates the multi-D
 MHD simulations not only with GR for 
handling the BH formation (thus full GR), but also with the multi-angle neutrino 
transfer for treating highly anisotropic neutrino radiation from the accretion disk.
To get a unified picture of massive stellar death, we need to draw a 
 schematic picture (like Table 2) also in the case of the BH-forming supernovae.
 We anticipate that our long-lasting focus
 (and experience) on the explosion dynamics of canonical CCSNe will be readily
 applicable to clarifying the origin of the gigantic explosion energy of hypernovae
 and also unraveling (ultimately) the central engine of long-duration GRBs. This  
 should provide yet another grand challenge in computational astrophysics 
 in the next decades.

 As repeatedly mentioned so far, all the numerical results in this article should 
 be tested by the next-generation calculations by which more sophistication 
 is made not only in the treatment of multi-D radiation transport (both of
neutrino and photon), 
but also in multi-D hydrodynamics including stellar rotation and magnetic fields in 
 full GR. From an optimistic point of view, our understanding on 
 every issue raised in this article
 can progress in a step-by-step manner at the same pace 
as our available computational resources will be growing bigger and bigger 
from now on. Since 2009, several neutrino detectors form 
 the Supernova Early Warning Systems (SNEWS) to broadcast the alert to astronomers
 to let them know the arrival of neutrinos \citep{anoto}. Currently, Super-Kamiokande,
 LVD, Borexino, and IceCube contributes to the SNEWS, with a number of other neutrino
 and GW detectors planning to join in the near future. This is a very 
 encouraging news towards the high-precision multi-messenger astronomy. The interplay 
between the detailed numerical modeling, the advancing supercomputing resources, 
 and the multi-messenger astronomy, will remain a central issue for advancing 
our understanding of the theory of massive stellar core-collapse for the
 future. A documentary film recording our endeavours to make practicable 
 the "multimessenger astronomy of CCSNe"
 seems not to show "{\it fin}"  immediately and is becoming even longer as far as  
  the three components evolve with time. To raise the edifice, it 
 is becoming increasingly important to bring forward a
 world-wide, multi-disciplinary 
collaboration among different research groups. We believe that such an approach
  would provide the shortest cut to get a deeper understanding 
 of a number of unsettled and exciting issues that we were only able to touch 
in this article.

\acknowledgements{K.K. is thankful to K. Sato for continuing 
encouragements. He also wishes to thank his collaborators, S. Yamada, M. Liebend\"orfer, 
 N. Ohnishi, K. Sumiyoshi, K. Nakamura, T. Kuroda, 
M. Hashimoto, S. Harikae, N. Yasutake, N. Nishimura, 
K.Shaku, and H. Suzuki. We would like 
 to acknowledge helpful exchanges and stimulating discussions with E. M\"uller, 
T. Foglizzo, H.-T. Janka, C.D. Ott, C. Fryer, J. Novak, S. Ando, P. Cerd\'a-Dur\'an, 
M. Obergaulinger, J. Murphy, R. Fern\'andez, E. O'Connor, M. Shibata, T. Font, and 
 J.M. Iba\~nez. We are also grateful to our anonymous referees
 who gave us a number of valuable comments to enhance the quality of this article.
Numerical computations were carried out in part on XT4 and 
general common use computer system at the center for Computational Astrophysics, CfCA, 
the National Astronomical Observatory of Japan.  This 
study was supported in part by the Grants-in-Aid for the Scientific Research 
from the Ministry of Education, Science and Culture of Japan (Nos. 19540309, 20740150,
 23540323, and 23340069) and by HPCI Strategic Program of Japanese MEXT.}


\end{document}